\documentclass[aps,prb,twocolumn,superscriptaddress]{revtex4-1}
\usepackage{graphicx}
\usepackage{amsmath,amssymb}
\usepackage{multirow}
\usepackage{dsfont}

\usepackage{color}

\newcommand{\e}{\epsilon}

\newcommand{\minisec}[1]{\noindent{\textit{#1 ---}}}

\begin{document}

\title{Topology vs. Anderson localization: non-perturbative solutions in one dimension
}

\author{Alexander Altland}

\affiliation{Institut f\"ur Theoretische Physik, Universit\"at zu K\"oln,
Z\"ulpicher Stra\ss e 77, 50937 K\"oln, Germany}

\author{Dmitry Bagrets}

\affiliation{Institut f\"ur Theoretische Physik, Universit\"at zu K\"oln,
Z\"ulpicher Stra\ss e 77, 50937 K\"oln, Germany}

\author{Alex Kamenev}

\affiliation{W. I. Fine Theoretical Physics Institute and School of Physics and Astronomy, University
of Minnesota, Minneapolis, MN 55455, USA}

\date{\today}

\begin{abstract}
We present an analytic theory of quantum criticality in quasi one-dimensional topological Anderson insulators.  We describe these systems  
in terms of two parameters $(g,\chi)$ representing localization and topological properties, respectively. 
Certain critical values of  $\chi$ (half-integer for $\Bbb{Z}$ classes, or zero for $\Bbb{Z}_2$ classes)  define phase boundaries between distinct
topological sectors. Upon increasing system size, the two parameters exhibit
flow similar to the celebrated two parameter flow of the integer quantum Hall insulator. 
However, unlike the quantum Hall system,
an exact analytical description of the entire phase diagram can be given in terms of the transfer-matrix solution of corresponding supersymmetric non-linear sigma-models.  In $\Bbb{Z}_2$ classes we uncover 
a hidden supersymmetry, present at the quantum critical point.  
\end{abstract}

\maketitle

\section{Introduction}

The discovery of topologically non-trivial band insulators has defined a whole
new research field addressing the physical properties of bulk insulating
matter. What distinguishes a topological insulator~\footnote{We here consider
a topological superconductor as a thermal insulator, i.e. our usage of the
term `insulator' encompasses both conventional insulators, and
superconductors.} (tI) from its topologically trivial siblings is the presence of
non-vanishing topological invariants characterizing its band structure. While
these indices are not visible in the system's band structure, their presence
shows via the formation of gapless boundary states -- the celebrated bulk-boundary 
correspondence. In the bulk, the indices can be obtained via
homotopic constructions based on the functional dependence of the system
Hamiltonian (or its ground state) on the quasi-momenta of the Brillouin zone \cite{Hasan2010,Bernevig2013}.

It is a widespread view that individual topological phases owe their stability 
to the existence of bulk band-gaps. A topological number may change via a gap closure  which represents a topological
phase transition point and is accompanied by the
transient formation of a Dirac like metallic point in the Brillouin zone.
However, as long as a bulk gap remains open, weak system imperfections
('perturbations weak enough to leave the gap intact') will not compromise the
topological number. In particular, tI is believed to be robust against
the presence of a ``weak''  disorder. Indeed, one may argue that
the adiabatic turning on of a small concentration of impurities in a system
characterized by an integer topological invariant does not have the capacity
to change that invariant. It is due to arguments of this sort that
disorder is often believed to be an inevitable but largely inconsequential
perturbation of bulk topological matter.

However, on second consideration it quickly becomes evident that disordering
does more to a topological insulator than one might have thought. The presence
of impurities compromises band gaps via the formation of mid-gap states. In this way, even
a weak disorder generates Lifshitz tails in the average density of states which leak into the gap region, at
stronger disorder the band insulator crosses over into a gapless regime, which
in low dimensions $d=1,2$ will in general be insulating due to Anderson
localization. In this context, the notion of `weak' and `strong' disorder lack
a clear definition. Moreover, close to a transition point of the clean system, where the band gap is small, even very small impurity concentrations suffice to close gaps, which tells us that
disorder will necessarily interfere with the topological quantum criticality
of the system. As concerns the integrity of topological phases, one may argue
that for a given realization each system is still characterized by an
integer invariant $n$ (for it must be possible to adiabatically turn off the
disorder and in this way adiabatically connect to a clean anchor point.)
However, that number will depend on the chosen impurity configuration. In
other words, the topological number becomes a statistically distributed variable with
generally non-integer configurational mean, $\chi \equiv \left\langle n
\right\rangle$. In the vicinity of transition regions, the distribution of $n$
becomes wide, and one may anticipate  scaling behavior of $\chi$. We finally
note that  a theory addressing non-translationally  invariant environments
should arguably not be based on the standard momentum space/homotopy
constructions of invariants\footnote{As a matter of principle, the momentum
space description can be maintained at the expense of extending the unit cell
from atomistic scales, $a$, to the system size, $L$ (a disordered system is
periodic in its own size.) However, the
price to be payed is that the topological information is now encoded in the
configuration dependent structure of $\mathcal{O}((L/a)^d)$ bands. While the
multi-band framework may still be accessible by numerical means, it is less
suited for analytical theory building.}.  Rather, one would like to start out
from a more real space oriented identification of topological sectors.

\begin{figure}[t]
\includegraphics[width=8.6cm]{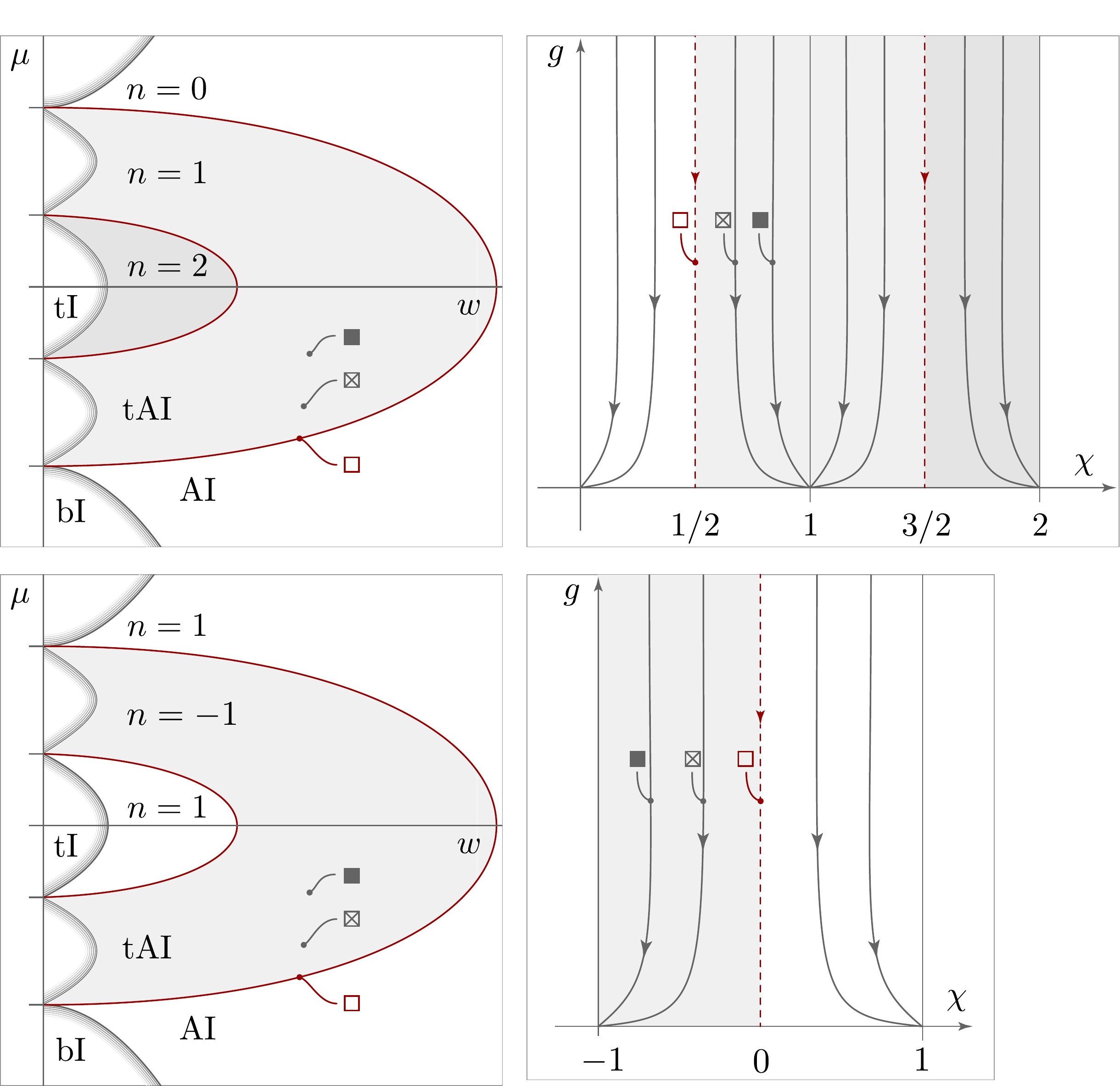}
\caption{ Schematic phase diagram of topological insulators. The $\Bbb{Z}$
(top) or $\Bbb{Z}_2$ (bottom) valued topological number of a clean topological insulator
can be controlled by a parameter $\mu$ (e.g. chemical potential, magnetic
field, etc.).  At the transition points separating distinct phases, band gaps
close. Disorder,   characterized by its strength 
$w$, induces a crossover from a clean band to an Anderson insulator (shaded
lines). The  amount of disorder required to close the band gap vanishes
at the transition points. The transition points themselves become end points
of transition \emph{lines} in the $(\mu,w)$ phase plane. Driving the system
through one of these lines via a parameter change implies IQH type
transition with a divergent localization length. The right panels
show the flow of the average topological number, $\chi$, and the conductance,
$g$ upon increasing the system size, $L$, starting from some non-universal bare values $(\tilde g,\tilde \chi)$, defined at a scale of the order of the mean-free path. In the $L\to \infty$ limit, the
insulating, $g=0$, and self averaging, $\chi=n$, Anderson topological
insulator configurations are generically approached. The critical surfaces
separating these regions are characterized by half integer $\chi=n+1/2$
($\Bbb{Z}$) or vanishing $\chi=0$ ($\Bbb{Z}_2$) values of the average topological number. }
\label{fig:PhasePortrait}
\end{figure}

The blueprint of a strategy to describe this situation can be obtained from
insights made long ago in connection with the integer quantum Hall effect
(IQH). In the absence of disorder, the IQH tI is
characterized by the highly degenerate flat band structure of the bulk Landau
level. Soon after the discovery of the quantized Hall effect it became
understood~\cite{Prange1981} that the smooth profiles of the observed data could not be
reconciled with the singular density of states of the clean system. The
solution was to account for the presence of impurities broadening the Landau
level into a Landau impurity band (thence washing out the system's band gaps.)
It was also understood, that the ensuing low temperature topological quantum
criticality could be described in terms of a two-parameter scaling approach~\cite{Khmelnitskii83}.
Its two scaling fields were the average longitudinal conductivity, $g\equiv
\sigma_{xx}$, a variable known to be central to the description of disordered
metals in terms of the `one-parameter scaling hypothesis'~\cite{Abrahams1979No}, and the transverse
conductivity $\sigma_{xy}$, which may be identified  with the configurational
average of the topological Hall number, $\sigma_{xy}=\chi$. The scaling of
these two parameters upon increasing system size and/or lowering temperature
(cf. Fig.~\ref{fig:PhasePortrait}) was first described on phenomenological grounds by
Khmelnitskii~\cite{Khmelnitskii83} and later substantiated in terms of field theory by Pruisken~\cite{Pruisken1984a,Pru1}.
Starting from bare values $\tilde g\gg 1$ characterizing a weakly
disordered metal, and the generally non-integer $\sigma_{xy}\equiv \tilde
\chi$ characterizing a diffusive finite size IQH system, the flow (upon increasing the system size)
is  towards two types of fixed points,
\begin{align}
\label{FlowGeneric}
(g,\chi)\longrightarrow \left\{
\begin{array}{ll}
	(0,n)&,\qquad \tilde \chi \notin n+\frac{1}{2},\crcr
	&\crcr
	(g^\ast,n+\frac{1}{2})&,\qquad \tilde \chi = n+\frac{1}{2},
\end{array}
\right.
\end{align}
i.e. generically, the flow approaches the Anderson localized fixed point, $g=0$, indexed by the
integer value $\chi=\sigma_{xy}=n$ of a quantum Hall configuration, where $n=[\tilde
\chi]$ is the integer arithmetically nearest to $\tilde
\chi$. Neighboring basins of attraction, $n$ and $n+1$, are separated by a
critical surface $\tilde \chi=n+1/2$, on which the flow is towards the IQHE fixed
point $(g,\chi)\to (g^\ast,n+1/2)$, where $g^\ast
=\mathcal{O}(1)$ is the critical value of the conductivity.  
The most natural way  
to access the topological parameter $\chi$  is via the
introduction of spatially non-local `topological sources'. As we will discuss
below, this idea is central to the description of topological invariants without
reference to the momentum space (and independent of a particular field theoretical
formalism).

Even before the advent of the clean topological band insulators, the above
quantum Hall paradigm was observed in other system classes, viz. the class C\cite{Gruzberg1999} and class D\cite{Senthil1999} quantum Hall effects. Similar physics showed up, but not understood as such, also in quasi one-dimensional
disordered quantum wires. Studies of  quantum wires in symmetry
classes, D\cite{Brouwer2000653}, DIII\cite{Brouwer2000,Gruzberg2005}, and AIII\cite{BMSA:98} describing disordered superconductors and
chiral disordered lattice systems, respectively had revealed unexpected de-localization
effects. Early observations of the phenomenon were subject to some
controversy, as it appeared to be tied to non-universal fine
tuning. The point  not understood at the time was that the
delocalized system configurations were actually topological insulators fine
tuned to a phase transition point conceptually analogous to the IQH transition. First parallels to QH physics and two parameter scaling  were drawn in
Ref.[\onlinecite{Gruzberg2005}], however the full framework of the underlying topology was probably not understood at that time.

The high degree of universality reflected in the above can be
understood from a simple argument first formulated in Ref. [\onlinecite{Motrunich2001}] (cf. Fig.~\ref{fig:PhasePortrait}): consider a
schematic phase plane of a topological insulator spanned by a parameter $\mu$
controlling the topological sector of the system (the chemical potential, a
magnetic field, etc.), and a parameter $w$ quantifying the amount of disorder.
In the clean system, $w=0$, the topological number jumps at certain values of
$\mu$ through topological phase transition points, characterized by a closure
of bulk band gaps. Turning on disorder at a generic value of $\mu$ generates a
crossover from the clean band insulator into a configuration characterized by
a non-vanishing density of states. In most symmetry classes --- for the
discussion of exceptional situations, see below --- Anderson localization will
turn the ensuing `impurity metal' into an Anderson insulator. The amount of
disorder required to drive this crossover vanishes at the clean system's gap
closing points. At the same time, the closing points mark points of quantum
phase transitions and the integrity of these cannot be compromised by the
\emph{crossover} from the band-- into the Anderson insulator regime. They
become, rather, endpoints of phase transition {\em lines} meandering through the
phase plane $(\mu,w)$. It is the existence of these lines that
distinguishes the `topological Anderson insulator' (tAI)  from a conventional Anderson insulator. At
the phase transition lines  the localization
length diverges and the system builds up a delocalized state. From an edge-oriented 
perspective, the delocalization accompanying a transition $n\to n-1$
means that a pair of edge states is hybridized across the bulk via a delocalized
state to disappear (i.e. move away from the zero energy level). Somewhat counter-intuitively,
the delocalization phenomenon can be driven by \emph{increasing} the amount of
disorder in the system, or by changing any other parameter capable of changing
the system's location in the phase plane. In  early works\cite{BMSA:98, Brouwer2000653,Brouwer2000}
delocalization was observed as a consequence of an `accidental' crossings of
phase transition lines. For other crossing protocols see Refs.~[\onlinecite{Groth:2009fv,Rieder2013}]. Notice that each
phase lobe in Fig.~\ref{fig:PhasePortrait} is characterized by an integer invariant. However, the
integerness of that value is tied to the limit of infinite system size
characterizing a thermodynamic phase. By contrast, the finite size
system will generally be described by a non-integer mean topological number, which leads to
perhaps counter-intuitive conclusion that Anderson localization actually
stabilizes the topological rigidity of disordered systems. The corresponding
flow $g\to 0$ (localization) and $\chi \to n$ (re-entrance of the topological number) is
described by the {\em two-parameter} flow diagram. 

For two-dimensional topological insulators the above argument has been made
quantitative, to varying degrees of completeness. In some cases (the class A
IQH, or the class AII quantum spin Hall effect) no rigorous theory describing
the strong coupling regime close to the fixed point exists, but the global
pattern of the flow can be convincingly deduced from a two-parameter effective
field theory pioneered by Pruisken\cite{Pruisken1984a} and Fu and Kane\cite{Kane-Fu2012}, respectively. In the class D or DIII system even the phase diagram is not fully understood, while the exact
equivalence of the class C insulator to a percolation
problem\cite{Gruzberg1999} implies  existence of exact solutions for the flow.  
Remarkably, in two-dimensional systems of chiral symmetry classes AIII, CII and BDI, 
which are not tI in 2d, the mechanism of Anderson localization is also controlled by the  
point-like topological defects\cite{Koenig:2012} (vortices) and in this way is analogous to 
class AII topological insulator studied by Fu and Kane.

In this paper, we will focus on the five families of topological multi-channel
quantum wires, AIII, CII, BDI, D, DIII. There are far reaching parallels between disordered
insulators in one and two dimensions: both show two-parameter scaling, and can
be described in terms of  field theories --- non-linear
$\sigma$-models containing a $\theta$-term/fugacity term measuring the action
contribution of smooth/point-like topological excitations in the
$\Bbb{Z}$/$\Bbb{Z}_2$ cases ---, the scaling variables are obtained from the
field theory via topological sources, and the bulk-boundary
correspondence establishes itself by identical mechanisms. However, unlike
the 2d systems, the 1d field theories are amenable to powerful
transfer matrix techniques. These methods can be applied to solve the problem
non-perturbatively, and to describe the results in terms of parameter flows
all the way from the diffusive regime into the regime of strong localization. 
Overall the situation in 1d is similar, but under much tighter theoretical control than in
2d.

\section{Main results} 

In this paper we describe five topologically non-trivial insulators in one dimension in terms of supersymmetric nonlinear $\sigma$-models with target spaces representing the
different symmetry classes. It provides a framework describing  non-translationally invariant topological
insulators in terms of a theory that: 
\begin{itemize}
	\item Is  universal, in that elements that are not truly essential to
	the characterization of topological phases, such as translational
	invariance, or band gaps, do not play a role. The theory, rather,
	describes the problem in a minimalist way, in terms of symmetry and
	topology. 
	\item Topological sectors are described in real space, rather
	than in terms of the more commonly used momentum space homotopy constructions. 
	To this end we study response of supersymmetric partition sum on a twisted boundary
	conditions. The later are given by a proper gauge transformations dictated by the 
	corresponding symmetry group and containing continuous as well as discrete (i.e. $\Bbb{Z}_2$) degrees of freedom.      
	\item These field theories differ from the ones describing conventional
	Anderson insulators  by the presence a topological contribution to the
	action. The latter weighs the contribution of smooth/point-like topological field
	excitations in terms of a  \mbox{$\theta$-term/fugacity term}
	depending on whether we are dealing with a  $\Bbb{Z}$/$\Bbb{Z}_2$
	insulator.
	\item At the bare (short distance) level, the field theories are described by
	two coupling constants $(\tilde g,\tilde \chi)$, where $\tilde g$ is the Drude
	conductance, central to the one-parameter scaling approach to conventional
	disordered conductors, and $\tilde \chi$ is the ensemble average  topological number.
	In the construction of the effective
	field theory these parameters are obtained from an underlying microscopic disordered lattice model by 
	a perturbative self-consistent Born approximation (SCBA). We provide numerical verification of this 
	approach, which appears to work well down to $N=3$-channel wires (though the theory is developed in $N\to\infty$ limit).  
	\item At large distance scales, these parameters exhibit the
	flow~(\ref{FlowGeneric}). Using transfer-matrix approach, we provide the exact quantitative description of this flow, including the strongly localized phase and quantum critical points.   For generic $\tilde \chi$, the fixed
	point configuration $(g,\chi)\to (0,n)$ is attained exponentially fast in
	the length, $L$, of the system. The fixed point value for the critical
	conductance, $g^\ast =0$, but the approach to this configuration is
	algebraic, $\sim L^{-1/2}$. The vanishing of the \emph{mean} conductance
	at criticality is manifestation of large sample-to-sample fluctuations in
	1d. In fact, it is known\cite{Mathur1997,BMSA:98} that a sub-Ohmic scaling $\sim L^{-1/2}$
	signifies the presence of a delocalized state in the system. (Some symmetry
	classes in 2d (D, AII, DIII) exhibit flow more complicated than that
	depicted in Fig.~\ref{fig:PhasePortrait} in that the critical surface broadens into a metallic
	phase. We return to the discussion of this point below.)
	\item The theory describes bulk-boundary correspondence  by a
	universal mechanism. In the  fixed points, $(0,n)$, the field
	theory becomes fully topological in the sense that its standard gradient
	term is absent. In the $\Bbb{Z}$-cases, the topological terms with integer
	coefficients become Wess-Zumino terms at the boundary, where they describe
	$n$ gapless boundary excitations. 
	\item In the $\Bbb{Z}_2$ cases the fermionic parts of the $\sigma$-model
	target spaces contain two      disconnected components\cite{Bocquet2000}.
	The topological quantum criticality turns out to be associated with the
	field configurations with kinks, switching between the two. The
	corresponding transfer-matrix evolution equation acquires a spinor form,
	which reveals a hidden supersymmetry (not related to  Efetov's
	supersymmetry of the underlying $\sigma$-models).  Its fermionic degree of
	freedom, creating kinks between the two sub-manifolds, is dual to the
	Majorana edge modes, residing on the boundaries between two topologically
	distinct phases.  Such supersymmetry may prove to be crucial for
	understanding of the bulk-boundary correspondence in the  2d $\Bbb{Z}_2$
	insulators, which has not yet been worked out.
\end{itemize}

\section{Solution strategy}
\label{sec:solution_strategy}

Before delving into more concrete calculations, it is worthwhile to
provide an overview of the key elements of our approach to the low energy physics of the five classes
of quantum wires:
\begin{enumerate}
	\item We find it convenient to model our wires as chains of coupled sites, or
	``quantum dots'', where each site carries an internal Hilbert space
	accommodating spin indices, multiple transverse channels, etc. The
	symmetries of the wire and its topological number are encoded in the intra-
	and inter-site matrix elements describing the system.
	\item Disorder is introduced by rendering some of those matrix elements
	randomly distributed. The choice of those random matrix elements is
	largely a matter of convenience, i.e. different models of disorder may
	alter the bare values of the two coupling constants entering the system's
	field theory, but not the universal physics.
	\item In the clean case, the topological sector of the system can be described
	in terms of the well established homotopy invariants constructed over the
	Brillouin zone. We will discuss how this information may be alternatively
	accessed by probing the response of the spectrum to either extended
	$(\Bbb{Z})$ or local $(\Bbb{Z}_2)$ changes in the inter-site hopping. The
	latter scheme generalizes to the presence of disorder.
	\item We describe this response in terms of supersymmetric Gaussian integrals.
	Upon averaging these integrals over disorder, the symmetries of the
	microscopic Hamiltonian turn into a `dual' symmetry of the corresponding
	functional integrals. (The mathematical concept behind this conversion is
	called `Howe-pair duality'~\cite{Howe:1989,Cheng:book}.) In practice this means that the Gaussian
	actions are invariant under  a group $G$ of  transformations whose
	symmetries are in one-to-one correspondence to that of the parent Hamiltonian.
	\item If the disorder is strong enough to close the gap, that symmetry gets
	spontaneously broken to a subgroup $H$. The ensuing Goldstone modes describe
	diffusive transport in the system. At large distance scales, these modes are
	expected to ``gap out'' due to Anderson localization. Within the field theoretical framework,
	Anderson localization manifests itself in a diminishing of the stiffness of
	Goldstone modes, and an eventual crossover into a disordered phase, not
	dissimilar to the disordered phase of a magnet. From yet another perspective
	one may understand this crossover in terms of a proliferation of topological
	excitations on the Goldstone mode manifold. At the strong disorder fixed
	point, which is characterized by a vanishing of longitudinal transport
	coefficients, the full symmetry of the system, $G$, is restored (once more in
	analogy to a magnet). However,
	\item It remains broken at the boundary points (or lines, in 2d) of the
	system. As one would expect on general grounds, the boundary Goldstone modes
	enjoy topological protection and describe the system's zero energy states.
	\item Methodologically, we describe the process of bulk disordering by a
	method conceptually allied to a real space renormalization group approach. In
	concrete terms, this means that we map the field integral description onto an
	equivalent transfer matrix equation which describes the dot-to-dot
	evolution along the system. The derivation of that equation does not rely on
	premature field-continuity assumptions. In face, we will observe that in the
	$\Bbb{Z}_2$ cases discontinuous changes of the field play a pivotal role.
	Evolution via the transfer matrix equation may be understood as a process
	whereby sites effectively fuse to larger sites, with renormalized parameters.
	However, rather than describing this process in explicit terms, we will
	analyze the eigenvalue spectrum of the transfer operator, and from there extract 
	the $L$-dependent flow of observables $(g(L), \chi(L))$.
	\item Within the field theoretical framework, the real space topological
	twists employed to access the system's topological numbers, become topological
	field excitations, smooth instantonic configurations/kinks for the
	$\Bbb{Z}/\Bbb{Z}_2$-insulators. The action cost of these configurations is
	quantified by a topological $\theta$-term/fugacity term. Localization can
	then be understood in terms of a proliferation of such topological
	excitations at large distance scales, and this process reflects in an
	effective flow of both the gradient term, and the coefficient of the
	topological term. However, at half-integer/zero bare topological coefficient, the
	contribution of such excitations gets effectively blocked, either in terms of
	a destructive interference of topological excitations (conceptually similar to
	what happens in a half-integer antiferromagnetic spin chain\cite{Haldane1983464})/or in terms of
	a vanishing fugacity.
\end{enumerate}

In the rest of the paper, we derive and solve the theory for the five families of topological quantum wires. The presentation is
self-contained, however, to keep the main text reasonably compact, details are
 relegated to appendices. We start out with a  preamble (section \ref{sec:solution_strategy}), in which we formulate the general strategy of our derivation. To avoid repetitions, we discuss two cases in more
detail, viz. the AIII $\Bbb{Z}$-insulator, section \ref{sec:symmetry_class_aiii}, and the class D
$\Bbb{Z}_2$ insulator, section~\ref{sec:FT_classD}. The theory for the  remaining classes, BDI,
CII, and DIII, largely parallels that of those two, and will be discussed in
more sketchy terms.

\section{$\Bbb{Z}$-Insulators}
\label{sec:symmetry_class_aiii}

In this section, we derive and analyze the effective theory for the
one-dimensional $\Bbb{Z}$-insulators. We start by discussing the simplest of
these viz. a `chiral' system lacking any other symmetries, class AIII, in a
fairly detailed manner. After that we turn to the time reversal invariant
chiral system, class BDI, whose theory will be described in more concise
terms, emphasizing the differences to the time reversal non-invariant case.
The theory of the third $\Bbb{Z}$ representative, CII, does not add
qualitatively new structures, and will be mentioned only in passing.

\subsection{Definition of the model}
\label{ssec:definition_of_the_model}

\begin{figure}
\centering{
\includegraphics[width=0.3\textwidth]{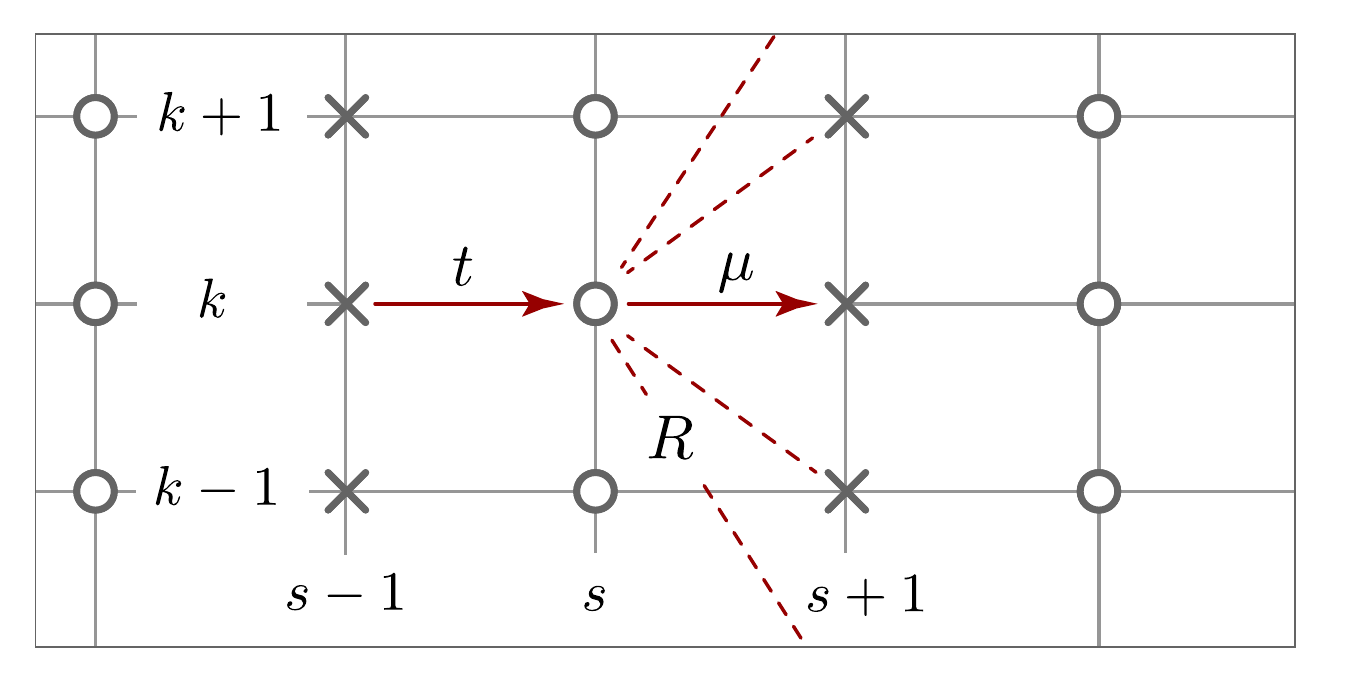}
}
\caption{Schematic of a multi-channel AIII quantum wire with staggered hopping of strength $t$, $\mu$, respectively, and random inter-chain hopping described by matrix elements $R^{kk'}_{ss+1}$.}
\label{fig:SSHChain}
\end{figure}

Consider a system of $N$-quantum wires, described by the Hamiltonian
\begin{align}
	\label{eq:AIIIHamiltonian}
	  H = \sum\limits_{s=1}^{2L} C^\dagger_{s} \left(t_{ss'}+R_{ss'}\right)C_{s'},
\end{align}
where $s$ is a site index, $C_s=\{C_{s}^k\}$, $k=1,\dots,N$ a vector of $N$
fermion creation operators, and $t$ a nearest neighbor hopping matrix defined
through $t_{s,s+1}=t_{s+1,s}=\mu$ if $s$ is even, $t_{s,s+1}=t_{s+1,s}=t$ if
$s$ is odd, and zero otherwise, cf. Fig.~\ref{fig:SSHChain}. In other words, the matrix $t$ implements a
staggered nearest neighbor hopping chain as realized, e.g., in a Su-Shrieffer-Heeger 
model\cite{SSH1979}. Randomness is introduced into the system through the Hermitean bond
random matrices $R_{ss'}$ as
\begin{eqnarray}
	\left\langle R^{kk'}_{s,s'}\right\rangle=0, \qquad
	\left\langle R^{kk'}_{s,s+1}
	R^{k'k}_{s+1,s}\right\rangle&=&\frac{w^2}{N},
	\label{eq:R_AIII}
\end{eqnarray}
where all other second moments of matrix elements vanish. To keep the model
simple, neighboring chains are only coupled
through randomness (one may switch on non-random hopping, at the expense of
slightly more complicated formulae).

To describe the symmetries of the system, we define the site parity operator,
$P_{s,s'}=(-)^s \delta_{ss'}$. The fact that the first quantized Hamiltonian
$H\equiv t+R$, defined through Eq.~\eqref{eq:AIIIHamiltonian}, is purely
nearest neighbor in $s$-space is then expressed by the anti-commutation
relation $\{H,P\}_+=0$. The absence of other anti-unitary symmetries makes $H$
a member of the chiral symmetry class AIII. To conveniently handle the
symmetry of the Hamiltonian, we switch to a two-site unit cell notation
through $c_{2s} \to c_{+,s}$, and $c_{2s-1} \to c_{-,s}$. In this
representation the Hamiltonian assumes a $+/-$ off-diagonal form,
$H=\left(\begin{smallmatrix} &h\cr h^\dagger &
\end{smallmatrix}\right)$, and $P=\sigma_3$ is represented by the Pauli matrix.
Anti-commutativity with $P$ implies the symmetry of $H$ under the continuous
but  transformation $H=THT$, where $T=\exp(i \theta P)$, and
$\theta$ is, in general, complex parameter.

\subsection{Topological invariants}

In the clean system ($R=0$), one may access the system's topological invariant by the
standard\cite{Bernevig2013} winding number construction. Turning to a Fourier representation with a wavenumber  
$q$ conjugate to $s$,
the block matrices $h$ become functions $h_{kk'}(q)=\delta_{kk'} (\mu + t e^{i q})$. We then
obtain the topological number $n$ as 
\begin{align}
	\label{AIIInTraditional}
	n=\frac{1}{2\pi}\mathrm{Im}\int\limits_0^{2\pi}\!dq \,\,\mathrm{tr}(h^{-1}(q) \partial_q h(q)).
\end{align}
For the simple model under consideration this becomes $n=N \Theta(|t|-|\mu|)$, where
$\Theta$ is the step function. A transverse coupling between the chains would
lift the degeneracy of this expression and turn $n$ into a function stepwise
diminishing from $N$ to $0$ upon changing system parameters.

We  aim to access the number $n$ in a manner not tied to the momentum space.
To this end, we consider a system of  $L$ unit cells, and close it to form a
ring. On this ring, we impose the non-unitary axial transformation $H_\phi=
T_\phi H T_\phi$, where $(T_\phi)_{s,s'}= \exp(-i\frac{s}{L} \phi P_{s,s'} \delta_{ss'})$. The transformation
$T_\phi$ changes the Hilbert space of the problem, and hence may affect its
spectrum. We will show that the sensitivity of the spectrum probes topological
sectors. To this end, we notice that the transformation affects the functions
$h$ as $h_{kk'}(q)\to h_{kk'}(q+\phi/L)$ and $h^\dagger_{kk'}(q)\to
h^\dagger_{kk'}(q-\phi/L)$. We next define the zero energy retarded Green
function $G_\phi=(i0 - H_\phi)^{-1}$  and compute its sensitivity to the insertion
of the flux as
	\begin{eqnarray}
	\,\,&&\frac{1}{4
	\pi}\,\ln\left(\frac{\det(G_{2\pi})}{\det(G_0)}\right)=\frac{1}{4
	\pi}\int_0^{2\pi}d
	\phi\,\partial_{\phi}
	\,\mathrm{tr}\ln(G_{\phi}) = \nonumber \\
    \,\,&&=\frac{1}{4\pi }\,\sum_q
	\int_0^{2\pi}\!\!\!
	d\phi\,\partial_{\phi}\,\mathrm{tr}\left( \ln h(q+\frac{\phi}{L}) + \ln 
    h^\dagger(q-\frac{\phi}{L})\right) \nonumber\\
	\,\,&&=\frac{i}{2\pi L}\mathrm{Im}\sum_q \int_0^{2\pi }d\phi  \,\partial_{q}\,\mathrm{tr}\left( \ln
	h(q+\phi/L)\right)=\crcr
	\,\,&&=\frac{i}{2\pi}\mathrm{Im}\int_0^{2\pi }dq  \,\partial_{q}\,\mathrm{tr}\left( \ln
	h(q)\right),     \label{eq:flux-insertion}
	\end{eqnarray}
where in the last line the general identity, $\sum_q \int_0^{2\pi}d\phi\,F(q+\phi/L)=L \int
dq F(q)$ was used. Comparison with Eq.~\eqref{AIIInTraditional} then shows that 
\begin{align}
	\frac{1}{4
	\pi}\mathrm{Im}\ln\left(\frac{\det(G_{2\pi
	})}{\det(G_0)}\right)=n.
\end{align}
This equation represents the topological invariant in terms of the `spectral
flow' upon insertion of one $2\pi$-twist under the axial transformation. 
To conveniently compute this expression, we define  the ``partition sum'',
\begin{align}
	\label{ZDefAIII}
	Z(\phi)\equiv \left\langle\frac{\det(G_{\phi_1}^{-1})}{\det(G_{-i\phi_0}^{-1})}\right\rangle,
\end{align}
where $\phi\equiv(-i\phi_0,\phi_1)^T$, and 
following Refs.~[\onlinecite{Nazarov1994,Lamacraft2004}] consider the generating function
\begin{equation}
\label{eq:F_phi}
{\cal F}(\varphi_0) = \partial_{\phi_1}Z(\phi)\bigl|_{\phi_1=-i\phi_0}=
\partial_{\varphi_1} \ln \det(G_{\varphi_1})\bigl|_{\phi_1=-i\phi_0},
\end{equation}
which contains the full information about the transport properties of the system.
From ${\cal F}$, our two variables of interest, $(g,\chi)$ can be accessed,
\begin{equation}
\label{AIIIObservableDef}
\chi=\frac{1}{4\pi} \int_0^{2\pi}d\phi_1\, \mathrm{Im}\,{\cal F}(i\phi_1); 
\qquad g = -i\partial_{\varphi_0} {\cal F}(\varphi_0)\bigl|_{\varphi_0=0}.
\end{equation}  
Here, the second equality expresses the conductance of the system in terms of
its sensitivity to a change in boundary conditions. The equivalence of this relation to the linear response representation of
the conductance is shown in Sec.~\ref{sec:Scattering}.

\subsection{Field theory representation}
\label{ssec:field_theory_representation}

We proceed by representing the ratio of determinants in~\eqref{ZDefAIII} as a
supersymmetric Gaussian integral, 
\begin{align}
	\label{GaussianIntegral}
	Z(\phi)=\left\langle \int d(\bar \psi,\psi)\,e^{i\bar \psi G_{\phi}^{-1}\psi
	}\right\rangle,
\end{align}
where $\psi=(\psi^{\rm b},\psi^{\rm f})$ and $\psi^\alpha$ are vectors of complex
commuting ($\alpha=\mathrm{b}$) or Grassmann variables ($\alpha=\mathrm{f}$) with components
$\psi^\alpha=\{\psi^\alpha_{\pm,s,k}\}$. Further $\bar
\psi^\mathrm{b}=\psi^{\mathrm{b}\dagger}$, while $\bar\psi^\mathrm{f}$ and $\psi^\mathrm{f}$ are independent, and
$G^{-1}(\phi)\equiv \mathrm{bdiag}(G_{-i\phi_0}^{-1},G_{\phi_1}^{-1})$ is a block
operator in bf-space. Gaussian integration over the super-field
$\psi^{\mathrm{b}/\mathrm{f}}$ produces the determinant/inverse determinant of
$G_{\phi_1}^{-1}/G_{-i\phi_0}^{-1}$, and in this way we obtain the partition sum $Z(\phi)$, Eq.~(\ref{ZDefAIII}).

The functional integral possesses a continuous symmetry under transformations
\begin{align}
	\bar \psi_+ \to \bar \psi_+ T_{\mathrm L},&\quad \psi_+\to T_{\mathrm R}^{-1} \psi_+\crcr
	\bar \psi_- \to \bar \psi_- T_{\mathrm R},&\quad \psi_-\to T_{\mathrm L}^{-1} \psi_-,
\end{align}
where $T_{\mathrm L,R}$ are $2\times 2$ supermatrices whose internal structure will be detailed below.  A
symmetry transformation of this type  generally spoils the adjointness
relation $\bar\psi^\mathrm{b}=\psi^{\mathrm{b}\dagger}$, but as long as we make sure not to hit singularities
it does not alter the result of the integration.

Denoting the set of these matrices by $\mathrm{GL}(1|1)$, we observe that the
action has a continuous symmetry under $G\equiv \mathrm{GL}(1|1)\times
\mathrm{GL}(1|1)$. This symmetry may be interpreted as the supersymmetric
generalization of the $\mathrm{GL}(n)\times \mathrm{GL}(n)$ symmetry under
unitary transformations of left- and right- propagating excitations in chiral
quantum systems; it is a direct heritage of the chiral symmetry of the
Hamiltonian.

Finally, notice that we may interpret the insertion of the chiral flux $\phi$ in
terms of a boundary condition changing chiral gauge transformation, $\bar \psi
G_\phi^{-1}\psi=\bar \psi' G^{-1}_0\psi'$, where  $\psi'_{\mathrm{L/R},s} \equiv
e^{(+/-)i\phi \frac{s}{L}}\psi_{\mathrm{L/R},s}$ are subject to the twisted boundary
condition $\psi_{\mathrm{L/R},L}'=e^{(+/-)i\phi } \psi_{\mathrm{L/R},0}'$, where
$\phi=\mathrm{diag}(-i\phi_0,\phi_1)^T$.

\subsection{Disorder average and low energy field theory}
\label{ssec:disorder_average_and_low_energy_field_theory}

We next average the theory over the distribution of the $R$-matrices and from
there derive an effective theory describing the physics at distance
scales larger than the elastic mean free path. There are two ways of achieving
this goal\cite{Altland2001511}, one being explicit construction, the other symmetry reasoning. For
an outline of the former route, we refer to Appendix~\ref{sec:field_theory_of_z_insulators}. Here, we discuss the
less explicit, but perhaps more revealing second approach. 

The averaging over disorder turns the infinitesimal increment $i0\to
i/2\tau$ of the retarded Green function into a finite constant, which
defines the inverse of the elastic scattering time. Its value may be
exponentially small or not, depending on whether the amplitude of the disorder,
$w$, exceeds the gap $\sim |t-\mu|$ of the clean system or not. This
criterion defines the crossover from the band insulator into the impurity
`metal'. The metallic regime is characterized by a globally non-vanishing
density of states, and finite electric conduction at length scales shorter
than the localization length to be discussed momentarily. In the metallic
regime, the  appearance of a finite diagonal term $i0\to i/2\tau$
in $\pm$-space `spontaneously breaks' the symmetry under $G$ down to the
diagonal group $H=\mathrm{GL}(1|1)$ defined by the equality $T_L=T_R$. (Within
the context of QCD this mechanism is known as the spontaneous breaking of
chiral symmetry by gauge field fluctuations, where in our context the role of the latter is played by impurity potential fluctuations.) We
expect the appearance of a Goldstone mode manifold,
$G/H=(\mathrm{GL}(1|1)\times \mathrm{GL}(1|1))/ \mathrm{GL}(1|1)\simeq
\mathrm{GL}(1|1)$. In mathematical terminology, that manifold is understood as
a Riemannian (super-)symmetric space, viz. the space A$|$A of rank 1. The
assignment AIII$_\text{Hamiltonian}\to$ (A$|$A)$_\text{field theory}$ is an
example of the symmetry duality mentioned in
section~\ref{sec:solution_strategy}.

We next identify the low energy `Ginzburg-Landau' action $S[T]$ describing the
Goldstone mode fluctuations, and its connection to physical observables.
Technically (cf. Appendix~\ref{app:classAIII}), the field $T=\{T_s\}$ appears
after averaging the theory over disorder and decoupling the ensuing
$\psi^4$-term  through a Hubbard-Stratonovich transformation. After
integrating over the $\psi$-fields, the partition function then assumes the
form ${Z(    \phi)=\int {\cal D}T \,\exp(-\tilde S[T])}$, where
\begin{align}
	\label{SAIIIStrLn}
	\tilde S[T]= \mathrm{str}\ln\left(\begin{matrix}
		i \Sigma_0 T & -h\crcr
		-h^\dagger & i \Sigma_0 T^{-1}
	\end{matrix}\right),
\end{align}
$\Sigma_0=1/2\tau$ is the impurity self-energy evaluated in the self-consistent Born approximation (SCBA) and
$h$ contains the disorder independent nearest neighbor hopping matrix
elements. Here $\mathrm{str}(A)= \sum_\alpha (-)^{\alpha}A^{\alpha\alpha}$ is
the so-called supertrace. We recall that the action must be symmetric under the action of the {\em full} symmetry group $G$.
Within the present context, the latter acts by transformation $T\mapsto T_L T
T_R^{-1}$, i.e. for constant (i.e. $s$-independent) transformations  our action must
be invariant under independent left- and right-transformation, and the fulfillment of 
this criterion is readily verified from the structure of the action. Specifically,
the action of a constant field $T$   vanishes, $\tilde
S[T]=\tilde S[\Bbb{I}]=0$. To obtain an effective action of soft fluctuations,
varying on length scales larger than the lattice constant, we replace  the
site index $s\to x$ by a continuous variable, $x$, and think of the hopping
operators as derivatives. Up to the level of two gradients, two
operators can be constructed from field configurations $T_s\to T(x)$: 
$\mathrm{str}(\partial_x T \partial_x T^{-1})$, and
$\mathrm{str}(T^{-1}\partial_x T )$\footnote{There exists one more term,
$[\mathrm{str}(T^{-1}\partial_x T )]^2$, which, however, does not play an
essential role in the present context\cite{Altland2001511}.}. A substitution of $T(x)$ into
Eq.~\eqref{SAIIIStrLn} followed by a straightforward expansion of the
logarithm (cf. Appendix~\ref{app:classAIII}) indeed produces the effective
action~\cite{Altland2001511}
\begin{equation}
\label{eq:AIIIAction}    
S[T]=\!
\int\limits_0^{L}  dx \left[-\frac{\tilde \xi}{4}\,\mathrm{str}(\partial_x T\partial_x T^{-1})
+ \tilde\chi\,\mathrm{str}(T^{-1}\partial_x T)\right],  
\end{equation}  
where $(\tilde \xi,\tilde \chi)$ are two coupling constants. In this
expression, the presence of the source  variable $\phi$ implies a twisted
boundary condition,
\begin{align}
	\label{AIIIFieldBoundaryTwist}
	T(L)=e^{i \phi}T(0)e^{i\phi}.
\end{align}
To make progress, we parameterize the fields $T$ as
as 
$T=U\left(\begin{smallmatrix} e^{y_0} & \crcr  & e^{iy_1} \end{smallmatrix}
\right)U^{-1}$, 
where $U=\exp \left(\begin{smallmatrix} & \nu \crcr \mu &  \end{smallmatrix}
\right)$ contains the Grassmann variables. The two {\em radial} coordinates $(y_0,y_1)$ (one non-compact and one compact)
parameterize the maximal domain for which the path integral over $T$  with the action~(\ref{eq:AIIIAction}) is convergent. 
Notice that the first derivative term, $\mathrm{str}(T^{-1} \partial_x
T)=\partial_x \,\mathrm{str}(\ln T)$ can formally (more on this point below) be expressed as a
surface
term, indicating that it is a topological $\theta$-term. In the absence of a boundary twist explicitly
breaking the symmetry between fermionic and bosonic integration variables,
that is for $\phi_1=i \phi_0$, the functional integral equals unity by
supersymmetry\cite{Efetov-book}, and $Z(\phi)=1$ by definition, i.e. the connection between
$Z(\phi)$ and the functional integral does not include normalization factors.

The interpretation of the two coupling constants $(\tilde\xi,\tilde\chi)$
appearing in the action can be revealed by taking a look at the short system
size limit $l<L<\tilde \xi$, where $l\sim t \tau$ is a short-distance cutoff
set by the elastic mean free path due to disorder scattering. In this limit,
field fluctuations are suppressed and we may approach the functional integral
by stationary phase methods. A straightforward variation of the action
$\delta_T S[T]=0$ yields the equation $\partial_x (T\partial_x T^{-1})$, and
the minimal solution consistent with the boundary conditions is given by
$T_s=e^{i2\phi \frac{s}{L}}$. Substituting this expression into the action and
ignoring quadratic fluctuations, we obtain the estimate,
$Z(\phi)\simeq\exp\left(-\frac{\tilde\xi}{L}(\phi_0^2+\phi_1^2)-2\tilde \chi
(\phi_0-i\phi_1)\right)$. Application of Eq.~\eqref{AIIIObservableDef}
then readily yields $\tilde \chi=\chi$ and $g=2\tilde \xi/L$. This identifies $\tilde
\chi$ as the bare value of the average topological number, and $\tilde \xi$ as
the localization length (for $L<\tilde\xi$, the conductance of the wire is
Ohmic, $g\sim \tilde\xi/L$). Within the explicit construction of the theory
outlined in Appendix~\ref{app:classAIII}, the coefficients ($\tilde\xi, \tilde \chi$) are obtained as
functions of the microscopic model parameters. For the specific model under
consideration one finds $\tilde\xi=N l$, and $ \tilde \chi=
\frac{i}{2}\mathrm{tr}(G^+ P \partial_k H)$, where $G^+$ is the Green
function subject to the replacement $i0\to i/2\tau$. In
parentheses we note that this expression can be identified with the
expectation value of velocity, or a `persistent current' flowing in response
to the axial twist of boundary conditions. Within our present model, one
obtains $\tilde \chi=\frac{N}{2}(1+\frac{(t-\mu)t}{w^{2}})$, see
Appendix~\ref{app:classAIII} for more details.

\subsection{Anderson localization}
\label{ssec:AIII_anderson_localization}

Before exploring in quantitative terms what happens at large scales,
$L>\tilde\xi$, let us summarize some anticipations. For generic values of
$\tilde \chi$ one expects flow into a `disordered' regime. At large distance
scales, the fields exhibit strong fluctuations and the `stiffness' term
$\propto \tilde \xi$ becomes ineffective. (Within an RG oriented way of
thinking, one may interpret this as a scaling of a renormalized localization
length $\xi(L)\to 0$.) On general grounds, we expect this scaling to be
accompanied by a scaling $\chi(L)\to n$. At the fixed point, the Goldstone
modes disappear from the bulk action, which we may interpret as a restoration
of the full chiral group symmetry $G$. The presence/absence of this symmetry
is a hallmark of localized/metallic behavior, the scaling is towards an
attractive \emph{bulk} insulating fixed point.

As for the boundary, the
fixed point topological term with quantized coefficient $n \int dx
\,\mathrm{str}(T^{-1}\partial_x
T)=n\,\left[\mathrm{str}\ln(T(L))-\mathrm{str}\ln(T(0))\right]\equiv S_\mathrm{b}[T]$, becomes a surface
term, where we temporarily assume our system to be cut open.
For generic values $\chi\not=n$ it actually is \emph{not} a surface term,
because $T=T(y_0,y_1)=T(y_0,y_1+2\pi)$ is  $2\pi$-periodic in the coordinates $y_1$ 
while $\exp(-\chi \,\mathrm{str}\ln T)=\exp(-\chi(y_0- i y_1))$ is
not. The requirement of a quantized coefficient reveals the surface terms
$n\,\mathrm{str}\ln T$ as zero dimensional variant of Wess-Zumino term. At
any rate, the $G$-symmetry at the boundary remains broken, and we will discuss
in section \ref{sec:DOS-AIII} how this manifests itself in the presence of protected surface
states. Notice how the protection of these states is inseparably linked to 
bulk localization. The latter plays the role of the bulk band gap in clean systems. 

The above picture can be made quantitative by passing from the functional
integral to an equivalent ``transfer matrix equation''\cite{Efetov-book,Altland2001511}.
The latter plays a
role analogous to that of the Schr\"odinger equation of a path integral.
Interpreting length as (imaginary) time, it describes how the amplitude
$\Psi(\phi,L)\equiv Z(\phi,L)-1$ defined by the functional integral at fixed
initial and final configuration $T(0)=\mathds{1}$, $T(L)=\exp(2 i \phi)$
evolves upon increasing $L$. (Since $Z(0,L)=1$, by its supersymmetric
normalization the function $\Psi$ is defined to describe the non-trivial
content of the partition sum.) This equation, whose derivation is detailed in
Ref.~[\onlinecite{Altland2001511}], is given by
\begin{equation}
\label{AIIISchrodinger}
-\tilde\xi\, \partial_x
\Psi(y,x)=\frac{1}{J(y)}(\partial_\nu - iA_\nu)J(y)(\partial_\nu - iA_\nu)\Psi(y,x),
\end{equation}   
where ${J(y)=\sinh^{-2}\left(\frac{1}{2}( y_0-i y_1)\right)}$ is the
Jacobian of the transformation to the radial coordinates $y_\nu$,
$\partial_\nu=\partial/\partial_{y_\nu}$, $A_\nu=\tilde\chi \,i^{1-\nu}$, and
the index $\nu=0,1$ is summed over. To understand the structure of this
equation notice that the action of the path integral~\eqref{eq:AIIIAction}
resembles the Lagrangian of a free particle, subject to a constant magnetic
field. One therefore expects the corresponding transfer matrix equation to be
governed by the Laplacian on the configuration space manifold
$\mathrm{GL}(1|1)$ of the problem. The differential operator appearing in
Eq.~\eqref{AIIISchrodinger} is the radial part of that Laplacian (much like
$r^{-2}\partial_r r^2 \partial_r$ is the radial part of the Laplacian in
spherical coordinates), i.e. the contribution to the Laplacian differentiating
invariant under angular transformations $U$. The
presence of the Jacobian $J(y)$ reflects the non-cartesian metric of the
manifold, and the vector potential, $A_\nu$, is proportional to the bare
topological parameter  $\tilde
\chi$.

It is straightforward to identify the eigenfunctions and eigenvalues of the transfer matrix
operator as  
\begin{align}
	\psi_{l}(     y) &=\sinh\left
(\frac{1}{2} ( y_0-i y_1)\right)e^{il_\nu y_\nu},\crcr
&\epsilon(l)=  (l_0-i\tilde\chi)^2+(l_1-\tilde \chi)^2,
\end{align}
where $l_0\in \Bbb{R}$, and $l_1\in \Bbb{Z}+\frac{1}{2}$ to make the
eigenfunctions $2\pi$-periodic in $\tilde y_1$. We may now  employ these
functions to construct a spectral decomposition $\Psi(\phi,L)=\sum_{l_1}\int
dl_0\,\mu(l) \,\psi_l(\phi)e^{-\epsilon(l) L/\tilde\xi}$. Using that
\begin{align}
	\left\langle \psi_l ,\psi_{l'}\right\rangle& \equiv \int\limits_0^\infty dy\int\limits_0^{2\pi} d\tilde y \,J(y)
\bar\psi_l(y)\psi_{l'}(y)=\crcr
&\quad=(2\pi)^2\delta(l_0-l_0')\delta_{l_1l'_1},
\end{align}
it is straightforward to obtain the expansion coefficients $\mu(l)$ by taking
the scalar product $\left\langle
\psi_l,\Psi(L\to 0)\right\rangle$. Upon substitution of the limiting value
$\Psi(\phi,L \to 0) \rightarrow -1$  (at any $\phi$, but $\phi=0$, where $\Psi(\phi,L \to 0) \rightarrow 0$) \footnote{There is a subtlety here.
The boundary condition $T(0)=\mathds{1}$ implies $Z(y,0)=\delta(y)\equiv
\lim_{\epsilon\to
0}\mathrm{str}(-\frac{1}{\epsilon}\mathrm{str}(T+T^{-1}))=\lim_{\epsilon\to
0}\mathrm{str}(-\frac{2}{\epsilon}(\cosh(y)-\cos(\tilde y)))$ describes the
functional integral upon approaching zero length. This is a `zero-dimensional'
$\delta$-function, in the sense that it does not contain a divergent pre-factor
$\sim \epsilon^{-d/2}$. In particular, it does not contribute to the spectral
decomposition of the function $\Psi=Z-1$.}, we obtain
$\mu(l)=\frac{1}{\pi}\frac{1}{l_1+il_0}$ and thus
\begin{align}
\label{eq:Z_phi_AIII}
 	Z(\phi)=1+ \sum_{l_1\in \Bbb{Z}+\frac{1}{2}} \int \frac{dl_0}{\pi}\, \frac{\psi_{l}(\phi)}{l_1+il_0}
 	\,
 	e^{-\epsilon(l)L/\tilde\xi}.
 \end{align} 
Differentiation of this result, according to Eq.~(\ref{AIIIObservableDef}), yields the two observables 
of interest\cite{Altland2014}
\begin{eqnarray}
 g&=&\sqrt{\frac{\tilde\xi}{\pi L}}\sum_{l_1\in \Bbb{Z}+1/2} e^{-(l_1-\tilde\chi)^2 L/\tilde\xi},\\
  \chi&=& n-\frac{1}{4}\!\sum_{l_1\in \Bbb{Z}+1/2}\! \left[\mathrm{erf}\Big(\sqrt{\frac{L}{\tilde\xi}}\,(l_1-\delta\tilde\chi)\Big)-(\delta\tilde\chi \leftrightarrow
  -\delta\tilde\chi)\right],\nonumber
 \end{eqnarray}
where $\delta\tilde\chi=\tilde\chi-n$ is the deviation of $\tilde\chi$ off the
nearest integer value, $n$. 

\begin{figure}
\centering{
\includegraphics[width=0.4\textwidth]{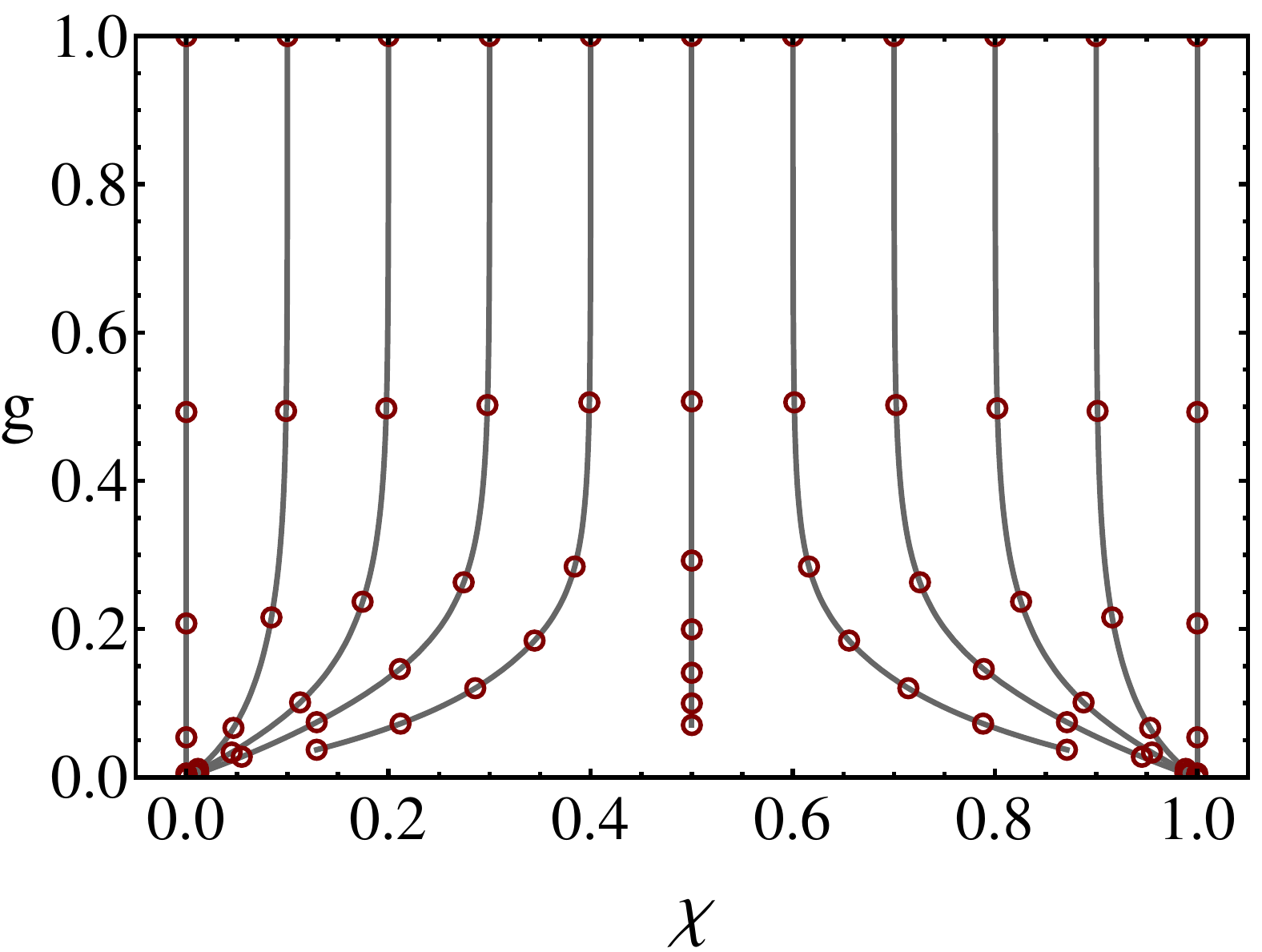}
}
\caption{Flow of the conductance $g$ and the topological parameter $\chi$ as a
function of system size for class AIII system. Dots are for values,
$L/    \tilde\xi=1,2,4\dots,32$.}
\label{fig:FlowGraph_AIII}
\end{figure}

These equations quantitatively describe the scaling behavior anticipated on qualitative grounds above: for generic bare values $(    \tilde\xi, \tilde
\chi)$ we obtain an exponentially fast flow of $(g(L),\chi(L))$  towards an
insulating state $(0,n)$. At criticality,
$(\tilde\xi,n+1/2)$, the topological number remains invariant, while
algebraic decay of the conductance $g(L)\approx \sqrt{\tilde\xi/\pi L}$
indicates the presence of a delocalized state at zero energy (i.e. in the center of the gap of a clean system). Introducing
the scaling form  $
\xi(\tilde\chi)=    \tilde\xi |\tilde \chi-n-1/2|^{-\nu}$ and comparing the ansatz, $g\sim
\exp(-L/\xi(\tilde \chi))$, with the result above, we obtain the correlation
length exponent $\nu=2$ describing the exponential decay of the average
conductance, $\langle g\rangle$. (This exponent differs from  $\protect{\nu=1}$ for the
typical correlation length, $\protect{\left\langle \xi
\right\rangle}= -L/\langle \ln g\rangle$.\cite{BMSA:98,Mondragon-Shem:2013bs}) The flow is shown graphically in Fig.~\ref{fig:FlowGraph_AIII}, and it represents the 1d
analogue of the two-parameter flow diagram~\cite{Khmelnitskii83} describing
criticality in the integer QH system.

\begin{figure}
\centering{
\includegraphics[width=7cm]{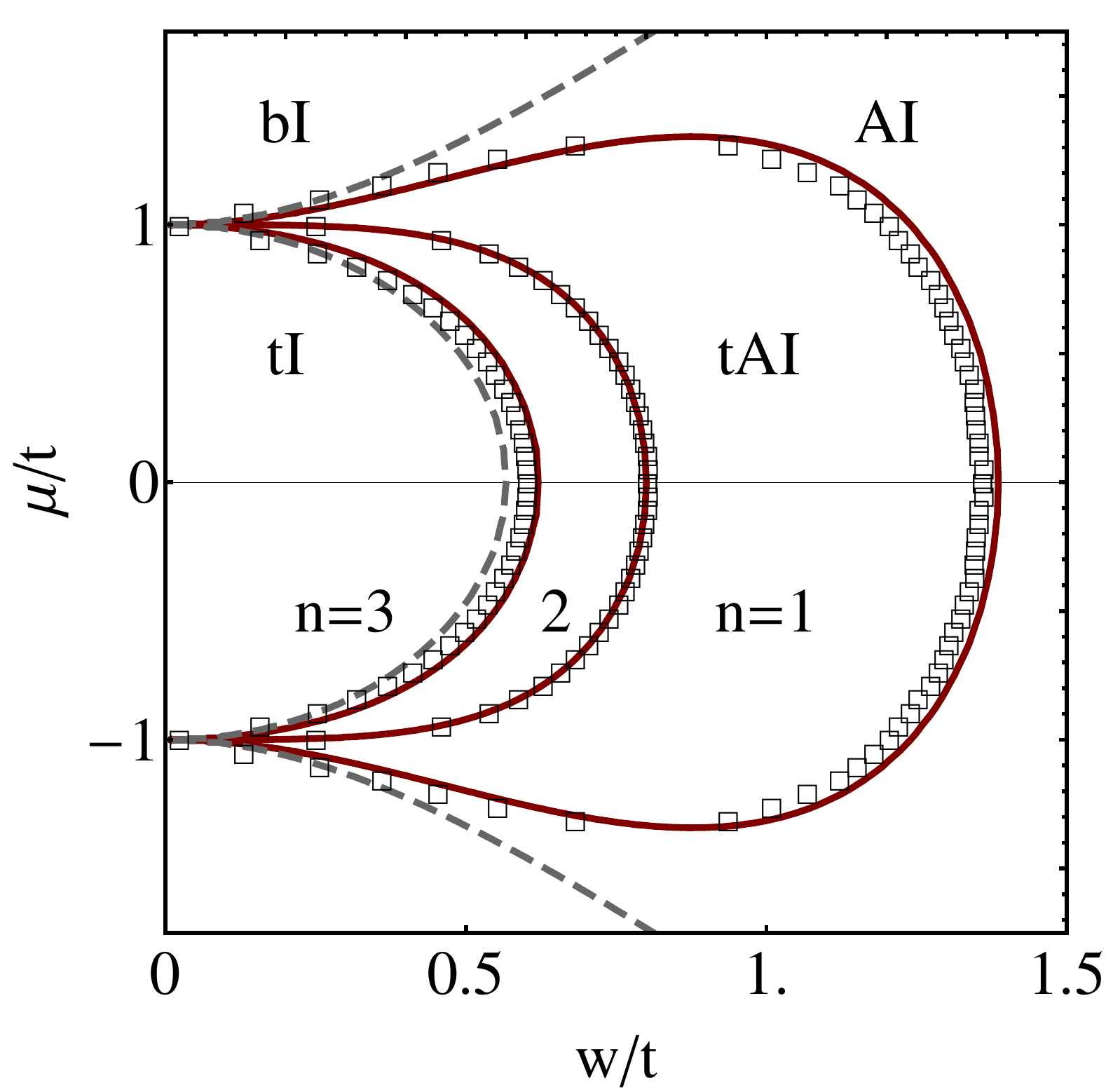}
}
\caption{Phase diagram of the $\mathrm{AIII}$ class 3-channel disordered wire.
Dashed lines show crossover regions between band insulator (bI) and Anderson insulator (AI) or tI and tAI phases, 
derived from the SCBA. Solid lines correspond to half-integer values of the SCBA computed  topological number $\tilde\chi$ and mark boundaries between phases of different $n$. 
bI and AI have $n=0$, while  for tI and tAI $n\neq 0$. Squares: data points, representing phase boundaries found from a numerical analysis of Lyapunov exponents, section \ref{sec:Scattering}.
}.
\label{fig:Diagram_AIII_N3}
\end{figure}

In Fig.~\ref{fig:Diagram_AIII_N3} we show the phase diagram of $N=3$--channel disordered AIII wire in 
the $(\mu, w)$ plane. The clean system, $w=0$, exhibits topological phase transitions at $\mu/t=\pm1$. 
Solid lines show  half-integer values of the SCBA computed  topological number $\tilde\chi$, see Appendix \ref{app:classAIII} for the details. Squares show  numerically computed\cite{Altland2014}  boundaries between 
regions with different number of negative Lyapunov exponents, section \ref{sec:Scattering}, of the transfer matrix.  Notice a very satisfactory agreement between numerical transfer-matrix calculation  and SCBA, even though the latter is justified only in $N\gg 1$ limit.

\subsection{Density of states}
\label{sec:DOS-AIII}

The critical physics discussed above also shows in the density of states of
the system. We here recapitulate a few results derived in  more detail in
Ref.~[\onlinecite{Altland2001511}]. At the insulating fixed points, $(0,n)$, the zero energy action of the
system with vacuum boundary conditions reduces to the boundary action
$S_\mathrm{b}[T]$, i.e. the $G$-symmetry remains broken at the metallic system
boundaries, which may be interpreted as `quantum dots' of size $\simeq
\tilde \xi$. At finite energies, $E$, the boundary action representing
the Green function $G_E\equiv (E+i0-H)^{-1}$ at, say, the left
boundary is given by\cite{Ivanov:2002}
\begin{align}
 	S_\mathrm{L}[T]=-n\,\mathrm{str}(\ln T)+ i\frac{\epsilon}{ 2}\,\mathrm{str}(T+T^{-1}),
 \end{align} 
where $\epsilon=\pi|E|/\Delta_\xi$, and $\Delta_\xi$ is the average single
particle level spacing of a wire segment of extension $\tilde \xi$. The
fact that the energy $\epsilon$ enters the action like a `mass term' for the
Goldstone modes reflects the explicit breaking of the chiral symmetry
$\{G^{-1}_E,P\}_+\not=0$. From this expression, the density of states at
the system boundaries is obtained as
\begin{align}
	\label{AIIIDos}
	\rho(\epsilon)=\frac{1}{2 \Delta_\xi}\left\langle
(T+T^{-1})^{\mathrm{bb}}\right\rangle.
\end{align}
The integral can be done in closed form, and as
a result one obtains\cite{Verbaarschot1993,Ivanov:2002}
\begin{align}
 	\rho(\epsilon)=\frac{1}{\Delta_\xi}\left(\pi n \delta(\epsilon)+\frac{\pi
 	\epsilon}{2}\left[J_n^2(\epsilon)-J_{n+1}(\epsilon)J_{n-1}(\epsilon)\right]\right).
 \end{align} 
The first term here represents the $n$ topologically protected zero energy
states, and the second describes the rest of the spectrum in terms of a
bathtub shaped function which remains strongly suppressed up to values
$|E|\sim n \Delta_\xi$. This suppression reflects the level
repulsion off the zero energy states in the chaotic scattering environment
provided by the disorder. For larger energies, the second term asymptotes to
unity, i.e. $\rho(\epsilon) \stackrel{\epsilon \gg 1}{\longrightarrow} \Delta_\xi^{-1}$. The
boundary DoS obeys the sum rule $\lim_{\Delta \epsilon\to \infty} \int_{-\Delta \epsilon}^{\Delta \epsilon}d\epsilon\, \rho(\epsilon)=2 \Delta\epsilon/\Delta_\xi$, i.e. the spectral weight $n$
sitting at zero all is taken from the bulk of the spectrum.

At criticality, the bulk of the system remains in a symmetry broken state. The
transfer matrix method discussed above may then be applied to compute the bulk
density of states~\eqref{AIIIDos} at observation points $\xi \ll x\ll L$ deep
in the system. The result\cite{Altland2001511}: $\nu(\epsilon)=- \frac{\nu_0}{\epsilon\ln^3(\epsilon)}$ shows a strong
accumulation of spectral weight at the band center. This spectral anomaly is
based on the same buildup of long range correlations that gives rise to the
delocalization phenomenon. Heuristically, one may interpret it as a `channel'
through which a left and a right boundary state hybridize at the critical point
to move away from the zero energy.

\subsection{Topological Sources}

Unlike the locally confined source terms commonly used to compute
observables from field theories, the phase variable $\phi$ employed above is a
`topological' source', i.e. one that twists boundary conditions and is defined
only up to local deformation. In view of our later consideration of other
symmetry classes we here briefly discuss the geometric principles behind this construction
and how to extract the variable pair $(g, \chi)$ from the field theory by boundary
twists generalizing the phase variable $\phi$ to other symmetry classes.

In all one-dimensional cases the relevant  fields are `maps'
$Q:S^1\to G/H$ from a circle (the quantum wire compactified to a ring) into a
Goldstone mode manifold realized as the quotient of a full symmetry group $G$ over
a group of conserved symmetry $H$, e.g. $G=\mathrm{GL}(1|1)\times
\mathrm{GL}(1|1)$ and $H=\mathrm{GL}(1|1)$ above. We are putting quotes in
`map' because it is essential to include  fields subject to boundary twist,
i.e. configurations that cannot be described in terms of smooth maps. Also, in
some cases, the Goldstone mode manifold includes a discrete Ising type sector
$\sim \Bbb{Z}_2$ which is non-smooth by itself. A more geometric way to think
of the fields would be in terms of sections of a bundle structure, where the
latter has $S^1$ as its base, and $G/H$ as fibers. Within this setting, the
emergence of boundary twist means that we will be met with `nontrivial
bundles', i.e. the ones that cannot be reduced to a product space $S^1\times G/H$.
This is another way of saying that in the presence of twist there are no
globally continuous fields. On the bundle structure, the group $G$ acts as a
local symmetry group, e.g. by the transformations $T\to T_L T T_R^{-1}$,
$(T_L,T_R)\in G=\mathrm{GL}(1|1)$, which makes our theory a gauge theory. The
source fields employed to compute observables are gauge transformations by
themselves, and they do cause boundary twist. In more mathematical language,
one would say that the bundle is equipped with a non-trivial connection, i.e.
a twisted way of parallel transportation. The absence of periodicity on the
twisted background can be equivalently described as the presence of
non-vanishing curvature or gauge flux. The theory responds to the presence of
such type of connection in terms of deviations of the partition sum $Z$ off
unity, and in this way the observable pair $(g,\chi)$ can be obtained. The
situation bears similarity to the quantum mechanical persistent current
problem, where the presence of a magnetic flux (or twisted boundary
conditions) leads to flux-dependence of the free energy (corresponding to our $Z$).
In that context, the insertion of a full flux quantum generates spectral flow,
i.e. a topological response (similar to our $\chi$), while the probing of
`spectral curvature', i.e. a second order derivative w.r.t. the flux generates
a dissipative response ('Thouless conductance', similar to our $g$).

The question then presents itself how the connection yielding the observables
should be chosen in concrete cases. (In view of the dimensionality $>1$ of the
target manifolds there is plenty of freedom in choosing twisted connections,
which nevertheless may yield equivalent results.) Below, we will approach this question
in pragmatic terms, i.e. we have an expression of the topological invariants
in terms of Green functions, these Green functions can be represented in terms
of Gaussian superintegrals (cf. Eq.~\eqref{GaussianIntegral}) subject to a
source, and that source then lends itself to an interpretation as a gauge
field acting in the effective low energy field theory (cf.
Eq.~\eqref{eq:AIIIAction}.)  While the concrete implementation of this
prescription depends on the symmetry class, and in particular on whether a
$\Bbb{Z}$ or a $\Bbb{Z}_2$ insulator is considered\footnote{In the $\Bbb{Z}_2$
cases, we will be  led to non-trivial $\Bbb{Z}_2$ connections, i.e. the
placement of a \emph{single} Ising like kink into a circular structure, a
configuration that clearly would not work on a trivial product space
$S^1\times \Bbb{Z}_2$ (on which only even numbers of kinks are permissible.)
In these cases, the gauge flux simply is the number of kinks mod 2.}  
the
general strategy always remains the same. Likewise, the extension of the
source formalism to one yielding the dissipative conductance is comparatively
straightforward, as discussed in the specific applications below. We finally
note that the global gauge formalism can be generalized to higher dimensions,
Pruisken's `background field method'\cite{Pruisken1984a} being an early example of a $d=2$
implementation. For further discussion of this point see
section~\ref{sec:comparison_to_2d}.

\subsection{Class BDI}
\label{sec:BDI}

We next extend our discussion to the one-dimensional $\mathds{Z}$-insulator in
the presence of time reversal, symmetry class $\mathrm{BDI}$. Class BDI can be
viewed as a time reversal invariant extension of class AIII discussed above.
Readers primarily interested in the much more profound differences between
$\Bbb{Z}$ and $\Bbb{Z}_2$ insulators, are invited to directly proceed to
section~\ref{sec:Z2}

\minisec{Model Hamiltonian}
 Systems of this
type are realized, e.g. 
as  $N$-channel lattice $p$-wave superconductors\cite{Kitaev2001}  with the Hamiltonian
\begin{equation}
\label{eq:BDI_model_H}
  H = \sum_{s=1}^L [C^\dagger_s 
H_{0,s} C_s+(C^\dagger_s  H_{1,s} C_{s+1}+\mathrm{h.c.})],
\end{equation}
where the spinless fermion operators $C_s=(c_{s,k},c_{s,k}^\dagger)^T$ are
vectors in channel and Nambu spaces with $s$ being site and $k=1, ..., N$
being channel indices. The on-site part of the Hamiltonian,  ${H_{0,s}= (\mu +
V_{s}) \sigma_3}$ contains the chemical potential $\mu$ and real symmetric
inter-chain matrices  $ V_s^{kk'}$. The Pauli matrices $\sigma_i$ operate in
Nambu space. The inter-site term, $H_{1,s}=-\frac12 t_s
\sigma_3+ \frac12 \Delta_s \sigma_2$, contains nearest neighbor
hopping, $t_s$,  and the order parameter, $\Delta_s^{kk'}$, here assumed to be
imaginary for convenience. Quantities carrying a subscript `$s$' may contain
site-dependent random contributions. The first quantized representation of
$  H$ obeys the chiral symmetry $\{  P,  H\}_+=0$, with $ 
P=\sigma_1$ and the BdG particle-hole symmetry $\sigma_1   H^T \sigma_1 = -  H$.
The combination of these two results in the effective time-reversal symmetry $H^T = H$. 
In what follows we consider the simplest model of disorder in which $t_s=\Delta_s=t$ 
are non-random and diagonal in the channel space while the matrices  
$V_s^{kk'}$ are Gaussian distributed as
\begin{equation}
\langle V_{s}^{kk'} V_{s}^{k''k'''} \rangle = (w^2/N) ( \delta_{k'k''} \delta_{kk'''} + \delta_{kk''} \delta_{k'k'''}),
\end{equation}
and the parameter $w$ sets the strength of the disorder.

\minisec{Field theory}

Due to the presence of both chiral and time-reversal symmetry the Goldstone mode manifold
of the effective low-energy field theory in the BDI class spans the coset space
$\mathrm{GL}(2|2)/\mathrm{OSp}(2|2)$~\cite{Zirnbauer1996} which can be parameterized in terms of
$4\times 4$ matrices $Q=T\bar T$, where the 'bar' operation is defined as $\bar T = \tau T^T \tau^T$ and 
$\tau={\cal P}^\mathrm{b}\otimes \tau_1 + {\cal P}^\mathrm{f}\otimes i\tau_2$. Here  ${\cal P}^\mathrm{b}$ and 
${\cal P}^\mathrm{f}$ are projectors on the bosonic and fermionic space while $\tau$-matrices operate 
in the so-called 'charge-conjugation' space. It is clear from this parametrization that all matrices 
$T$ obeying $\bar T = T^{-1}$ form the subgroup $K=\mathrm{OSp}(2|2)$ in the larger group $G=\mathrm{GL}(2|2)$
and do not contribute to the $Q$-field, which thereby spans the coset $G/K$.
By considering rotations in the fermionic sector only, one finds that 
$T_{\rm ff} \in {\rm U}(2)/{\rm Sp}(2) \simeq {\rm U}(1) \simeq S_1$. 
The non-trivial homotopy group $\pi_1(S_1) = \mathds{Z}$ implies the presence of winding 
numbers in the low-energy field theory.

For our subsequent discussion we will need the parametrization of the Goldstone manifold spanned
by 8 coordinates, three of which, $(y_0, y_1, y_2)$ with $y_0 \in \mathds{R}$, $y_1 \in [0,2\pi[$,
 and $y_2 \in \mathds{R}^+$, play the role analogous to the radial coordinates 
of the $\mathrm{AIII}$ manifold. It reads
\begin{equation}
\label{eq:Q_BDI}
Q = e^{\cal W} {\cal Q} e^{-{\cal W}},
\quad {\cal Q} = \left(
\begin{array}{cc}
{\cal Q}^{\rm bb} & \\
& {\cal Q}^{\rm ff}
\end{array}\right)^{\rm bf}
\end{equation}
where the ff-block ${\cal Q}^{\rm ff} = e^{2i y_1}\tau_0$ is parametrized by a compact radial variable $y_1$ and 
the bb-block is parametrized by two hyperbolic radial variables $y_{0,2}$
and one angle $\alpha$, 
\begin{equation}
{\cal Q}^{\rm bb} = e^{2 y_0}\times e^{i\alpha\tau_3} e^{2 y_2\tau_1} e^{-i\alpha\tau_3}.
\end{equation}
The off-diagonal rotations mixing bosonic and fermionic sectors have the form 
\begin{equation}
{\cal W} =\left(
\begin{array}{cc}
&{\cal B} \\
\tilde{\cal B} & 
\end{array}
\right)^{\rm bf}, \quad
{\cal B} = \left(\begin{array}{cc}
\xi & \nu \\
\mu & \eta
\end{array}\right),
\end{equation}
where ${\cal B}$ is a matrix in charge-conjugation space depending solely on Grassmann angles and 
$\tilde{\cal B} =  i\tau_2 {\cal B}^T \tau_1$.

The field theory action of the BDI disordered system has the same form as in the class $\mathrm{AIII}$,
\begin{equation}
S[Q]=\!
\int\limits_0^{L} \!\! dx \left[-\frac{\tilde \xi}{16}\,\mathrm{str}(\partial_x Q\partial_x Q^{-1})
+ \frac{\tilde\chi}{2}\,\mathrm{str}(Q^{-1}\partial_x Q)\right],  
\label{eq:DBIAction}    
\end{equation}  
and a sketch of its derivation is outlined in Appendix~\ref{app:BDI}. The topological coupling constant is given by
$\tilde \chi = \frac{i}{2}{\rm tr}(G^+ P \partial_k H)$, where the retarded Green's function $G^+$ has to be calculated within the SCBA. The concrete dependence of $\tilde \chi$ on the parameters defining the model~\eqref{eq:BDI_model_H} will be discussed below.

The 'partition sum' of the $\mathrm{BDI}$ system is again given be Eq.~(\ref{ZDefAIII}), and its path integral representation reads  
$Z(\phi)=\int {\cal D} Q \exp(-S[Q])$,
where the integral is over all smooth realizations of the $Q$-field with fixed initial and final configuration, 
$Q(0)= \mathds{1}$ and 
$Q(L)={\rm diag}(e^{2\phi_0}, e^{2i\phi_1})^{\rm bf}$.
As in the $\mathrm{AIII}$ system its non-trivial content $\Psi(\phi, L)\equiv Z(\phi,L)-1$ 
can be found from the solution of the transfer matrix equation~(\ref{AIIISchrodinger}), which is now defined for 
three radial coordinates $y=(y_0,y_1,y_2)$, with  Jacobian  
\begin{equation}
J(y)=\frac{\sinh(2y_2)}{16 \sinh^2(y_0-i y_1+y_2)\sinh^2(y_0-i y_1-y_2)},
\end{equation}
and vector potential  $A=2\tilde\chi(i,1,0)^T$.
The partition sum is obtained from the solution of the equation at the radial configuration  
$y=\phi\equiv (\phi_0, \phi_1, 0)$. 

The spectrum of the transfer matrix operator can be found by analyzing the asymptotic
of the eigenfunctions $\psi_l(y)$ at large values of variable $y_{0,2}$. In this regime the $\sinh$-functions
simplify to exponentials and the eigenfunctions $\psi_l(y)$ show the same exponential profile. In this way we find
\begin{equation}
\label{eq:spectrum_BDI}
\epsilon(l_0, l_1, l_2) = 1+ (l_0-2i\tilde\chi)^2 + (l_1-2\tilde\chi)^2 + l_2^2, 
\end{equation}
with $l_1 \in 2\mathds{Z}$ and $l_{0,2} \in \mathds{R}$.

Obtaining the initial value solution $\Psi(\phi,L)$ requires the application
of more elaborate techniques. The key is to extend the super-Fourier analysis
of Ref.~[\onlinecite{Mirlin1994}] for the three standard Dyson
symmetry classes to the symmetry classes presently under consideration.
Relegating an exposition of  mathematical details to a subsequent publication,
we here state  only the main results. For any set of radial coordinates
$y=(y_0,y_1,y_2)$ the partition sum can be written as a spectral sum analogous
to Eq.~(\ref{eq:Z_phi_AIII}) for the class $\mathrm{AIII}$ system
\begin{equation}
\label{eq:Psi_y_BDI}
\Psi(y,L) = \sum_{l_1 \in 2\mathds{Z}} \int \frac{dl_0 dl_2}{(2\pi)^2} \,\mu(l)\psi_l(y)e^{-\epsilon(l) L/2\tilde \xi}. 
\end{equation}  
Here $l=(l_0, l_1, l_2)$ denotes the set of quantum numbers, and the measure $\mu(l)$ is found to be 
\begin{equation}
\mu(l) = 
\frac{(\pi l_2/8) \tanh(\pi l_2/2)}{\left[ l_2^2 + (il_0 + l_1 - 1)^2\right]\left[ l_2^2 + (il_0 + l_1 + 1)^2\right]}.  
\end{equation}
The functions $\psi_l(y)$ appearing in the Fourier
expansion~(\ref{eq:Psi_y_BDI}) are the generalized spherical eigenfunctions of
the Laplace-Beltrami operator on the coset space $G/K$. They do not depend on
the vector potential $A\sim \tilde \chi$. As in the AIII case, the topological parameter enters the solution $\Psi(y,L)$ only   through the $\tilde\chi$-dependence
of the spectrum $\epsilon(l)$, Eq.~(\ref{eq:spectrum_BDI}).

While for arbitrary $y$ the wave function $\psi_l(y)$ cannot be written in
closed form, an integral representation due to Harish-Chandra~\cite{Helgason}
exists. The analysis of this representation greatly simplifies for the
configuration of interest, $y=(\phi_0,\phi_1,0)$.
Using the Harish-Chandra integral representation for $\psi_l(\phi)$ we  obtain the generating function ${\cal F}(\phi)$, Eq.~(\ref{eq:F_phi}), as
\begin{equation}
\label{eq:F_phi1_BDI}
{\cal F}(\phi) = 4\, \partial_{\phi_1}\sum_{l_1 \in 2\mathds{Z}}\int \frac{dl_0 dl_2}{\pi^2}\,\mu(l)
e^{(il_1 -l_0)\phi_1}
e^{-\epsilon(l) L/2\tilde \xi}.
\end{equation}  
From this result the asymptotic values of $\chi(L)$ and $g(L)$ in the limit $L/\tilde\xi \gg 1$ can be extracted,
and we obtain results qualitatively similar to those of the $\mathrm{AIII}$ system. For example, 
far from criticality  
keeping the dominant terms in the Fourier series~(\ref{eq:F_phi1_BDI}) we find 
\begin{eqnarray}
\chi(L) &\simeq& n  + \frac{1}{4}\, {\rm sign}(\delta\chi)\,  \sqrt{\frac{\xi(\tilde\chi)}{\pi L}} \,
e^{-L/\xi(\tilde\chi)},
\nonumber \\
g(L) &\simeq& \frac{1}{2}\sqrt{\frac{\tilde\xi}{\pi L}} \,
e^{-L/\xi(\tilde\chi)},
\end{eqnarray}  
where as before $\tilde\chi = n + \delta \chi$ and the localization length 
$\xi(\tilde\chi)=\tilde\xi|\tilde \chi - n - 1/2|^{-2}$.

\begin{figure}
\centering{
\includegraphics[width=7cm]{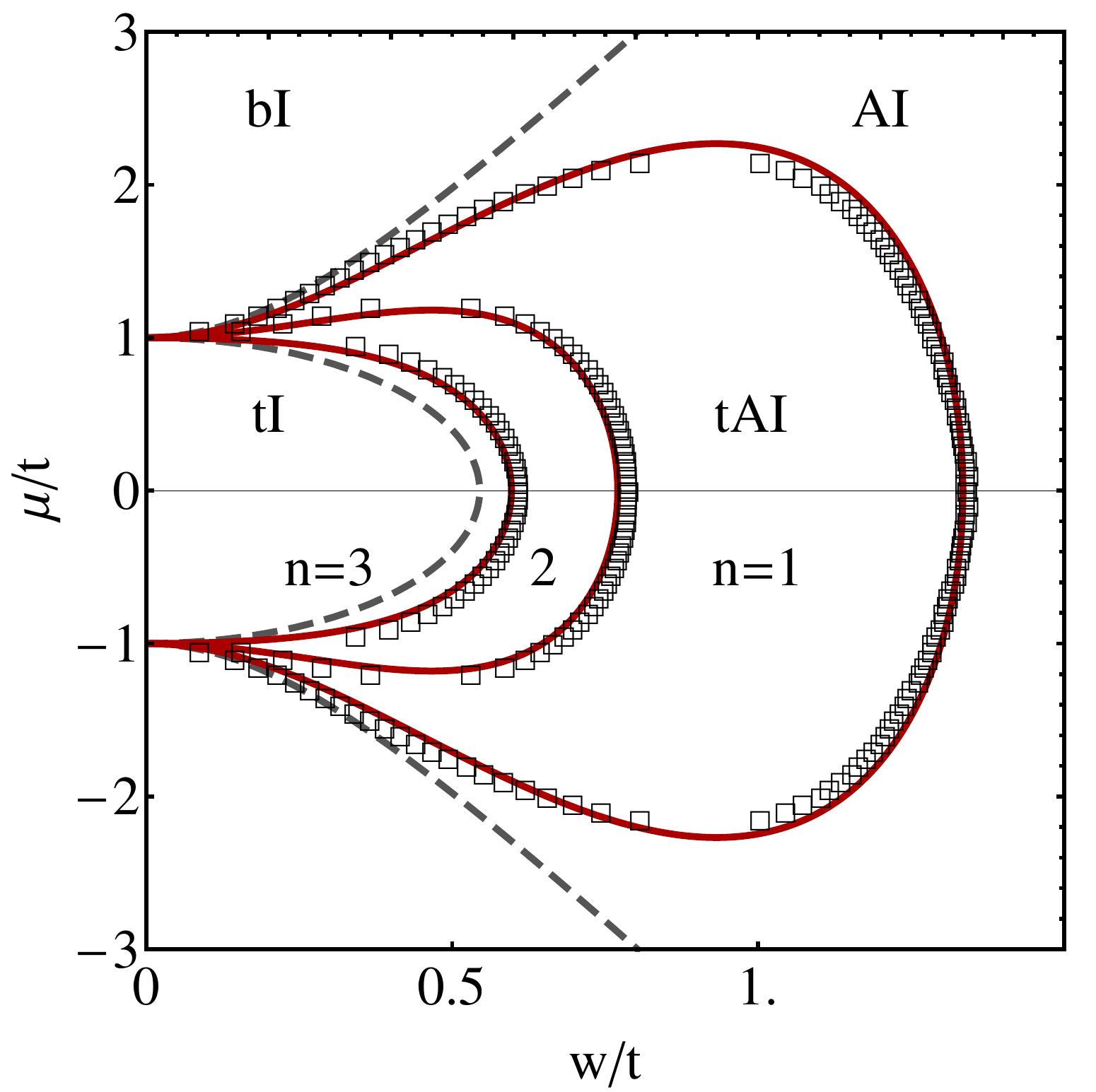}
}
\caption{Phase diagram of the $\mathrm{BDI}$ class 3-channel disordered $p$-wave superconducting wire.
Dashed lines show crossover regions between band insulator (bI) and Anderson insulator (AI) or tI and tAI phases, 
derived from the SCBA. Solid lines correspond to half-integer values of the SCBA computed  topological number $\tilde\chi$ and mark boundaries between phases of different $n$. 
bI and AI have $n=0$, while  for tI and tAI $n\neq 0$. 
}.
\label{fig:Diagram_BDI_N3}
\end{figure}

\minisec{Phase diagram} 

For the model of the $N$-channel $p$-wave wire
defined above, the constant $\tilde \chi=\tilde\chi(\mu,w)$ takes values in
the interval $(0, N)$. Its explicit form can be found analytically in limiting
cases. Specifically, in the low energy limit $|\mu - t| \ll t$ we obtain
\begin{equation}
\label{eq:chi_BDI_1}
\tilde\chi(w,\mu) =  \frac{N}{4} \left( 3 + \frac{(t-\mu) t}{w^2} \right),
\end{equation}
while in the limit $\mu \to 0$ and for any disorder strength $w$,
\begin{equation}
\label{eq:chi_BDI_2}
\tilde\chi(\mu=0,w) =  N t^2/2 w^2.
\end{equation}
Localization is avoided if $\tilde\chi(\mu,w) = n+1/2$, 
with integer $n \in [0, N-1]$, and the corresponding contour lines in the $(\mu,w)$-plane define  
boundaries between  different phases of the tAI with indices $n$ and $n+1$. The ensuing phase diagram at $N=3$ is shown in 
Fig.~\ref{fig:Diagram_BDI_N3}. Using Eq.~(\ref{eq:chi_BDI_2}) we find that the phase transition
points on the $(\mu = 0;w)$ line are located at 
\begin{equation}
w_n=t(2N/(2n+1))^{1/2},
\end{equation} 
where $0 \leq n <N$ (cf. also Ref.\onlinecite{Rieder2013}). Similarly, employing relation~(\ref{eq:chi_BDI_1})
we find that the degenerate phase transition point $(\mu,w) = (t; 0)$ on the clean system ordinate
splits into the set of $N$ critical parabolas 
\begin{equation}
\mu_n(w)=t+\frac{w^2}{t}\left(3-\frac{4n+2}{N}\right).
\end{equation}
We have also compared the profiles of the $\chi=n+1/2$ contour lines obtained
by SCBA evaluation of the topological parameter against numerical transfer
matrix method.  The excellent agreement  was found, in spite of the fact that strictly speaking
the field theory approach requires $N\gg 1$. The diagram in
Fig.~\ref{fig:Diagram_BDI_N3} supports the qualitative discussion of the
introductory section. In particular, one observes that somewhat
counterintuitively the {\em increase} of disorder strength $w$ at fixed
chemical potential $\mu$ may induce the quantum phase transition (and thus
{\em delocalization}!) from the trivial Anderson insulator ($n=0$) to the tAI
($n=1$)\cite{Fulga2011,Fulga2012}.

We conclude by noting that the physics of the class CII quantum wire, governed
by a chiral time-reversal invariant Hamiltonian, with broken spin-rotation
invariance, is essentially similar to that of the AIII and BDI systems. A quantitative
solution along the lines of the ones discussed above can be formulated, but it
does not add qualitatively new information and we do not discuss it here.

\section{$\Bbb{Z}_2$ insulators}
\label{sec:Z2}

In many ways, the effective theories of the five classes of
topological quantum wires resemble each other. All five can be described in terms of  two
parameter nonlinear $\sigma$-models, and in all cases critical flows characterized by the fixed point
structure~\eqref{FlowGeneric} are predicted. (Similar parallels are observed in 2d, cf.
section \ref{sec:comparison_to_2d}.) However, there are also important differences notably  between
the $\Bbb{Z}$ and $\Bbb{Z}_2$ representatives. The rule of thumb is that while
the topological textures responsible for the flow in the $\Bbb{Z}$-insulators
are \emph{smooth} -- phase windings in 1d, instantons in 2d -- they are
singular in the $\Bbb{Z}_2$ systems -- point defects in 1d, line defects in 2d. The 
$\Bbb{Z}_2$-wires considered below are driven into a localized regime by a proliferation of
kinks. Much like vortices in 2d (which have been seen\cite{Kane-Fu2012} to play a similar role
there), kinks are topological in nature, however they cannot be
described in terms of a gradient-topological term. Its role is taken, rather,
by a fugacity term, i.e. a term describing the action cost of individual
kinks. The fugacity coefficient, $\ln(\tilde \chi)$ assumes the role of the
$\theta$-angle in the $\Bbb{Z}$-insulators.

In the following we describe the construction and solution of the theory on
the example of the class D quantum wire, i.e. the spin-rotation and time-
reversal symmetry broken system currently under intense experimental and
theoretical scrutiny. We then generalize the treatment to class DIII system. 

\subsection{Definition of the model}

The Bogoliubov-deGennes Hamiltonian of  class D superconductor obeys the symmetry
relation $H^T=-\sigma_1^\mathrm{ph} H \sigma_1^\mathrm{ph}$, where the Pauli
matrices act in particle-hole space. One may perform unitary transformation to
`real' superpositions of particle and hole degrees of freedom -- the
`Majorana' basis -- in which the symmetry assumes the simple form $H=-H^T$. We
will work in this basis throughout, and model our system as a chain of $L$ coupled 
`dots', $s=1,\dots, L$ where each dot represents a disordered superconductor. The corresponding Hamiltonian reads
\begin{align}
	\label{D-DotHamiltonian}
	H = \sum_{ss'}C_s^\dagger \left[{\cal H}_s \delta_{ss'}+iW(\delta_{ss'+1}-\delta_{s's+1})\right]C_{s'},
\end{align}
where ${\cal H}_s=-{\cal H}_s^T$ is a matrix ${\cal H}_s= \{{\cal
H}_s^{kk'}\}$ with random contributions, and the inter-dot coupling matrices,
$W=W^T$, are assumed to be non-random. Without loss of generality, we may
choose a basis in which $W=\mathrm{diag}(w_1,\dots,w_{2N})$ is diagonal.

\subsection{Topological invariant}

In the clean case,  the $\Bbb{Z}_2$ invariant carried by the system is defined
as\cite{Kitaev2001}  $\mathrm{sgn}\left(\mathrm{Pf}(H_\pi)/\mathrm{Pf}(H_0)\right)$, where
$H_q$ is the first quantized Hamiltonian defined by the bilinear
form~\eqref{D-DotHamiltonian}, and $q$ is the wavenumber conjugate to the index $s$.
The definition may be generalized to one working in the presence of
disorder\cite{Groth:2009fv}  by interpreting the $L$-site chain as one giant unit-cell of an
infinitely extended system. Within this interpretation, the system is
described by a complicated Hamiltonian $H'$ containing $\propto L$ bands,
whose Brillouin zone is given by the cutoff momentum $2\pi/L$. The invariant
is now given by $\mathrm{sgn}(\mathrm{Pf}(H'_{\pi/L})/\mathrm{Pf}(H'_0))$. We
may imagine the system compactified to a ring, in which case the ratio is that
of Pfaffians of Hamiltonians in the presence/absence of a half magnetic flux
quantum threading the ring. That flux picture is gauge equivalent to one where
the phase $\pi$ picked up upon traversal of the ring is concentrated on one of
its links, i.e. we may obtain the invariant by taking the ratio
$\mathrm{sgn}(\mathrm{Pf}(H_\pi)/\mathrm{Pf}(H_0))$, where $H_0\equiv H'$ 
and $H_\pi$ differs from $H_0$ by the sign inversion of one of the bond
matrices, e.g. $W\to -W$, say, at the bond $0\to 1$. We will use this representation throughout.

\subsection{Field theory}
\label{sec:FT_classD}

In this section we introduce a partition sum for the class $\mathrm{D}$ wire which is able to generate
the conductance and the $\mathds{Z}_2$ topological
invariant. To this end we consider the super-Gaussian integral 
\begin{align}
	\label{eq:Zsum_D}
	Z=\left\langle \int D\Psi\,e^{i\bar \Psi G^{-1}\Psi
	}\right\rangle,
\end{align}
where $\Psi=\{\Psi_{sk}^{\alpha,t}\}$ is a super-vector field carrying site
indices $(s,k)$, a super-index $\alpha=\mathrm{b},\mathrm{f}$ distinguishing between commuting
and anti-commuting indices, and a two-component `charge conjugation' index
$t=1,2$. The vectors $\bar \Psi$ and $\Psi$ are mutually dependent through the
symmetry relation
\begin{align}
	\label{eq:DPsiSym}
	\Psi=\tau\bar\Psi^T, \qquad \tau=\mathcal{P}^\mathrm{b}\otimes (i\tau_2)+\mathcal{P}^\mathrm{f}\otimes \tau_1,
\end{align}
where  $\tau_i$ are matrices acting in charge-conjugation space. Further,
$G^{-1}\equiv i0\tau_3-H$ comprises the retarded and advanced Green functions.
In Appendix~\ref{app:D}, we discuss the relation of the symmetry
structure~\eqref{eq:DPsiSym} to the anti-symmetry of the Hamiltonian, $H=-H^T$.

As it stands, $Z=1$ is unit normalized by supersymmetry. To obtain useful information from
the integral, we couple it to a gauge field $a=(\phi,\sigma)\in \Bbb{R}\times
\Bbb{Z}_2$ comprising a $\mathrm{U}(1)$ phase variable $\phi$ and a
$\Bbb{Z}_2$ variable $\sigma=\pm $. The former acts only in the bosonic
sector of the theory, $\alpha=0$, and the latter in the fermionic sector
$\alpha=1$. The field $a$ is non-vanishing only on one link of the lattice, which we choose to be the $0\leftrightarrow 1$ link. On this link, we replace the hopping operator $ i W|0\rangle \langle 1| +\mathrm{h.c.} $ by 
\begin{align}
	\label{eq:DSourceLink}
 	iW \left( \mathcal{P}^\mathrm{b} e^{\phi \tau_1}+\mathcal{P}^\mathrm{f}\tau_1^{(1-\sigma)/2} \right)|0\rangle \langle 1|
+\mathrm{h.c.}\bigl|_{\phi \to -\phi}.
 \end{align} 
We denote the Green function modified in this way as $G_a\equiv
G_{(\phi,\sigma)}$. Notice that up to a unitary transformation diagonalizing
$\tau_1\to \tau_3$, the fermionic sector of $G_{(0,-)}$ comprises an
unperturbed Green function (the eigenvalue $+1$ of $\tau_3\leftrightarrow
\tau_1$) $G_0$ and one $G_\pi$ that contains a sign-inverted hopping matrix
element on the $0\leftrightarrow1$ link (the eigenvalue $-1$).  Denoting the partition function defined for the
supersymmetry broken Green function $G_a$, $a\not=(0,+)$ as $Z(a)\equiv Z^{(\sigma)}(\phi)$, it is
straightforward to verify that
\begin{equation}
	\label{eq:Dclass_chi}
	\chi \equiv Z^{(-)}(0)=\left\langle \frac{\mathrm{Pf}(G_{\pi}^{-1}) }{\mathrm{Pf}(G_{0}^{-1}) }.
\right\rangle
\end{equation}
Indeed, the integral over the bosonic variables sandwiching the unperturbed
Green function at $\phi=0$ produce a factor $\det(G_0^{-1})=\mathrm{Pf}(G_0^{-2})$
in the denominator, while the integral over the Grassmann variables gives a
factor $\mathrm{Pf}G_0^{-1}\mathrm{Pf}G_\pi^{-1}$ in the numerator where the
two factors come from the distinct eigenvalue sectors mentioned above, and the
integration over Grassmann variables produces Pfaffians (rather
than determinants) because  $\bar\Psi$ and $\Psi$ contain
the same integration variables\footnote{To see it
explicitly let us represent $H=H' + W'$, with $H'$ being the Hamiltonian of
the auxiliary system with the deleted single bond $0 \to 1$ and $W'$ is the hopping matrix on this bond. 
Then the fermionic action with the
inserted source reads $S_{\rm f} = \frac{1}{2}\bar\Psi^1 (\tau_0 \otimes H' +  \tau_1 \otimes W' )\Psi^1 = 
\frac{1}{2}(\Psi^1)^T ( \tau_1 \otimes H' +  \tau_0 \otimes W' ) \Psi^1$, where we have used
$\bar \Psi^1 =(\Psi^1)^T \tau_1$. We can now apply the orthogonal $(\pi/4)\tau_2$ rotation
in the 'charge-conjugation' space to the spinor $\Psi$ in order to transform the action
into the form $S_{\rm f} = \frac{1}{2} (\tilde\Psi^1)^T (\tau_3 \otimes H' +  \tau_0 \otimes W' )\tilde\Psi^1=
\frac{1}{2} (\tilde\Psi^1)^T 
\left(\begin{smallmatrix} H & \\ & -H_\pi \end{smallmatrix}\right)\tilde\Psi^1$
with $H_\pi=H'-W'$  
being the Hamiltonian  of the system with a sign inversion on the bond $0 \to 1$. 
All Grassmann fields in the rotated spinor $\tilde \Psi$ are independent. Hence, performing the Gaussian integration
we find that the modified fermion partition sum reads $Z(0,-1) = {\rm Pf}(G_{0}^{-1}) {\rm Pf}(G_{\pi}^{-1})$.}. Factor $\mathrm{Pf}G_0$ in numerator and denominator
cancels out, and we are left with the expression above.

In section~\ref{sec:Scattering} we show that 
\begin{equation}
	\label{eq:Dclass_g}
	g=\partial^2_{\phi}\big|_{\phi=0}Z^{(+)}(\phi),
\end{equation}
i.e. the conductance is obtained by probing sensitivity of the partition
function w.r.t. a phase twist in the bosonic sector (and 
unperturbed fermionic sector). The underlying transformation, too, is
`topological' in that it changes the boundary conditions in a way that cannot
be removed by unitary transformation. To summarize, the observable pair
$(g,\chi)$ can be obtained by exposing the partition sum to a boundary
changing gauge transformation which is continuous/discrete in the
bosonic/fermionic sector.

What makes the source a genuine gauge field is its compatibility with the
symmetry transformations of the theory. The action is invariant under
space-uniform  transformations $\Psi
\to T \Psi$, $\bar \Psi \to \bar \Psi T^{-1}$, where compatibility with the
symmetry of the $\Psi$-field requires that $T^T=\tau T^{-1}\tau^{-1}$. This is
the defining relation for the super-group $G=\mathrm{SpO}(2|2)$, where the
notation indicates that $(T^\mathrm{bb})^T = (i\tau_2)^{-1}(T^\mathrm{bb})^{-1}(i\tau_2)$ is
in the non-compact group of real-symplectic  $2\times 2$ matrices, while
$(T^\mathrm{ff})^T = \tau_1^{-1}(T^{\mathrm{ff}})^{-1}\tau_1$ is in the compact  group of
$2\times 2$ orthogonal matrices. We will see momentarily, that on the level of
the effective low energy theory, the symmetry group $G$ of transformations $T$
gets broken to the group $H$ of transformations commutative with $\tau_3$,
i.e. the Golstone mode manifold is $G/H$, and
$(G/H)^{11}=\mathrm{O}(2)/\mathrm{SO}(2)\simeq \Bbb{Z}_2$ reduces to a
discrete set. On this fermion-fermion sector, the source $\sigma$ acts as a
$\Bbb{Z}_2$ gauge field. In the boson-boson sector, the gauge source is
continuous. Later on, we will see that the gauge conformity of the sources
with the symmetries of the theory plays an important role in the solution of
the latter.

\subsection{Disorder average and low energy action}
\label{sec:action-D}

Following the same logic as in  section \ref{ssec:disorder_average_and_low_energy_field_theory} we now perform averaging over the Gaussian 
disorder and introduce  Goldstone Hubbard-Stratonovich field to decouple the ensuing $\Psi^4$-term. Referring for 
technical details to 
Appendix~\ref{app:D}, we here motivate the emerging effective theory by symmetry considerations, 
conceptually analogous to that of section \ref{ssec:disorder_average_and_low_energy_field_theory}. 
The immediate
consequence of the disorder averaging is that the $G$ symmetry gets broken by an emergent self energy $i0
\tau_3 \to \frac{i}{2\tau}\tau_3$ to the subgroup $H=\mathrm{GL}(1|1)$ of
transformations commutative with $\tau_3$ matrix. The resulting Goldstone
mode manifold may be parameterized by $Q=T \tau_3 T^{-1}$, where $T\in G$. This manifold has the
topologically important property of disconnectedness. To see this, we span  the fermionic 
$\mathrm{bb}$-block of the symmetry group $G^\mathrm{ff}\simeq \mathrm{O}(2)$ by two
disconnected set of matrices parameterized, respectively, as $T^{(\pm)}\equiv
\tau_1^{\frac12 (1\mp 1)}e^{i\phi\tau_3}$. This implies that the $(11)$-sector of
the Goldstone mode manifold contains only the two elements $T^{(\pm)}\tau_3
T^{(\pm)-1}=\pm\tau_3$. One may switch from one configuration to the other  by
the symmetry group element $\tau_1$. These observations indicate that the
field theory contains $\Bbb{Z}_2$ kink excitations, which switch between the
two disconnected parts of the Goldstone manifold\cite{Bocquet2000}.

For later reference, we note that a complete parameterization of the two Goldstone mode submanifolds is given by  
\begin{equation}
                                                      \label{eq:Q_par}
Q^{(\pm)} = e^{\cal W} {\tilde Q}^{(\pm)} e^{-{\cal W}},
\end{equation}
where $\tilde Q^{(\pm)}=\tilde  Q^\mathrm{b}\otimes \mathcal{P}^\mathrm{b} +\tilde  Q^{\mathrm{f}(\pm)} \otimes
\mathcal{P}^\mathrm{f}$ is block-diagonal in bf space with  ff-block $\tilde Q^{\mathrm{f}(\pm)}=\pm
\tau_3$ and a  bb-block parametrized by one hyperbolic radial variable $y$
and one angle $\alpha$ as
\begin{equation}
\tilde Q^{\mathrm{b}} = e^{i\alpha\tau_3}e^{y\tau_1}\tau_3e^{-y\tau_1}e^{-i\alpha\tau_3}.
\end{equation}
The boson-fermion  rotations  are given by 
\begin{equation}
                                                    \label{eq:B_blocks}
{\cal W} =\left(
\begin{array}{cc}
&{\cal B} \\
\tilde{\cal B} & 
\end{array}
\right)^{\rm bf}, \quad {\cal B} = \left(\begin{array}{cc}
\xi &  \\
 & \eta
\end{array}\right), 
\end{equation}
where ${\cal B}$ is a matrix in $\tau$-space, 
$\tilde{\cal B} =  - \tau_1 {\cal B}^T i\tau_2$, and $\xi,\eta$ are Grassmann variables.

After integration over the $\Psi$-fields the Goldstone mode partition function assumes the form $Z=\int DQ \,\exp\{-\sum_{s=1}^L  S(Q_s,Q_{s+1})\}$, where $S(Q,Q') =\tilde S(Q,Q')+\ln \sigma(Q,Q')$, where $\sigma$ is a sign factor to be discussed momentarily,  
\begin{equation}
\label{eq:Action_D}
	\tilde S[Q,Q']=\frac{1}{4}\sum_{k=1}^{2N} \mathrm{str}\ln\left(1+\frac{ t_k^2}{4}\big(\{Q,Q'\}_+-2\big)\right),
\end{equation}
$\{\;,\;\}_+$ is the matrix anti-commutator, $ t_k^2 = 4(\pi\nu w_k)^2/(1 +
(\pi\nu w_k)^2)^2$ is the $k$th of $2N$ intra-dot transmission coefficients and
$\nu$ is the DoS in the dot. Actions of this architecture universally appear
in the description of granular (chain of dotes)
matter~\cite{Efetov-book,Nazarov-book,Kamenev-book}. A feature that sets
the action apart from that of an ordinary quantum dot action is the presence
of the sign $\exp(\ln \sigma(Q,Q'))\in \{1,-1\}$. The sign originates in the
fact that the integration over Grassmann variables actually produces a
Pfaffian of the antisymmetric operator  in site, channel, and charge
conjugation space defined by the Gaussian action. That Pfaffian differs from
the square root of a determinant (the action $\tilde S$) if (i) the system
is in a topological phase, and (ii) the matrices $Q_s$ and $Q_{s+1}$
neighboring the link belong to different parts of the manifold, i.e. if there
is a kink sitting on the link. If the system is topological, each such kink
produces a sign in the Pfaffian relative to the (positive) sign of the
determinant. 

Keeping this subtlety in mind, we now turn to the discussion of
the action contribution $\tilde S$. The presence of kinks in the system
invalidates an expansion of the logarithm in `smooth fluctuations'. To compute
the action cost of a kink on a link between the sites $s_0$ and $s_0+1$, we
consider a piecewise constant configuration with field variables $Q^{(+)}$ and
$Q^{(-)} \equiv \tilde T Q^{(+)}\tilde T^{-1}$ at sites $s\leq s_0$ and $s\ge
s_0+1$, respectively, where $\tilde T=\mathcal{P}^0+\mathcal{P}^1\otimes \tau_1$
Substitution of this profile into the action then gives a vanishing
contribution from all links other than $s_0\to s_0+1$. The discontinuity
itself yields $\tilde S_s(Q^{(+)},Q^{(-)})=\frac{1}{2}\sum_k \ln(r_k^2)$,
where $r_k^2=1-t_k^2$ afford an interpretation as squared reflection
amplitudes (cf. section~\ref{sec:Scattering}). In the
topologically non-trivial case, the sign of the products of reflection
coefficients is negative\cite{Fulga2012}, $\prod_k r_k<0$. This means that the sign factor
$\exp(\ln \sigma_s)$, equally negative in the topological case, in $S =\tilde
S + \ln \sigma$ can be accounted for by writing $S(Q^{(+)},Q^{(-)})=\sum_k
\ln r_k$; for negative product of the $r_k$'s, this adds factor $\pm i\pi$ to
the positive action $\tilde S$, as required. Summarizing, the kink action
yields a constant $S_k$ defined through
\begin{align}
	e^{-S_\mathrm{k}}\equiv \prod_{k=1}^{2N} r_k=\det\hat r \equiv \tilde \chi,
\end{align}
Notice that $|\tilde \chi|\leq
1$. In the topological (non-topological) case, $\tilde \chi<0$ ($\tilde \chi>0$).

The identification of the exponentiated kink action, or kink \emph{fugacity},
with the bare value of the topological variable, $\tilde \chi$, can be
understood by representing the reference field configuration as $\dots Q^{(+)}
\leftrightarrow Q^{(+)} \leftrightarrow Q^{(+)} \leftrightarrow Q^{(-)}
\leftrightarrow Q^{(-)} \leftrightarrow Q^{(-)} \leftrightarrow \dots$. This can be
identically rewritten as $\dots Q^{(+)} \leftrightarrow Q^{(+)}
\leftrightarrow Q^{(+)} -\tilde T-Q^{(+)} \leftrightarrow Q^{(+)} \leftrightarrow Q^{(+)}
\leftrightarrow \dots$, i.e. the kink amounts to the appearance of a
$\tau_1$-matrix in the Grassmann sector on the link $s_{0}\leftrightarrow s_0+1$. This, on
the other hand, is equivalent to the substitution of the `topological
source'~(\ref{eq:Dclass_chi}) into the action. The source was designed in such
a way that in its presence the partition sum remains unchanged (trivial
superconductor), or changes sign (topological superconductor).  In the
disordered case, the two options $\tilde
\chi \in ]0,1]$ and  $\tilde \chi \in [-1,0[$ are realized according to a
certain distribution, i.e. we expect the presence of a source to generate a
real valued coefficient $\tilde \chi$. The critical value $\tilde \chi =0$
means a complete blocking of kinks. At any rate, the action cost of an
individual kink is given by $S_\mathrm{k}\equiv-\ln(\tilde
\chi)=-\ln(|\tilde \chi|)+\Theta(-\tilde \chi)i\pi$ where the phase $i\pi$ is
absent (present) in the trivial (topological) case. The phase will be seen
below to be crucial to  the formation of boundary states in the topological
phase.

The gauged partition function $Z(a)=Z^{(\sigma)}(\phi)$ is obtained by evaluating
 the  path integral subject to the twisted boundary condition $Q(0)=\tau_3
 \otimes \mathds{1}^{\rm bf}$, $Q(L) = {\rm diag}( \tau_3 e^{2\phi\tau_1},
 (-1)^{(1-\sigma)/2}\tau_3)^{\rm bf}$. This implies that the path integral in
 the presence/absence of the external $\Bbb{Z}_2$ charge is the sum over
 trajectories with an odd/even number of kinks.

Away from the kinks, the field configurations  are smoothly fluctuating, and a straightforward expansion of the logarithm in Eq.~(\ref{eq:Action_D}) in long wavelength  fluctuations leads to 
\begin{align}
	\label{SDContinuum}
 	S[Q]=-\frac{\tilde \xi}{16}\int dx\,\mathrm{str}(\partial_x Q\partial_x Q)+ \ln{\tilde \chi} \times n_\mathrm{k}, 
 \end{align} 
where the discrete index $s$ is replaced by a continuum variable $x$,
$n_\mathrm{k}$ is the number of kinks and the first term describes the action
of smooth field fluctuations in kink-free regions of the system. Here, the 
'bare' dimensionless localization length, 
$\tilde \xi = g = \sum_{k=1}^{2N}  t_k^2$,
measured in units of the inter-dot spacing coincides with the dot-to-dot to
conductance. We note, however, that the above action is symbolic in that it
does not specify boundary conditions at the terminal points of segments were
kinks occur. To consistently treat the latter, one needs to  retain the
discrete representation (\ref{eq:Action_D}), as detailed in the next section.

For later reference, we notice that the action cost of a configuration with $n$ kinks,  $\ln{\tilde \chi}\times n_\mathrm{k}$ can be represented as 
\begin{equation}
\label{eq:S_top_classD}
S_{\rm top}[Q] = i\frac{\ln(\tilde \chi)}{\pi} \int_0^L dx\, \partial_x {\rm str} \ln (T).
\end{equation}
Indeed, a multi-kink configuration with kinks at $x_i,i=1,\dots, n_\mathrm{k}$
can be parameterized as $T(x)\equiv \tilde T^{f(x)}T^{(+)}(x)$, where
$T^{(+)}$ generates fluctuations in the $(+)$ sector of the manifold, and
$f(x)=\sum_{i=1}^{n_\mathrm{k}}\Theta(x-x_i)$. Since $\det T^{(+)}=1$ we have
$\mathrm{str}\ln (T)=\mathrm{str}\ln \tilde T^f=-\ln (-1)^{f}=-i\pi f$, and
hence $S_\mathrm{top}[Q]=\ln \tilde \chi \times n_\mathrm{k}$, as required.

The structure of the continuum representation~\eqref{SDContinuum} makes the
parallels and differences to the description of the $\Bbb{Z}$-insulators
manifest. In all cases, the system is described by a two-parameter field
theory comprising a standard gradient operator (the first term), and a
`topological term' determining the action cost of topological excitations.
However, unlike with the smooth phase winding excitations of the
$\Bbb{Z}$-insulators, the latter are singular topological point defects, which
means that the role of the topological $\theta$-terms is now taken by the
fugacity counting term. (A similar structure is found in 2d, cf.
section \ref{sec:comparison_to_2d}.) As with the AIII system, the bare values of the coupling
constants may be identified by probing the response of a short system $\tilde
\xi \ll L$ to the presence of sources. Substitution of a single kink into the
system generates $Z^{(-)}(0)=\chi\stackrel{L\ll \xi}=\tilde \chi$ as discussed
above. Likewise, the substitution of a minimal configuration
$Q(x)=\mathcal{P}^\mathrm{b}\otimes
\left(e^{\phi\frac{x}{L}\tau_1
}\tau_3e^{-\phi\frac{x}{L}\tau_1}\right)+\mathcal{P}^\mathrm{f}$ consistent with the
source-twisted boundary condition defining $Z^{(+)}(\phi)$ leads to
$S[Q]=\frac{1}{2}\phi^2 \frac{\tilde \xi}{L}$. Differentiating $Z^{(+)}(\phi)\simeq
\exp(-S[Q])$, according to Eq.~(\ref{eq:Dclass_g}),  one finds $g\simeq
\frac{\tilde \xi}{L}$, which connects $\tilde \xi$ with the Drude conductance of a short chain.

\subsection{Anderson localization}

We now proceed to investigate how multiple-kink field configurations affect
properties of long wires, $L\gg \tilde \xi$.  To this end, let $Z_s(Q)$ denote
the partition function for the wire of length $s$ with a fixed boundary field
$Q_s=Q$. (The boundary condition at the other end of the wire is set to
$Q_1=\Lambda\equiv \tau_3 \otimes \mathds{1}^{\rm bf}$. For $Q\not=Q_1$ this
setup equivalently describes a ring subject to boundary twist.) Since $Q_s$ at
the $s$-th dot may be on  the either part of the manifold, the partition
function can be identified with a two-component spinor  $Z_s=(Z^{(+)}_s,
Z^{(-)}_s)^T  $. Provided one knows the partition function for the system of
length $s$, the one for length $s+1$ is obtained as
\begin{equation}
                                                      \label{eq:transfer-matrix-class_D}
Z^{(\sigma)}_{s+1}(Q^{(\sigma)}) = \int {\cal D}\tilde Q^{(\tilde\sigma)}\,\, e^{-S(Q^{(\sigma)}, \tilde Q^{(\tilde\sigma)})}\,
Z^{(\tilde\sigma)}_{s}(\tilde Q^{(\tilde\sigma)})\,,
\end{equation} 
where $\sigma,\tilde\sigma=\pm$. As a result, the transfer matrix operator acquires a
structure of $2\times 2$ matrix in the space of the two sub-manifolds\cite{Gruzberg2005}
(in addition to acting on the $Q$-field coordinates). 

Its diagonal parts describe evolution of the field confined to the $(+)$ or
$(-)$ sub-manifolds, respectively. For multichannel wires with $g\gg  1$ this
evolution is slow on the scale of one dot, and one may pass to the continuum
representation $s\to x$ and $Z_s(Q)\to Z(Q,x)$. In this approximation the
diagonal parts of the transfer matrix operator are the familiar
Laplace-Beltrami heat-kernel operators. For a particular set of coordinates on
the two sub-manifold given by Eqs.~(\ref{eq:Q_par})--(\ref{eq:B_blocks}), the
latter  takes the form of Eq.~(\ref{AIIISchrodinger}) with a single radial
coordinate $y$. The corresponding Jacobians are evaluated in Appendix
\ref{app:D_1} and are given by
\begin{equation}
                                               \label{eq:Jacobian-D} 
J^{(+)}(y)= 2\coth y, \quad\quad  
J^{(-)}(y)=2\tanh y.
\end{equation}
They depend on the hyperbolic radial variable $y$, but not on the angles
$\alpha,\xi,\eta$. Since the initial condition is isotropic in angular
variables,  one may restrict oneself to a radial partition function $Z(Q,y)\to
Z(y,x)$.  We also note that in the absence of twisted boundary conditions,
$ y=0$, the supersymmetric normalization of the functional integral implies
$Z(0,x)=(1,0)^T$.

The off-diagonal parts of the the transfer-matrix equation require a separate derivation, which may be found in 
Appendix \ref{app:D_2}. The resulting  transfer-matrix problem for the two-component spinor $\Psi(y,x) \equiv Z(y,x)-(1,0)^T$ takes the following form:
\begin{equation}
-\tilde\xi\, \partial_x \Psi = \left(\!\! \begin{array}{cc}
 {1\over {2J^{(+)}}} \partial_ y J^{(+)} \partial_ y  &    -\tilde\chi \sqrt{\frac{J^{(-)}}{J^{(+)}}} \, \partial_ y  \\
 -\tilde\chi \sqrt{\frac{J^{(+)}}{J^{(-)}}}\, \partial_ y   & {1\over {2J^{(-)}}} \partial_ y J^{(-)} \partial_ y
 \end{array}\!\!\right)
\Psi, 
 \end{equation}
where  
$\partial_x \Psi=\Psi_{s+1}-\Psi_s$ and $\Psi= \left(\Psi^{(+)},\Psi^{(-)}\right)^T$. Notice that the kink-generating off-diagonal operator is anti-Hermitian. 
Following the same strategy as in
section~\ref{ssec:AIII_anderson_localization}, one needs to identify the
(right-)eigenfunctions $\Psi_l( y)$ and eigenvalues $\epsilon(l)$ of the
transfer operator. To this end it is convenient to perform the Sutherland
substitution, $\Phi^{(\pm)}(y,x)=\sqrt{J^{(\pm)}(y)}\Psi^{(\pm)}(y,x)$, which
leads to the following  compact formulation of the transfer-matrix problem:
\begin{equation}
                                        \label{eq:transfer-matrix-D}
-\tilde\xi\, \partial_x  \hat \Phi = \left[ \frac12\,{\hat{\cal B}}^2 +i\,\tilde \chi\, {\hat{\cal B}}\right] \hat\Phi , 
\end{equation}
where the $2\times 2$ first-order Hermitian operator $\hat {\cal B}$ is defined as 
\begin{equation}
                                        \label{eq:Dirac-D}
{\hat{\cal B}} = \left(\begin{array}{cc}
  &    B^\dagger  \\
B       & 
 \end{array}\right).
 \end{equation}
Here we defined  $B^{(\dagger)}=-i\partial_y \pm i A(y)$, where $A(y)=-1/\sinh
 2y$.  Since the `potential' $A$ decays at $y\to \infty$, the
 eigenfunctions may be labelled by their asymptotic behavior $\Phi_l(y)\sim
 e^{ily}$ at $y\to \infty$ (their exact form is given in Appendix
 \ref{app:D_2}), where $l\in \Bbb{R}$. The corresponding spectrum is given by:
\begin{equation}
\epsilon(l)=\frac{1}{2}\, l^2+i\, \tilde\chi\, l,
                                  \label{eq:spectrum-D-1}
\end{equation}

The key feature of the transfer-matrix problem (\ref{eq:transfer-matrix-D}) is
that it assumes the form of a {\em supersymmetric} imaginary time
Schr\"odinger equation. (This supersymmetry is `genuine' and should not to be
confused with the boson-fermion structure used to facilitate the average over
disorder.) In the parlor of supersymmetric quantum mechanics, the operator
$B$ is a ladder operator, and $A$ the corresponding super-potential.  The
fact that the latter is an odd function indicates that the supersymmetry is
unbroken. As a result, the operator $ B^\dagger B$ must have a zero energy
eigenvalue $\epsilon(0)=0$, which is responsible for the absence of
localization, if  $\tilde\chi=0$. We conclude that the criticality of the
$\tilde\chi=0$ class D model may be attributed to its hidden SUSY structure
(\ref{eq:transfer-matrix-D}). It is an intriguing prospect if this disorder-induced supersymmetry in
$\Bbb{Z}_2$ symmetry classes is related to the one recently found\cite{Grover:2014} in connection with the 
dynamic fluctuations of the order parameter in the quantum critical points of some clean tI's.

  Also notice that the  diagonal part of the transfer matrix operator
 (\ref{eq:transfer-matrix-D}) consists of Hermitian operators $ B^\dagger B$
 and $B B^\dagger $, which have the form of the generalized P\"oschl-Teller
 Hamiltonians\cite{Poschl-Teller}: 
\begin{equation}
                                                                                             \label{eq:PT1}
-\partial_ y^2 -  \frac{\lambda(\lambda-1)}{\cosh^2 y}+  \frac{\lambda(\lambda+1)}{\sinh^2 y},
\end{equation}
with $\lambda=\pm1/2$, respectively. One may now show that the eigenfunctions
of the full problem (\ref{eq:transfer-matrix-D}) are not affected by the
finite fugacity $\tilde\chi$ of the kinks (the spectrum
(\ref{eq:spectrum-D-1}) is, of course, sensitive to it). The situation is
 exactly parallel to that in the $\Bbb{Z}$ symmetry classes, where the
topological term affects the spectrum, but {\em not} the eigenfunctions. To
show this denote the eigenfunctions of the two P\"oschl-Teller operators
(\ref{eq:PT1}) as $\Phi_l^{(\pm)}(y)$, for their exact expressions in terms of
hypergeometric functions see Appendix \ref{app:D_2}. It is easy to see that
$B^\dagger\Phi_l^{(-)} =il \Phi_l^{(+)}$ and  $B\Phi_l^{(+)} =-il
\Phi_l^{(-)}$ (multiply the equation $B^\dagger B \Phi^{(+)}_l
=l^2 \Phi^{(+)}_l $ from the left by the operator $B$, to obtain $BB^\dagger
(B\Phi^{(+)}_l) =l^2 (B\Phi^{(+)}_l) $, which means that the function
$B\Phi^{(+)}_l $ is proportional to an eigenfunction of the supersymmetric
partner operator $BB^\dagger$ with the same eigenvalue $l^2$, that is to
$\Phi_l^{(-)}$).  As a result the spinor $\Phi_l\equiv
(\Phi^{(+)}_l,\Phi^{(-)}_l)^T$ solves the full eigenvalue
problem~(\ref{eq:transfer-matrix-D}) with the eigenvalue
(\ref{eq:spectrum-D-1})  for any fugacity $\tilde\chi$.

The proper solution of the transfer-matrix equation may now be represented in
terms of a spectral decomposition as $\Psi( \phi,L)=\sum_l \mu(l) \,
\Psi_l( \phi)\, e^{-\epsilon(l)L/\tilde\xi }$, where the expansion coefficients $\mu(l)
= \left\langle \Psi_l|\Psi(0)\right\rangle=-\left\langle
\Psi_l|(1,0)^T\right\rangle=-\langle \Psi^{(+)}_l|1\rangle$ are determined by the
constant offset $(1,0)^T$ (cf. the corresponding remarks in
section~\ref{ssec:AIII_anderson_localization}). Using the explicit form of the
eigenfunctions $\Phi_l^{(\pm)}(y)$ we find in Appendix~\ref{app:D_3} that
\begin{equation}
\mu(l) = -\sqrt{\frac{\pi}{l\tanh\frac{\pi l}{2}}}.
\end{equation}
We finally recall that $Z(\phi,L)=\Psi( \phi,L)+(1,0)^T$ to obtain
the  partition sum as
\begin{equation}
Z(\phi,L)=\left(\begin{matrix}
1\crcr 0	
\end{matrix}\right) + \int \frac{dl}{2\pi} \,\mu(l)\Psi_l(\phi)\, e^{-\epsilon(l) L/\tilde\xi}.
\end{equation}

\begin{figure}
\centering{
\includegraphics[width=0.4\textwidth]{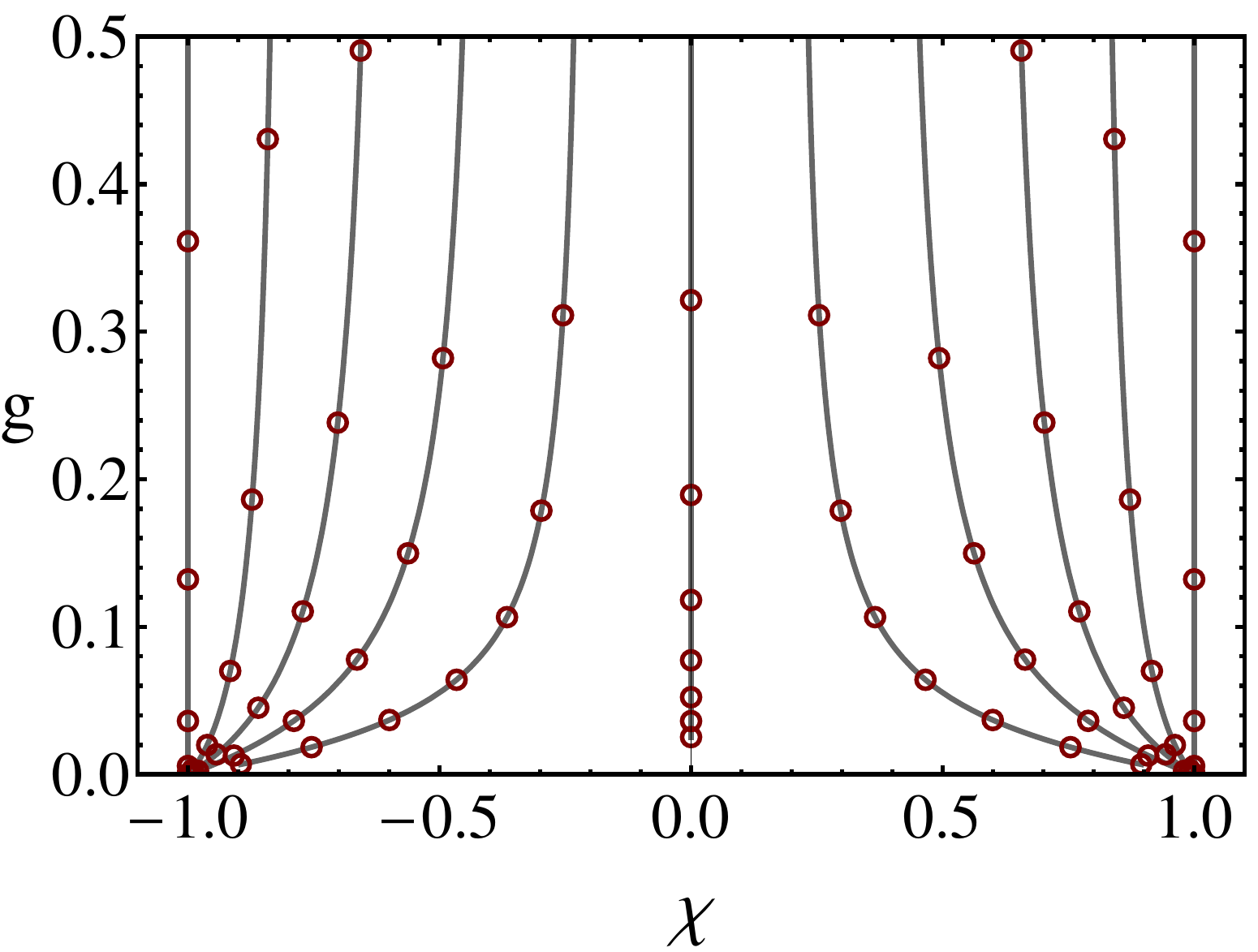}
}
\caption{Flow of the conductance $g$ and the kink's fugacity $\chi$ as a
function of system size for class D system. Dots are for values,
$L/ \tilde\xi=1,2,4\dots,32$.}
\label{fig:FlowGraph_D}
\end{figure}

From this expression and using the explicit form of the eigenfunctions (Appendix~\ref{app:D_3}), observables may now readily be extracted. The topological number $\chi(L)$ is given 
by Eq.~(\ref{eq:Dclass_chi}), as $\chi(L) = Z^{(-)}(0,L)$, resulting in:
\begin{equation} 
                                        \label{eq:v-D}
\chi (L) = {1\over 2} \int \!dl\,\coth\left({\pi l}/{2}\right) \sin(\tilde\chi lL/\tilde\xi)e^{- l^2L/2\tilde\xi}. 
\end{equation}
One notices that at $L\to \infty$, the variable $\chi$ approaches $\mathrm{sign}(\chi)=\pm 1$ exponentially fast, indicating the stabilization of a topologically trivial or non-trivial phase, respectively. 
The conductance is obtained by differentiation of the  partition sum $Z^{(+)}(\phi,L)$ with respect to the boundary twist $\phi$ according to Eq.~\eqref{eq:Dclass_g}. As a result, one finds
\begin{equation}
                                         \label{eq:g-D}
g(L)  = \frac{1}{8}
\int dl\,l\coth\left({\pi l}/{2}\right) \cos(\tilde\chi lL/\tilde\xi)e^{- l^2L/2\tilde\xi}.
\end{equation}
From this expression it is straightforward to verify that for  $\tilde\chi\neq 0$ the conductance decreases exponentially with the system size $g(L)\propto \frac{1}{\sqrt L}\,e^{-\tilde\chi^2L/(2\tilde\xi)}$.
This shows that the effective localization length 
\begin{equation}
                                           \label{eq:localiz-D}
\xi=2\tilde\xi / \tilde\chi^{2}
\end{equation} 
diverges towards the critical point $\tilde\chi=0$. At the threshold between
the ordinary Anderson insulator and its topological sibling, $\tilde\chi=0$,
the system is in the critical delocalized state with $g(L)\sim 1/\sqrt{L}$.
The overall flow diagram in the $(g(L), \chi(L))$ plane is shown in
Fig.~\ref{fig:FlowGraph_D}.

Finally, it is  interesting to note that the effective localization length may
be exponentially large close to criticality. To show this  consider a
tunneling limit of dot-to-dot couplings such that all $ t_k^2 \ll 1$ and at
the same time $g = \sum_{k=1}^{2N}  t_k^2 \gg 1 $. In this case $\tilde\chi
=e^{\sum_k\ln(1- t_k^2)^{1/2}} \approx e^{-g/2}$ and we obtain $ \xi \sim
\tilde\xi \, e^g$, where $g \propto  \tilde\xi \propto N$. The fact that the
localization length in class D is exponentially large in the number of
channels was first realized by Gruzberg, Read  and Vishveshwara\cite{Gruzberg2005} in the
context of the transfer matrix DMPK treatment. They have also gave a treatment
in terms of the supersymmetric spin chain and realized that the corresponding
transfer matrix equation acquires a two-spinor form. It can be verified (although this key point was not discussed in the original reference) that their transfer-matrix
equation, too, encodes a supersymmetry. 

\subsection{Boundary density of states}

In the $L\to \infty$ critical points $\chi=-1/1$ the system does/does not
support a Majorana state at its ends. In this limit, the bulk
theory (by which we mean the bulk theory off criticality) becomes purely
topological: the gradient term in~\eqref{SDContinuum} has scaled to zero,
$\xi\to 0$. In the trivial phase of the $\chi=1$ -- Anderson insulator, the story ends here. In
the tAI phase $\chi=-1$ we are left with a term counting kink fugacities in terms of a
phase action~(\ref{eq:S_top_classD}), at the fixed point, $\tilde \chi =-1$, the coefficient simplifies as $i\ln(\tilde \chi)/\pi=-1$ and the topological action may be written as $S_\mathrm{top}[Q]=
\mathrm{str}\left(\ln T(0) -\ln T(L)\right)$, i.e. as the sum of two boundary actions. These 
actions describe the boundary Green function at zero energy. Generalization to
finite energies, $E$, is straightforward and leads to the left boundary action
(analogously for the right)
\begin{align}
	S_\mathrm{L}[Q]= -\frac{\epsilon}{2}\, \mathrm{str}(Q\tau_3)+\left(\frac{1-\chi}{2}\right) \mathrm{str}\ln(T),
\end{align}
where $\epsilon=\pi |E|/\Delta_\xi$ as in the AIII system and $\chi = \pm 1$
so that the form of the action is correct on both (AI and tAI) localized
phases. The density of state deriving from this description has been computed\cite{Ivanov:2002,Bagrets2012}        and reads as
\begin{align}
	\chi=1:&\qquad \rho(\epsilon)=\frac{1}{\Delta_\xi}\left(1+\frac{\sin \epsilon}{\epsilon}\right),\crcr
	\chi=-1:&\qquad \rho(\epsilon)=\frac{1}{\Delta_\xi}\left(1-\frac{\sin \epsilon}{\epsilon}+\delta(\epsilon)\right).
\end{align}
The $\delta$-function in the second line is the topological Majorana state.
Notice, however (see Ref.~[\onlinecite{Bagrets2012}] for further discussion), that in either case, 
$\int_{-\epsilon_0}^{\epsilon_0}d\epsilon\,\rho(\epsilon)\stackrel{\epsilon_0\gg
1}{=}2\epsilon_0+\frac{1}{2}$, i.e. the boundary accumulates an excess
spectral weight of $1/2$, which in the non-topological case is the consequence
of disorder generated quantum interference, and in the topological case due to
the Majorana partially `screened' by a negative interference contribution.

\subsection{Class DIII}
\label{sec:DIII}
\minisec{Model Hamiltonian} Similarly to class D, class DIII describes particle-hole symmetric, spin
rotation non-invariant superconductors. The difference with class D is in the presence of time-reversal symmetry in class DIII. The Hamiltonian obeys a particle-hole symmetry,
$H^T=-\sigma_1^\mathrm{ph}H\sigma_1^\mathrm{ph}$. Time reversal invariance requires $H^T=\sigma_2^\mathrm{sp} H \sigma_2^\mathrm{sp}$. These symmetries can be combined to obtain the chiral symmetry $P^\dagger H P = - H$
with $P = \sigma_1^{\rm ph}\otimes \sigma_2^{\rm sp}$. In the basis defined by this chiral structure, the Hamiltonian assumes the off-diagonal form
\begin{equation}
H' = \left(
\begin{array}{cc}     
 & D\\
D^\dagger & 
\end{array}\right), \quad D^T = -D.
\label{DIIIgeneric}
\end{equation}
A generalization of the granular Hamiltonian~(\ref{DDotHamiltonian}) to the DIII symmetric situation reads   
\begin{align}
	\label{DDotHamiltonian}
	  H = \sum_{ss'}C_{-,s}^\dagger \left({\cal H}_s \delta_{ss'}+i W \delta_{ss'+1}\right)C_{+,s'} + 
{\rm h.c.},
\end{align}
where $C_{\pm,s}^\dagger$ is a vector of creation operators on grain $s$ and the indices
$(+/-)$ refer to the chiral structure. The $2N \times 2N$ matrix ${\cal H}= - {\cal H}^T$ 
is assumed to be random Gaussian distributed,
while the $2N \times 2N$ hopping symmetric matrix, $W=W^T$, is translationally invariant and 
defines the non-random part of the Hamiltonian describing the inter-grain couplings.

\minisec{Topological number}
The definition of a topological number follows the lines of the construction in class D.
We imagine the system closed to a ring and select one particular bond where
$  W' = iW |+,1\rangle \langle -, 0|$ 
is the hopping
matrix associated to this bond. Representing the off-diagonal block of the Hamiltonian $H$ in the chiral 
basis~(\ref{DIIIgeneric}) as $D=D' +   W'$, the matrix $D_{\pi} = D' -   W'$ represents a system
with sign inverted hopping across the bond. The topological number can be now defined as
${\rm sgn}\left[{\rm Pf}(D_{\pi})/{\rm Pf}(D)\right]$. We show in Appendix~\ref{app:Scattering_DIII} 
that in the limit $L \to \infty$
this ratio of Pfaffians is a real number equal to $\pm 1$. 

\minisec{Field theory}The construction of a field theoretical partition sum parallels that of Section~\ref{sec:FT_classD} 
for the class D wire. Our starting point is a quadratic action $S[\bar
\Psi,\Psi]=\frac{1}{2}\bar \Psi(i0\tau_3 -   H)\Psi$, where $\Psi$ is an eight
component field obeying the symmetry  $\Psi = \sigma_1 \otimes \tau\,
\bar\Psi^T$, the matrix $\tau$ is defined as $\tau=i\tau_2 \otimes {\cal
P}^\mathrm{b}+\tau_1\otimes {\cal P}^\mathrm{f}$, and Pauli matrices $\sigma_i$ act in the
`chiral' space defined by Eq.~\eqref{DIIIgeneric}.
Resolving the chiral structure through $\Psi=(\Psi_+, \Psi_-)$, we have a continuous symmetry under transformations
\begin{align}
\label{eq:T_LR_DIII}
	\bar \Psi_+ \to \bar \Psi_+ T_{\mathrm L},&\quad \Psi_+\to T_{\mathrm R}^{-1} \Psi_+\crcr
	\bar \Psi_- \to \bar \Psi_- T_{\mathrm R},&\quad \Psi_-\to T_{\mathrm L}^{-1} \Psi_-.
\end{align}
Here $T_{\mathrm L,R}$ are $4\times 4$ matrices which act in the direct
product of ${\rm bf}$ and ${\rm cc}$ spaces and belong to the group ${\rm
SpO}(2|2)$ formed by all super-matrices of the type $T^{-1}=\tau
T^T\tau^{-1}$. The full continuous
symmetry group of the class $\mathrm{DIII}$ action  thus is $G={\rm SpO}(2|2)
\times {\rm SpO}(2|2)$, which as in the chiral class AIII will be broken to a
single copy ${\rm SpO}(2|2)$ upon disorder averaging. A key feature of that
manifold is that, as in class D, it is disconnected: the compact (fermionic) sector
$\mathrm{O}(2)$ comprises fluctuations with determinant $\pm 1$, which cannot
be continuously connected. Accordingly, the gauge symmetry group $G$ acting on
the Goldstone mode manifold again contains a $\Bbb{Z}_2$ degree of freedom,
generating kinks between the two disconnected components.

As before, we probe the system by insertion of topological gauge sources defined on one link $0 \leftrightarrow 1$ only. To this end, let us generalize the operator connecting the corresponding sites as
\begin{align}
\label{W_phi}
\left( \begin{array}{cc} &   W   \\ 
  W^\dagger & \end{array} \right)
	\to\left( \begin{array}{cc} 
	&   T_L(a) W  \\ 
  W^\dagger T_R^{-1}(a) &
  \end{array} \right).
\end{align}
Here, $T_L(a)$,
where $a=(\phi,\sigma)$, $\phi=(-i\phi_0,\phi_1)^T$, and $\sigma=\pm $. While the general form of the transforming matrices is defined as
\begin{align}
	T_L(a)&=e^{-\phi_0}\mathcal{P}^\mathrm{b}+e^{-i\phi_1}\tau_1^{(1-\sigma)/2}\mathcal{P}^\mathrm{f},\crcr
	T_R(a)&=e^{\phi_0}\mathcal{P}^\mathrm{b}+e^{i\phi_1}\mathcal{P}^\mathrm{f},
\end{align}
we will later apply the specific configurations $T_L(0,-)=\tilde T\equiv \mathcal{P}^\mathrm{b}+\mathcal{P}^\mathrm{f}\otimes \tau_1$, $T_R(0,-)=\mathds{1}$, and $T_R(\phi,+)=T_L^{-1}(\phi,+)=\mathrm{diag}(e^{\phi_0},e^{i\phi_1})^\mathrm{bf}$. 
Notice that the transformed operator lacks hermiticity, which indicates that the gauge transformations behind the insertion of the source are non-unitary chiral transformations. Indeed, a link modified as above can be generated by a transformation
\begin{align}
	T_\mathrm{L/R,s}&=T_\mathrm{L/R}(a) \Theta(s)+ \mathds{1}\Theta(-s).
\end{align}
The lattice-discontinuity across the $0\leftrightarrow 1$ link then generates the modified hopping operator. For a finite ring with periodic boundary conditions, the transformation above does not exist, implying that a single source link can be shifted through the system (by a boundary-consistent two-kink transformation), but not removed. The gauge transformation is described by two continuous variables $\phi$, and one $\Bbb{Z}_2$ variable $\sigma=\pm$, and we denote the corresponding partition function by $Z(a)\equiv Z^{(\sigma)}(\phi)$.   

The sources above are constructed in such a way, that for $a=(0,-)$, $T_\mathrm{L}(0,-)=\mathcal{P}^\mathrm{b}+\mathcal{P}^\mathrm{f}\tau_1$, while $T_R=\mathds{1}$, i.e. in the fermionic sector the sourced link is replaced by $\tau_1$. Arguing as in the class D case, the corresponding partition function $Z^{(-)}(0)$ generates a product of Pfaffians with off-diagonal sector, $D$ and $D_\pi$, respectively (see previous section). In other words,
\begin{equation}
	\label{eq:DIIIclass_chi}
	\chi = Z^{(-)}(0)=\left\langle \frac{\mathrm{Pf}(D_{\pi}) }{\mathrm{Pf}(D) }
\right\rangle.
\end{equation}
By contrast, for $a=(\phi,+)$, $T_L = e^{i\phi}$. The link modified by the
continuous parameters $\phi$ is to yield the conductance as in
Eq.~\eqref{AIIIObservableDef}, i.e. with ${\cal F}(\phi_0) =
\partial_{\phi_1}Z^{(+)}(\phi)\bigl|_{\phi_1=-i\phi_0}$, we have $g =
-i\partial_{\phi_0} {\cal F}(\phi_0)\bigl|_{\phi_0=0}$.

\minisec{Low-energy action}
The field manifold $G/H={\rm SpO}(2|2)$ comprises a non-compact bosonic sector $\mathrm{Sp}(2)$, and the two-component $\mathrm{O}(2)$ in the fermionic sector. We parameterize the full manifold as
\begin{equation}
                                                      \label{eq:Q_par_DIII}
T^{(\pm)} = e^{\cal W} {\tilde T}^{(\pm)} e^{-{\cal W}}, \qquad {\tilde T}^{(\pm)} = \left(
\begin{array}{cc}
{\tilde T}^{\mathrm{b}} & \\
& {\tilde T}^{\mathrm{f}(\pm)}
\end{array}
\right)^{\rm bf}.
\end{equation}
The bosonic part is parametrized by one hyperbolic radial variable $y_0$ and two angles $\rho$ and $\alpha$
\begin{eqnarray}
&&{\tilde T}^{\mathrm{b}} = e^{ i(b\tau_+ + b^*\tau_-)} e^{y_0\tau_3} 
e^{ -i(b\tau_+ + b^*\tau_-)} \\ \nonumber  && =
\left(\begin{array}{cc}
\cosh y_0 + \cos 2\rho \sinh y_0& -i e^{i\alpha}\sin 2\rho \sinh y_0 \\
i e^{-i\alpha}\sin 2\rho \sinh y_0    & \cosh y_0 - \cos 2\rho \sinh  y_0
\end{array} \right), 
\end{eqnarray}
where $\rho=\sqrt{b^* b}$ and $e^{i\alpha}=\sqrt{b/b^*}$, and matrices
$\tau_i$ act in charge conjugation space. The fermionic part is parameterized
by a single compact radial variable $y_1$ and may be specified on the two parts
of the group manifold
\begin{equation} 
{\tilde T}^{\mathrm{f}(+)} =e^{i y_1\tau_3}\,, \quad\quad 
{\tilde T}^{\mathrm{f}(-)} =\tau_1 e^{i y_1\tau_3}\,.
\end{equation}
Notice that only the $(+)$ manifold contains unit element and thus constitutes
a subgroup. The boson-fermion  rotations  are parameterized by four Grassmann
angular variables $\xi,\eta,\nu,\mu $ and restricted by the particle-hole
symmetry ${\cal W} = - \tau {\cal W}^T \tau^T$,
\begin{equation}
                                               \label{eq:B_blocks_DIII}
{\cal W} =\left(
\begin{array}{cc}
&{\cal B} \\
\tilde{\cal B} & 
\end{array}
\right)^{\rm bf}, \quad 
{\cal B} = \left(\begin{array}{cc}
\xi &  \nu \\
\mu  & \eta
\end{array} \right)^{\rm cc}, \quad \tilde{\cal B} = 
\left(\begin{array}{cc}
\eta &  -\nu \\
\mu & -\xi
\end{array} \right)^{\rm cc}, 
\end{equation}
where $\tilde{\cal B} =  - \tau_1 {\cal B}^T i\tau_2$. A straightforward if
somewhat lengthy calculation yields the Jacobians of the transformation to the
above system of polar integration variables as
\begin{equation}
                                                    \label{eq:Jacobians-DIII}
J^{(+)}={\sin 2 \rho \over 2}\, \frac{ \sinh^2 y_0  }{(\cosh y_0-\cos y_1)^2}\,,\quad 
J^{(-)}={\sin 2 \rho \over 2}\,.
\end{equation}
Finally, the supersymmetric action of the array, written in terms of on-site
group elements $T_s$ acquires the form (cf.
Appendix~\ref{app:Scattering_DIII})
\begin{equation}
                                     \label{eq:action-DIII}
S[T]={1\over 4} \sum\limits_{s,k=1}^{L,4N}  \mathrm{str}\ln \left[\openone +
\frac{t_k^2}{4} \left(T_s^{-1}T_{s+1} + T_{s+1}^{-1}T_{s}-2\right)\right],
\end{equation}  
where $t_k$ are transmission matrix eigenvalues. This action is a counterpart
of Eq.~\eqref{eq:Action_D} for class $\mathrm{D}$. When subjected to a symbolic
gradient expansion it takes the form $S[T]=-(\tilde \xi/8)\int\! dx\,
\partial_x T^{-1}\partial_x T+ \ln\tilde \chi\times n_\mathrm{k}$, where
$\tilde\xi = g/2$, $g = \sum_{k=1}^{4N} t_k^2$ is the inter-dot conductance,
and the second term represents the kink-action to be discussed momentarily.
However, as with the class D system, a consistent treatment of kinks forces us
to work with the granular action  (\ref{eq:action-DIII}). Following the same
recipe as in the class D case, we start by considering a configuration with one
kink, where $T_s=\openone$, while $T_{s+1}={\tilde T}\equiv
\mathcal{P}^\mathrm{b}+\mathcal{P}^\mathrm{f}\otimes \tau_1$. The corresponding contribution to
the partition function acquires the form
\begin{equation}
                                     \label{eq:fugacity-DIII}
e^{S(\mathds{1},\tilde T)}=\prod\limits_{k=1}^{2N}r_k \!\! =\mathrm{Pf}\, (\hat r P ) \equiv \tilde\chi\,,
\end{equation} where we took into account that all $4N$ eigenvalues $r_k = \pm
(1-t_k^2)^{1/2}$ of the transmission matrix $\hat r$ are Kramers degenerate
and thus its Pfaffian may be defined as the product of $2N$ non-degenerate
eigenvalues.

As in class D, we define a spinor partition function $Z_s=(Z^{(+)},
Z^{(-)})_s^T$, where $Z^{(\pm)}$ describes the evolution of configurations
starting at $T_0=\tau_3$ and ending at $T^{(\pm)}_s$ belonging to the same/opposite connectivity component. The evolution of the two-component $Z_s$ is described by the
equation
\begin{equation}
                                                      \label{eq:transfer-matrix-DIII}
Z^{(\sigma)}_{s+1}(T^{(\sigma)}) = \sum_{\tilde\sigma=\pm} \int {\cal D}\tilde T^{(\tilde\sigma)}\,\, e^{-S(T^{(\sigma)}, \tilde T^{(\tilde\sigma)})}\,
Z^{(\tilde\sigma)}_{s}(\tilde T^{(\tilde\sigma)})\,,
\end{equation} 
We now again pass to the continuum limit, $Z_s\to Z(x)$ in which the diagonal
blocks  of the $2\times 2$ transfer operator become the standard
Laplace-Beltrami operators $\sum_{\nu=\rho,y_0,y_1}(J^{\pm})^{-1}\partial_\nu
J^\pm \partial_\nu$ on the corresponding sector of the field manifold with Jacobians
given by Eq.~(\ref{eq:Jacobians-DIII}). The off-diagonal parts are somewhat more
intricate. It turns out that the action $S(T^{(+)}, \tilde T^{(-)})$ is
independent of the compact radial variables $y_1, \tilde y_1$ (while it still exhibits 
conventional Gaussian confinement $\sim g( y_0-\tilde y_0)^2$ in the
non-compact direction). As a result, the transfer matrix operator becomes a non-local integral operator in the compact $ y_1$-direction.  After  Sutherland
substitution $\Psi^{(\pm)}(y,x)=\Phi^{(\pm)}(y,x)/\sqrt{{\cal J}^{(\pm)}(y)}$
with ${\cal J}^{(\pm)}(y) = \partial_\rho J^{(\pm)}\bigl|_{\rho=0}$, it takes
the form:
\begin{equation}
\label{eq:TM_DIII}
-\tilde\xi \,\partial_x \Phi = 
\left(\!\!\begin{array}{cc}
\partial_{y_0}^2 + \partial_{y_1}^2   &    -\displaystyle\tilde\chi\sqrt{g\over 2\pi}\,   \partial_{y_0} \int\! d\tilde y_1\\
- \displaystyle\tilde\chi\sqrt{g \over 2\pi} \,\partial_{y_0}  \int\! d \tilde y_1     &  \partial_{y_0}^2+ \partial_{y_1}^2
 \end{array}\!\!\right) \Phi,
\end{equation}
where we have denoted $\Phi=(\Phi^{(+)}, \Phi^{(-)})^T$ and the length scale $\tilde\xi = g/2$ (in units of
inter-dot spacing).
This operator acts in the space of $2\pi$ periodic functions of $y_1$ (it is important that $\sqrt{{\cal J}^{(\pm)}(y)}$ are periodic), which may be written as $\Phi^{(\pm)}( y)=\sum_{l_1}\Phi^{(\pm)}_{l_1}( y_0)e^{il_1y_1}$, with integer $l_1$. 
The $l_1\neq 0$ components are {\em not} affected by the kinks at all. The corresponding eigenvalues are 
$\epsilon(l_0,l_1\neq 0) =l_0^2+l_1^2$, where $l_0\in \mathds{R}$ is a real quantum number from the non-compact direction $ y_0$. On the other hand, the $\Phi^{(\pm)}_0( y_0)$ spinor obeys the supersymmetric quantum mechanics (\ref{eq:transfer-matrix-D}) with $B^{(\dagger)}=-i\partial_{y_0}$, i.e. with zero super-potential $A=0$, and renormalized fugacity 
$\tilde\chi \to \tilde\chi\sqrt{2\pi g}$.   
The corresponding $l_1=0$ eigenvalue is  $\epsilon(l_0,0) =l_0^2+i l_0\, \tilde\chi\sqrt{2\pi g}$. The eigenfunctions are the plane waves $\Phi_{l_0,0}^{(\sigma)}( y_0)=\frac{1}{\sqrt 2} \, e^{il_0 y_0} (1,\sigma)^T$, where $\sigma=\pm$,  and again do {\em not} depend on the topological number $\tilde \chi$.

We now use the solution of the transfer matrix problem to extract observables. The  measure is given by 
the overlap of the conjugated wave function with the initial conditions,
$\mu(l)=-\langle \hat\Psi_{l}|(1,0)^T\rangle$. It leads to
$\mu(l) = -2il_0/(l_0^2+l_1^2)$,  $l_1 \neq 0$,
and $\mu(l_0,0) = \sqrt{2}i/l_0$ in case of $l_1=0$. 
It is clear that all components but $l_1=0$ decay exponentially on the scale given by the bare localization length $\tilde \xi \sim g$. Hereafter we thus focus exclusively on $l_1=0$ component, relevant for longer wires $L\gg \tilde\xi$.  The corresponding  partition sum spinor is  given by
\begin{equation}
\label{eq:ZL_DIII} 
Z(\phi,L)=(1,0)^T + \int \frac{dl_0}{2\pi} \mu(l_0,0) \Psi_{l_0,0}( 2\phi)e^{-\epsilon(l_0,0) L/\tilde\xi},
\end{equation}
where we put the radial coordinates to be $y=2\phi$, i.e. given by the boundary condition at $x=L$.
Its $(-)$ component yields the renormalized fugacity as $\chi(L) = Z^{(-)}(0,L)$, Eq.~(\ref{eq:DIIIclass_chi}). 
Referring for the details to Appendix~\ref{app:DIII_3}, we give below only the final result 
\begin{equation} 
                                        \label{eq:v-DIII}
\chi(L)= \mathrm{erf} \left(\tilde\chi\sqrt{\pi L}\right).
\end{equation}
For $L\to \infty$ the renormalized fugacity exponentially  approaches $\chi\to \mathrm{sign}(\tilde\chi)=\pm 1$, indicating topologically trivial and non-trivial phases correspondingly. 
In turn, the conductance evaluated with the help of relation~\eqref{AIIIObservableDef} 
takes the form
\begin{equation}
\label{eq:g_DIII}
g(L) = 4\,\sqrt{\frac{g}{\pi L}}\, e^{-\pi g\, \tilde\chi^2 L/2\tilde\xi }.
\end{equation}
We see that the average localization length, defined as
\begin{equation}
\label{eq:xi_DIII}
\xi=\frac{2\tilde\xi}{\pi g\, \tilde\chi^{2}} = \frac{1}{\pi \tilde \chi^{2}},
\end{equation}
diverges towards the critical point $\tilde\chi=0$, resulting in the critical delocalized state with $g(L)\sim 1/\sqrt{L}$. At the same length scale $\xi$ the mean topological number $\chi(L)$, Eq.~(\ref{eq:v-DIII}), 
approaches its quantized values $\pm 1$.
The corresponding flow diagram on $(g(L),\chi(L))$ plane is qualitatively identical to the one in
the class $\mathrm{D}$ shown in Fig.~\ref{fig:FlowGraph_D}. 

\section{Scattering theory approach}
\label{sec:Scattering}

In this section we discuss a relation of the SUSY partition function to the scattering matrix approach for
mesoscopic wires. More precisely, we establish a formal equivalence of $Z(\phi,L)$ to the generating function of transmission eigenvalues for the wire of finite length $L$ connected to external leads.  In classes $\mathrm{AIII}$
and $\mathrm{DIII}$, this enables us to get additional insights on localization/delocalization 
phenomena in topological 1d wires 
which go beyond the studies of conductance and average topological number.  This section
contains only the detailed statement of our results while its derivation is relegated to Appendix~\ref{app:Scattering}.   

We start with a summary of the relevant definitions made in previous sections. 
Our main object of study, the partition sum $Z(\phi)$, was defined as the
disorder averaged ratio of the fermionic vs. bosonic determinants,
\begin{equation}
\label{eq:ZF_ZB}
Z^{(+)}(\phi) = \left\langle \frac{Z_F(\phi_1)}{Z_B(\phi_0)} \right\rangle= 
\left\langle \frac{{\rm det}^{1/\nu}\left( i0^+\hat\tau- H(\phi_1)\right)}{{\rm det}^{1/\nu}\left( i0^+\hat\tau- H(-i\phi_0)\right)}\right\rangle.
\end{equation}
The phase dependent Hamiltonian here is the result of non-local gauge transformation,
\begin{equation}
 H_{s s'}(\phi) = e^{ - i \phi_s \hat j_S} H_{ss'} e^{  i \kappa \phi_{s'} \hat j_S}, \quad \kappa= \pm 1,
\end{equation}
where the sign $\kappa =+1$ for class $\mathrm{D}$ and $\kappa=-1$ (i.e. the chiral gauge transform) 
is to be chosen for other classes 
($\mathrm{AIII}$, $\,\,\mathrm{BDI}$ and $\mathrm{DIII}$). 
For $\mathds{Z}_2$-insulators $\hat\tau=\tau_3$ and $\nu=2$. In the case of 
$\mathds{Z}$-insulators, $\hat\tau=1$ and $\nu=1$.
The appearance of $1/\nu$ power stems from the doubling procedure  which was required for the proper construction of the path integral in the
case of BdG classes $\mathrm{D}$ and $\mathrm{DIII}$.
The generator $\hat j_S$ is related to the conserved current which choice depends on the symmetry class.
The phase $\phi_s = 0$ for sites with $s \leq 0$ and $\phi_s = \phi$ if $s\geq 1$. 
For the lattice model with the nearest neighbor hopping the phase dependent part of 
$H(\phi)$ is localized on a single link $0\leftrightarrow 1$.  The generator $j_S$ of the symmetry current reads
\begin{eqnarray}
\hat j_S &=& P, \qquad\quad\,\, \mbox{classes AIII, BDI}; \nonumber \\
\hat j_S &=& \tau_1^{\rm cc}, \qquad\quad\!\! \mbox{class D}; \\
\hat j_S &=& P \tau_3^{\rm cc}, \qquad \mbox{class DIII}.  \nonumber
 \end{eqnarray}
Here $P$ always denotes the parity operator, $\{P,H\}_+ =0$, and $\tau$ matrices operate in charge-conjugation space.
The 'minus' component of the partition sum relevant for $\mathds{Z}_2$-insulators will be discussed later.  

The partition sum~(\ref{eq:ZF_ZB}) cast into the language of supersymmetric
functional field integral was studied above. Our goal here is to relate
$Z(\phi)$ to the scattering matrix of the disordered wire. More precisely, we
assume that the system is now open to the external world rather than  closed
into the ring, meaning that the wire is connected to left/right leads. Then
for any given realization of disorder potential the scattering ($\hat S$) and
transfer ($\hat M$) matrices of dimension $2N'\times 2N'$ can be defined with
$N'$ being the number of scattering channels. For classes $\mathrm{AIII}$,
$\mathrm{BDI}$ and $\mathrm{D}$ we have $N' = 2N$ where the factor $2$ is due
to sub-lattice ($\mathrm{AIII}$) or particle-hole index (if classes
$\mathrm{BDI}$ and $\mathrm{D}$ refer to spinless fermions).   In the spinfull
case and for class ${\mathrm{DIII}}$ one has $N' = 4N$ due to spin and p/h
quantum numbers. With the use of DMPK theory both $\hat S$ and $\hat M$ can be
reduced to the canonical form\cite{BMSA:98,Brouwer2000,Gruzberg2005}. For the
transfer matrix this representation takes the form
\begin{equation}  
\label{eq:M_DMPK}
\hat M = \left(\begin{array}{cc} V & \\ & V'^\dagger \end{array} \right)
\left(\begin{array}{cc} \cosh\lambda & \sinh\lambda \\ 
\sinh\lambda & \cosh\lambda
\end{array} \right)
\left(\begin{array}{cc} U' & \\ & U^\dagger \end{array} \right),
\end{equation}
where $\lambda = {\rm diag}(\lambda_1, \dots, \lambda_{N'})$ is the set of so-called Lyapunov exponents and $U,V, U', V' \in \mathrm{U}(N')$ 
are unitary matrices.  
Each exponent $\lambda_k$ defines the transmission coefficient $t_k^2 =1/\cosh^2\lambda_k$ of the
$k$-th transport channel.  

Note that the flux conservation condition, $ M^\dagger \sigma_3^{RL} M = \sigma_3^{RL}$, which stems from the 
hermicity of the underlying Hamitonian $\hat H$, is met by such form of $\hat M$.  The presence of other non-unitary symmetries (time-reversal, particle-hole or parity) in the system 
puts additional constrains on the transfer matrix $\hat M$. They are specified by the basis
dependent matrices $U_T$, $U_C$ and $P$. To proceed one 
should augment original spinors by the
left/right grading, $\Psi = (\psi_R, \psi_L, \bar \psi_L^T, \bar \psi_R^T)$, to define the scattering states. 
This in turn requires the corresponding
extension of the two symmetry matrices. Namely, ${\cal U}_T = \sigma^{RL}_1 \otimes U_T$ becomes the proper 
matrix for time-reversal symmetry and ${\cal P} = \sigma^{RL}_1 \otimes P$ should stand for the parity operator.
Then time-reversal and particle-hole symmetries require $ {\cal U}_T \hat M \,{\cal U}_T^{-1} = \hat M^*$ and
$ U_C \hat M \, U_C^{-1} = \hat M^*$, respectively, while the parity symmetry implies $ {\cal P} \hat M \,{\cal P} = \hat M$, see e.g. Ref.~[\onlinecite{Brouwer2000}].

In application to the DMPK decomposition the chiral symmetry in $\mathds{Z}$-classes leads to relations $U' = U^\dagger P$ and $V' = V^\dagger P$.
Besides that in chiral class $\mathrm{BDI}$ as well in class $\mathrm{D}$ 
it is advantageous to work in Majorana basis where the
particle hole symmetry is the transposition ($H^T = - H$) with
$U_C = \mathds{1}$. This reduces all rotation matrices to orthogonal ones, $\{U,V,U',V'\} \in \mathrm{SO}(N')$. 
In class $\mathrm{DIII}$ the ancestor BdG Hamiltonian has the symmetry matrices 
$U_C = \sigma_1^{\rm ph}\otimes \mathds{1}^{\rm sp}$, $U_T = \mathds{1}^{\rm ph} \otimes \sigma_2^{\rm sp}$ 
and $P = U_C U_T^{-1}$.
The choice of Majorana representation transforms them to $U_C = \mathds{1}$ and 
$P = U_T = \sigma_3^{\rm ph} \otimes \sigma_2^{\rm sp}$. 
In such basis channels' transformations $V,U'$
become orthogonal matrices, $\{V,U'\} \in \mathrm{SO}(N')$, defining the other two rotations according to
$V'^T = P\, V P$ and $U^T = P\, U' P$.
It is worth mentioning here that the eigenvalues of any transfer matrix, $e^{\pm \lambda_k}$, 
always occur in inverse pairs. 
Moreover, in class $\mathrm{DIII}$ they are 2-fold degenerate pairs (Kramers' degeneracy).

The symmetries of $\hat M$ listed above are strictly valid only at zero energy. 
Hence, when evaluating the determinant of the inverse Green's function $G^{-1}_\phi = i\delta \hat \tau - H(\phi)$, we have to assume
that a symmetry breaking term $i\delta \hat\tau$ is present only in the leads. This means that $\delta_s \equiv 0$ for cites
with $0\leq s \leq L$ (inside the wire) and $\delta_s \to 0^+$ otherwise. With such regularization at hand ${\rm det}( G^{-1}_\phi)$ is expressible
via the set of Lyapunov exponents $\{\lambda_k\}$.  Following Nazarov~\cite{Nazarov1994} we have accomplished this program using the method of quasiclassical  
Green's functions with the 'twisted' boundary conditions in the leads. 
The details of the calculations are presented in Appendix~\ref{app:Scattering}, here we proceed further with the discussion of results. 

For two $\mathds{Z}$-classes we have found
\begin{equation}
\label{eq:Part_sum_Z}
Z(\phi) = \left\langle \prod_{k=1}^{N'} \frac{\cosh(\lambda_k + i \phi_1)}{\cosh(\lambda_k + \phi_0)}\right\rangle,
\end{equation}
while for $\mathds{Z}_2$ topological wires the analogous 'plus' partition sum reads
\begin{equation}
\label{eq:Part_sum_Z2}
Z^{(+)}(\phi) = \left\langle \prod_{k=1}^{N'} \left( \frac{\cosh(\lambda_k + i \phi_1)\cosh(\lambda_k - i \phi_1)}
{\cosh(\lambda_k + \phi_0)\cosh(\lambda_k - \phi_0)}\right)^{1/2}\right\rangle.
\end{equation}
In order to extract the physical observable we introduce the generating function (GF)
\begin{eqnarray}
\label{eq:F_phi_1}
{\cal F}(\phi_0) &=& \partial_{\phi_1}Z^{(+)}(\phi)\bigl|_{\phi_1=-i\phi_0} \\
&=&
\nu^{-1}\langle {\rm tr}( (-\partial_{\phi_1} H) G_{\phi_1}) \rangle\bigl|_{\phi_1=-i\phi_0}, \nonumber
\end{eqnarray}
where the very last equality follows directly from Eq.~(\ref{eq:ZF_ZB}).
We see that GF is the zero energy expectation value of the symmetry current defined on a link $(0,1)$
with the help of the operator $\hat {\cal I}_S = -\partial_{\phi_1} H(\phi_1)$. It is worth mentioning here 
that due to gauge invariance the position of the source $\phi_0$ can be shifted to any link.
In this way one can define the current $\hat {\cal I}_S$ at each point of the wire.
Thus obtained $\hat{\cal I}_S$ is the {\it conserved} Noether's symmetry current
and its average value is some analytical function of $\phi_0$ which is independent of the actual 
choice of the link. 

Using now Eqs.~(\ref{eq:Part_sum_Z}, \ref{eq:Part_sum_Z2}) we find 
\begin{eqnarray}
\label{eq:F_phi_Z}
{\cal F}(\phi_0) &=& i\sum_{k=1}^{N'} \Bigl\langle \tanh(\phi_0+\lambda_k)\Bigr\rangle, \quad \mathds{Z}\,\mbox{-classes}.\\
{\cal F}(\phi_0) &=& i\sum_{k=1}^{N'} \left\langle \frac{\sinh 2\phi_0}{ \cosh 2\lambda_k + \cosh 2\phi_0}\right\rangle, 
\quad \mathds{Z}_2\,\mbox{-classes}. \nonumber
\end{eqnarray}
Following further the recipe~(\ref{AIIIObservableDef}) we first of all check that
\begin{equation}
g(L) = -i{\cal F}'(0) = \sum_{k=1}^{N'} \left\langle \frac{1}{\cosh^2 \lambda_k}\right\rangle
\end{equation}  
is the (thermal) conductance at scale $L$, as expected. In application to the SUSY $\sigma$-model
calculations such procedure works for three classes with the $P$-symmetry: 
$\mathrm{AIII}$, $\,\mathrm{BDI}$ and $\mathrm{DIII}$. In class $\mathrm{D}$ the low-energy field theory does
not have the continuum phase $\phi_1$ in the fermionic sector and the evaluation
of the generating function ${\cal F}$ introduced in Eq.~\eqref{eq:F_phi_1} becomes problematic. 
However, by setting $\phi_1=0$ one finds from Eq.~(\ref{eq:Part_sum_Z})
the series expansion
\begin{equation}
Z^{(+)}(\phi_0, \phi_1=0) = 1 + \frac{1}{2} g(L) \phi_0^2 + {\cal O}(\phi_0^4),
\end{equation}
which proves that Eq.~(\ref{eq:Dclass_g}) is the conductance in this symmetry class.

As for the topological number $\chi$, we start from
$\mathds{Z}$-insulators. According to the definition~(\ref{AIIIObservableDef}) we find
\begin{equation}
\label{eq:chi_L_BDI_AIII}
\chi(L) = \frac12 {\rm Im}\,{\cal F}(0) = \frac12 \sum_{k=1}^{N'} \Bigl\langle \tanh \lambda_k \Bigr\rangle
\end{equation}
In the trivial AI phase the number of positive and negative $\lambda$'s is equal and hence $\chi=0$. 
In the tAI phase all Lyapunov exponents satisfy $|\lambda_n| \gg 1$ and the topological number approaches an integer. 
In this phase the number of positive $\lambda$'s does not coincide with the number of negative ones.  
An integer $\chi$ changes by $\pm 1$ if any of $\lambda$'s
changes sign when crossing the line of quantum phase transition. Exactly at the transition point
the minimal Lyapunov exponent is zero, $\lambda_{\rm min} = 0$, and hence $\chi$ takes a half-integer value.

The situation is more intricate for $\mathds{Z}_2$-wires where we have to know $Z^-$ partition
sum to find the  topological number $\chi$. In class $\mathrm{D}$ one can identify
\begin{equation}
\label{eq:Z_kink_D}
Z^{(-)} = Z^{(+)}(\phi_0=0, \phi_1=\pi/2),
\end{equation}
as the partition function of a kink's configuration. The proof can be found in Appendix~\ref{app:Scattering_D}.
With the use of Eq.~(\ref{eq:Part_sum_Z}) this argument leads to:
\begin{equation}
\label{eq:chi_L_Z2}
\chi(L) = \Bigl\langle \prod_{k=1}^{N'} \tanh \lambda_k \Bigr\rangle.
\end{equation}
In the class $\mathrm{DIII}$ we have obtained the same result with the only difference that the product
now is taken over $N'/2=2N$ Lyapunov exponents, where each Kramers' degenerate eigenvalue $\lambda_k$ is
taken into account only once (cf. Appendix~\ref{app:Scattering_DIII}). 

We note that Eqs.~(\ref{eq:chi_L_BDI_AIII}) and (\ref{eq:chi_L_Z2}) can be written in the basis independent form,
\begin{align}
\label{eq:chi_r_Z}
\chi(L) &= {1\over 2} \langle {\rm tr} (\hat r P) \rangle,  \quad \mbox{classes AIII, BDI}; \\
\label{eq:chi_r_D}
\chi(L) &= \langle{\rm det} (\hat r) \rangle,  \quad\quad \mbox{class D}; \\
\label{eq:chi_r_DIII}
\chi(L) &= \langle{\rm Pf} ( \hat r P)\rangle,  \quad\quad\!\! \mbox{class DIII},
\end{align}
where $\hat r$ is the reflection matrix of the wire. Deep in the localized phase, $L \gg \xi$ ,
the average  topological number $\chi$ saturates to integer value. In this limit it coincides with the topological 
number ${\cal Q}$ introduced by Beenakker and co-workers~\cite{Fulga2011}. 
Relations~(\ref{eq:chi_r_Z}-\ref{eq:chi_r_DIII}) follow
from the DMPK decomposition of the scattering matrix
\begin{equation}  
\hat S = \left(\begin{array}{cc} U & \\ & V \end{array} \right)
\left(\begin{array}{cc} \tanh{\lambda}  & (\cosh{\lambda})^{-1} \\ 
(\cosh{\lambda})^{-1} & -\tanh{\lambda}
\end{array} \right)
\left(\begin{array}{cc} U' & \\ & V' \end{array} \right),\nonumber \\
\end{equation}
which gives for the reflection matrix $\hat r = U \tanh{\lambda}\, U'$. Indeed, employing the symmetry constrains
discussed previously we can state that $\hat r P = U \tanh{\lambda}\, U^\dagger $ in classes $\mathrm{AIII}$ and $\mathrm{BDI}$ and thereby Eq.~(\ref{eq:chi_r_Z}) reduces to~(\ref{eq:chi_L_BDI_AIII}).
In class  $\mathrm{D}$ one has ${\rm det U} = {\rm det U'}=1$, by construction, so that 
Eq.~\eqref{eq:chi_r_D} is in agreement with Eq.~\eqref{eq:chi_L_Z2}. Finally, in class 
$\mathrm{DIII}$ there is a basis where the parity operator should factorize into the product 
$P = \bigotimes_{k=1}^{2N} \Sigma_y$ where in the $k$-th block the corresponding Lyapunov exponent $\lambda_k$ 
is double degenerate. 
Hence $\hat r P = U \left(\bigotimes_{k=1}^{2N} \lambda_k \Sigma_y \right) U^T$ and the
Pfaffian of this antisymmetric matrix simplifies to Eq.~\eqref{eq:chi_L_Z2} where one substitutes $N' \to N'/2 = 2N$.

\begin{figure}[t]
\includegraphics[width=8.0cm]{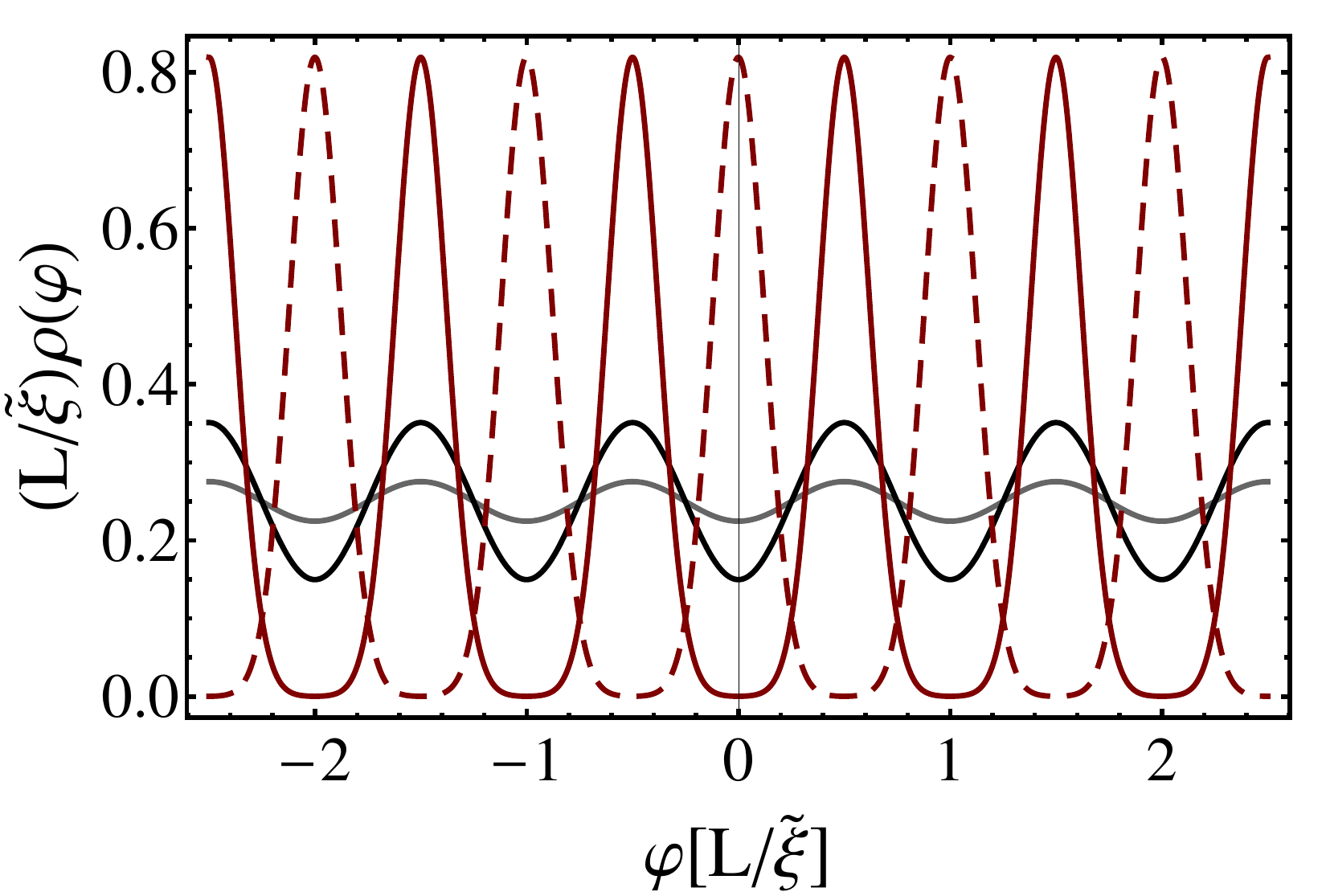}
\caption{Average density of Lyapunov exponents $\rho(\phi)$ in the class $\mathrm{AIII}$ disordered wire in the case of 
$\tilde\chi=0$ shown for $L/\tilde\xi = 1$ (weak localization --- grey line),  $L/\tilde\xi = 4$ (black line) and  
$L/\tilde\xi = 32$ (strong localization --- solid read line) and in the case of $\tilde\chi=1/2$ for
$L/\tilde\xi = 32$ (dashed read line).}
\label{fig:rho_AIII}
\end{figure}

The generating function ${\cal F}(\phi_0)$ can be used to recover the average density
of Lyapunov exponents\cite{Lamacraft2004}, 
\begin{eqnarray}
\label{eq:rho_lambda}
\rho(-\phi) &=& \sum_{k=1}^{N'} \biggl\langle  \delta(\phi+\lambda_k)\biggr\rangle \nonumber \\
&=&\frac{1}{2\pi}\Bigl( {\cal F}(\phi - i \pi/2) - {\cal F}(\phi + i \pi/2) \Bigr).
\end{eqnarray}
From our SUSY calculations $\rho(\phi)$ is available analytically in two symmetry classes, $\mathrm{AIII}$
and $\mathrm{DIII}$.  In the class $\mathrm{AIII}$ disordered wire, 
using the Poisson resummation formula, we have obtain 
from Eq.~(\ref{eq:Z_phi_AIII}) the following result,
\begin{eqnarray}
\rho(\phi;L) &=& \frac{\tilde \xi}{4L} + \sum_{n>0} (-)^n \frac{e^{- \pi^2 n^2 \tilde\xi/L}}{2\pi^2 n} \nonumber \\
&\times& \cos\bigl( 2\pi n ( \phi \tilde \xi/L - \tilde\chi)\bigr) \sinh\frac{\pi^2 n \tilde \xi}{L}.
\end{eqnarray}
The plot of this density, which shows the typical crystallization of Lyapunov exponents, is shown in
Fig.~\ref{fig:rho_AIII}. We note the periodic dependence on $\tilde \chi$. At $\tilde\chi=n+1/2$ 
the central peak is located at $\phi=0$ signaling the delocalized critical state. 
In the above result one should assume that the phase variable 
is limited to $\phi \in (-NL/\xi, NL/\xi)$, where $N$ is the number of channels.  

\begin{figure}[t]
\includegraphics[width=8.0cm]{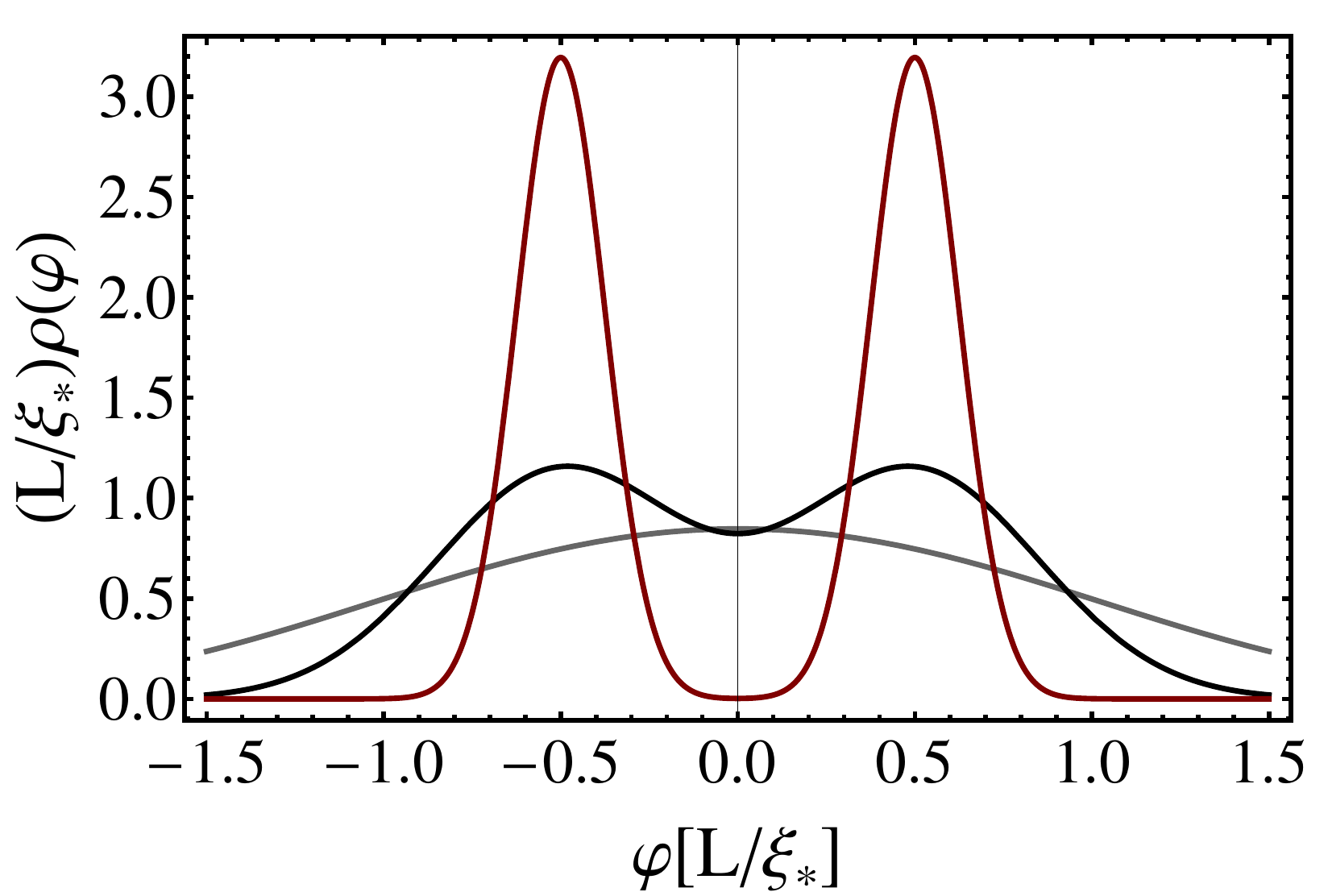}
\caption{Average density of two minimal Lyapunov exponents $\rho(\phi)$ in the class $\mathrm{DIII}$ 
wire shown for the bare value of the renormalized fugacity $v=0.2$ (see the text) and $L/\tilde\xi = v, 1, 8$
(grey, black and read lines, resp.).
In the limit $L \gg \xi\gg\tilde\xi$ (strong localization --- red line) the minimal exponents 'crystallize' around
values $\pm L/2\xi_*$.  }
\label{fig:rho_DIII}
\end{figure}

In the class $\mathrm{DIII}$, using Eq.~\eqref{eq:ZL_DIII}, one finds for the two minimal Lyapunov exponents
the following average distribution
\begin{eqnarray}
\label{eq:rho_DIII}
\rho(\phi;L) &=& 2 \int  \frac{dl_0}{\pi^2}  \frac{\sinh \pi l_0}{l_0} \, 
\cos(2 l_0\phi_0)\nonumber\\
&\times&\cos\!\left(\!\displaystyle\sqrt{2\pi g}\,\tilde\chi\, l_0 \times 
\frac{L}{\tilde\xi} \right)\e^{- l_0^2L/\tilde\xi},
\end{eqnarray}
which is normalized to $\int \rho(\phi;L) d\phi  =2$.
To quantify the crossover from the weak to strong localization in this distribution it is useful to introduce the renormalized bare fugacity $v = \tilde \chi\,\sqrt{2\pi g}$.   The average localization length~\eqref{eq:xi_DIII}
then reads as $\xi = 4 \tilde \xi/ v^{2}$, which is parametrically longer than $\tilde \xi$ in the limit $v\ll 1$.
Figure~\ref{fig:rho_DIII}  shows that at $L \gg \xi$ the 'crystallization' of two minimal Lyapunov exponents occurs, 
with the most probable $\phi$'s being $\pm L/\xi_*$. Interestingly, the new length scale, $\xi_* = \tilde \xi/v$
(typical localization length),
is parametrically different from $\xi$ (the average localization length) if $v\ll 1$. 
In the localized regime the length scale $\xi_*$ is defined by $ \langle \ln g(L) \rangle \simeq  - L/\xi_*$.
The main contribution to the average of conductance, $\langle g(L) \rangle  = \int d \phi\, \rho(\phi;L)$, 
however, come from the tails of the distribution $\rho(\phi;L)$ around $\phi=0$ 
resulting in the different length scale $\xi$ for the exponential decay $\langle g \rangle \sim e^{-L/\xi}$, 
such that in the close proximity to criticality one has $\xi \gg \xi_*$. 
The same conclusion has been reached in Ref.~\onlinecite{Gruzberg2005} on the basis of DMPK approach.
Equation~\eqref{eq:rho_DIII} also shows that for $v=0$ the maximum of the distribution is always 
located at $\phi=0$, corresponding to a perfectly transmitting channel, $t=1$. This is the origin of the delocalization in the absence of kinks. On the contrary, at $\tilde\chi\neq 0$ and large $L$ the perfect transmission is exponentially suppressed, signifying the localization. 

\section{Comparison to $2d$}
\label{sec:comparison_to_2d}

\subsection{The $\Bbb{Z}$-insulator classes: A,C, D}

Historically, the first tI under
consideration was the $2d$ integer quantum Hall effect. The importance of the static disorder in IQHE class A
tI was emphasized right from the beginning (the reason being that the singular
spectral of the clean bulk Landau level would not sit comfortably with
observed data).  The interplay of topological quantization and Anderson
localization in the system found a powerful description in terms of Pruisken's
low energy field theory\cite{Pruisken1984a} already in 80th. In its subsequent supersymmetric
formulation\cite{Zirnbauer1994} the Pruisken theory is described by the effective action,
\begin{align}
        \label{PruiskenAction}
        S[Q]=\frac{1}{8}\int d^2x\,\left[-\tilde g\,\mathrm{str}(\partial_\mu Q\partial_\mu Q
        )+\tilde \chi \epsilon_{\mu\nu}\mathrm{str}(Q\partial_\mu Q\partial_\nu Q)
        \right],
\end{align}
where $\tilde g=\sigma^0_{xx}$ and $\tilde \chi =\sigma_{xy}^0$ are the bare (
Drude)
values of longitudinal and Hall conductivities, respectively and $Q=T\tau_3 T^{-1}$
takes values in the super-Riemannian space
$\mathrm{U}(2|2)/\mathrm{U}(1|1)\times \mathrm{U}(1|1)$.

The Pruisken theory shows striking parallels to that of the $1d$ class 
$\Bbb{Z}$ -- tI's discussed above: (i) the theory assumes the form of a nonlinear $
\sigma$-model containing a topological $\theta$-term (the second term in the
action). (ii) Upon increasing length scales, the two parameters in the action
renormalize according to Eq.~\eqref{FlowGeneric}. (iii) The renormalization of
the topological parameter $\chi$ is driven by a proliferation of topological
excitations in the system, which in $2d$ are instantons on the 
fermionic target space, $Q^{\mathrm{ff}}\in \mathrm{U}(2)/\mathrm{U}(1)\times
\mathrm{U}(1 )\simeq S^2$, the two-sphere. These excitations assume the role
of the `phase' windings' in the $1d$ context. (iv) For generic values $\tilde
\chi \notin\Bbb{Z }+\frac{1}{2}$, the flow towards an insulating configuration
$g=\sigma_{xx }=0$ implies a restoration of the full symmetry under
$G=\mathrm{U}(2|2)$ in the bulk. At the same time, the $\theta$-term at its
fixed point coupling $\chi\in \Bbb{Z}$ becomes a boundary action of Wess-
Zumino type\cite{Zirnbauer1994}, which describes the gapless propagation of  boundary
modes. (v) The observables $(g,\chi)$ can be extracted from the theory by
coupling to topological sources, Pruisken's background field method\cite{Pruisken1984a}.

However, unlike the $1d$ $\Bbb{Z}$-systems, the critical physics of the
IQHE system has not been quantitatively described beyond the weak coupling 
(large $g$) perturbative regime. The theory describing the QH fixed point, must
be conformally invariant, a feature the Pruisken model lacks. One therefore
expects a metamorphosis towards a conformally invariant fixed point theory along
the  flow on the $(g,\Bbb{Z}+1/2)$ critical surfaces. However, both the fixed
point theory and the conversion mechanism were not identified so far.

The situation is different in the case of the class C topological insulator. This
system, too, is described by a theory with Pruisken type action~\eqref{PruiskenAction}, 
the difference lying in the target space which now is $\mathrm{OSp}(2|2)/
\mathrm{U}(1|1)$. All points (i)-(v) above remain valid as they stand, with the
added feature that the critical point is under control: it has been shown\cite{Gruzberg1999}
that the class C quantum Hall transition belongs to the percolation
universality class, which implies that its CFT is under control. By contrast,
the class D system, another $2d$ class $\Bbb{Z}$ topological insulator, is not
fully understood: as discussed above the class D field manifold contains two
disjoint sectors. In $2d$, the ensuing $\Bbb{Z}_2$ leads to the emergence of
Ising type criticality. It is believed\cite{Bocquet2000}, that this leads to the formation of
a tri-critical point separating a topological and a non-topological insulator
phase, and a (thermal) metal phase. For further discussion of this system, we
refer to the literature\cite{Bocquet2000,Mirlin:2007} 

\subsection{The $\Bbb{Z}_2$-insulator classes: AII, and DIII} 

Above we found that  the $1d$ insulators of type $\Bbb{Z}$ and $\Bbb{Z}_2$,
respectively were different in the nature of their topological structures: smooth
instanton excitations vs. topological point defects. A similar dichotomy
appears to be present in $2d$. While the class $\Bbb{Z}$ quantum Hall
insulators of class A, C, D admit instantons as discussed in the previous
subsection, a pioneering study\cite{Kane-Fu2012} of Fu and Kane (KF) on the $\Bbb{Z}_2$ spin
quantum Hall effect suggests that vortices with the point-like singularity in the middle are the relevant
topological excitations of the system. The role of the $\theta$-term is  taken by a
contribution to the action $\ln \tilde \chi \times n_\mathrm{v}$, where
$n_\mathrm{v}$ is the number of vortices. At criticality, the vortex fugacity
vanishes $\tilde \chi\to 0$, and the resulting zero-vortex theory, $2d$ nonlinear
$\sigma$-model belonging to the symplectic (AII) symmetry class exhibits 
delocalized  behavior, expected of the critical system. The situation is a
little more complicated due to the fact that a $2d$ system with vortices admits a
Kosterlitz-Thouless transition. The analysis of KF indeed suggests that the KT
transition points are positioned relative to the $\tilde \chi\to 0$ axis in
such a way as to extend the metallic critical surface to a {\em metallic phase} (for
further discussion of this point, we refer to the original reference [\onlinecite{Kane-Fu2012}]). 
Further, the KF analysis is based on a replica framework, in which the
vortices are phase windings between select replicas. In view of the 
non-perturbative nature of these excitations,  an adaption of the approach to
the mathematically more rigorous framework of supersymmetry seems worthwhile.
However, the topological structure of the corresponding supersymmetric field
manifold, does {\em not} seem to support vortices, at lest not in the most obvious sense. Further
work is required to better understand this point. In this context, it is worth mentioning
the pivotal role of vortices played in metal-insulator transition governed by the topological Anderson 
localization in the two-dimensional disordered fermionic systems of chiral symmetry classes, 
as it was recently suggested by K\"onig {\it et al}.~\cite{Koenig:2012} We finally note that the
$2d$ DIII system --- the time reversal invariant but spin rotation non-
invariant $\Bbb{Z}_2$ topological superconductor -- has not been addressed so
far. However, the structure of its field manifold suggests that as in class
AII topological point defects will be present.

We conclude that the list (i)-(iv) of $1d$--$2d$ analogies formulated above essentially generalizes to the $4$ $\Bbb{Z}_2$ systems. The main difference between the $\Bbb{Z}$ and the $\Bbb{Z}_2$ insulators is that the role of smooth instantons is taken by point defects, and that of field theoretical $\theta$-terms by fugacity terms. Without further discussion, we also note that the smooth/point-like dichotomy pertains to the realization of the topological sources, required to read out observables. 
		
\section{Acknowledgments}  

Discussions with P. Brouwer and N. Read are gratefully acknowledged.
AA and DB were supported by SFB/TR 12 of the Deutsche Forschungsgemeinschaft. 
AK was supported by NSF grant DMR1306734. AK also kindly acknowledges the hospitality of the 
International Faculty  program at the University of K\"oln.  

\appendix
\label{a.1}

\section{Field theory of $\mathds{Z}$-insulators}
\label{sec:field_theory_of_z_insulators}

We outline here the derivation of the field theory actions for classes $\mathrm{AIII}$ and $\mathrm{BDI}$ 
and discuss the self-consistent Born approximation (SCBA).

\subsection{Class $\mathrm{AIII}$}

\label{app:classAIII}
The supersymmetric action of the $N$-channel disordered quantum wire 
corresponding  to the microscopic model with the Hamiltonian~(\ref{eq:AIIIHamiltonian}) written
in terms of the spinor $\psi=(\psi_+, \psi_-)$,
\begin{eqnarray}
S &=&- \sum_s \left( \mu\, \bar \psi_{-,s} \psi_{+,s}  + t\, \bar \psi_{-,s-1} \psi_{+,s} + {\rm h.c.} \right)
+ S_{\rm dis},  \nonumber \\
S_{\rm dis} &=& -\sum_{s,kk'} \left( \bar \psi_{s+1,k}  R^{kk'}_{s+1,s} \psi_{s,k'}  
+ {\rm h.c.}
\right),
\end{eqnarray} 
where $\psi_s = \psi_{+,s/2}$ for $s$ even and $\psi_s = \psi_{-,(s-1)/2}$ for $s$ odd.
Averaging $e^{-S_{\rm dis}}$ over the Gaussian fluctuations of disorder~(\ref{eq:R_AIII}) 
one obtains the effective action containing the spatially local quartic term,
\begin{equation}
\tilde S_{\rm dis} = \frac{2 w^2}{N}\sum_s {\rm str} \left( g^{++}_s g^{--}_s \right),
\end{equation}
where we have introduced bilinears $g^{\alpha\beta}_s = \sum_k \psi_{\alpha,s,k} \otimes \bar\psi_{\beta,s,k}$.
This term can be decoupled with the use of Hub\-bard-Stratonovich transformation by introducing two
auxiliary $2\times 2$ matrix fields $Q^\pm_s = Q_{1,s} \pm i Q_{2,s}$ operating in the ${\rm bf}$-space.
Integrating further over the $\psi$-fields, the partition sum assumes the form 
$Z=\int {\cal D}Q^\pm  \,\exp(- S[Q])$, where
\begin{equation}
S[Q] = \frac{N}{2w^2} \sum_s {\rm str} (Q_s^+ Q_s^-)\, +\, 
{\rm str}\ln\! \left( \!
\begin{array}{cc}
i0 - Q^+ & - \hat h_+ \\
- \hat h_- & i0 - Q^- 
\end{array} \!\!\!\right)\!.
\end{equation}
Here $\hat h_\pm$ are block matrices of the disorder independent Hamiltonian which in momentum space become
$h_\pm (q) = \mu + t e^{\pm i q}$. 

Derivation of the sigma-model proceeds by subjecting the effective action $S[Q]$ to 
the saddle point analysis and identifying  the soft Goldstone modes around the saddle point. 
Saddle point equations are known to be equivalent to the self-consistent Born approximation (SCBA) with the saddle
point playing the role of the self-energy. In the present case they can be easily resolved by the ansatz 
$Q^\pm=-i \Sigma_0 \sigma_0$, where $\Sigma_0 \in \mathds{R}$. The solution of the SCBA equation,
\begin{eqnarray}
\label{eq:SCBA_AIII}
-i \Sigma_0 &=& w^2\, {\rm tr}\, \langle s | ( i \Sigma_0 - \hat H_0 )^{-1}  |s\rangle,
\end{eqnarray}
is analyzed in the next subsection. Here $\hat H_0 = \hat h_+ \sigma_+ + \hat h_- \sigma_-$, and the trace is taken with respect to the chiral space. Solving this equation, results in 
$\Sigma_0(\mu/t, w/t)$, which is a function of microscopic parameters in $H_0$ and the disorder strength $w$.

To identify the soft modes we substitute 
$i Q_s^+ = \Sigma_0 T_s$ and $i Q_s^- = T_s^{-1} \Sigma_0$ to find that
\begin{eqnarray}
\label{eq:ST_AIII}
\!\!\!\!{\cal S}[T] &=& {\rm str} \ln\left( 
\begin{array}{cc}
i \Sigma_0 T & - \hat h_+ \\
- \hat h_- & i T^{-1} \Sigma_0 
\end{array} \right) \nonumber \\
&=& {\rm str} \ln\left( 
\begin{array}{cc}
-i \Sigma_0\quad  &  \hat h_+ + T^{-1} [\hat h_+, T]\\
 \hat h_- & -i \Sigma_0 
\end{array} \right),
\end{eqnarray}
where the last equality results from the gauge transform of the ${\rm Sdet}$.
Notice that for a uniform in space $T$-field the action ${\cal S}[T]$ 
is extremal and thus $\mathrm{GL}(1|1)$ is the Goldstone manifold of our problem 
and the field $T_s$, when slowly varying in space, is the soft-mode fluctuation. One also
observes that ${\cal S}[T_L T T_R^{-1}] = {\cal S}[T]$ which shows that the full
symmetry group of the initial problem is $G=\mathrm{GL}(1|1) \otimes \mathrm{GL}(1|1)$,
which is then broken to $G/\mathrm{GL}(1|1)$ after disorder averaging.

At the next step we expand the action ${\cal S}[T]$ in a gradient $\partial_x T(x)$ assuming 
this field is changing slowly on the lattice scale and thus passing to the continuum limit $s \to x$. 
To deal simultaneously with momentum and
coordinate dependence of operators under the ${\rm str}$ we use the Moyal formula ('star' product)
$(A \star  B) (x,q) = A \exp\{ i(\overleftarrow{\partial_x}\overrightarrow{\partial_q} -\overleftarrow{\partial_q}\overrightarrow{\partial_x} )/2\} B$  to evaluate the convolution of 
any two operators. Expanding the 'star' product in gradients we get
\begin{equation}
T^{-1} [h_+, T] \simeq  -iv_+(q)\, T^{-1} \partial_x T + {\cal O} (\partial_x^3 T)
\end{equation}  
with the complex 'velocities' 
$v_\pm(q) = \partial_q h_\pm(q)\bigl|_{\Delta=t} = \pm i t\, e^{\pm i q}$.
The further expansion of ${\cal S}[T]$ up to second order in the field $\Phi_x = T^{-1}\partial_x T$ gives the low-energy field theory\cite{Altland2001511} with the action~(\ref{eq:AIIIAction}).

The microscopic value of the bare localization length $\tilde\xi$  is expressed in terms of velocity correlation
function. Its exact value is of no importance for our consideration. In the limit of the weak disorder,
$w \ll t \sim \mu$, one can estimate $\tilde\xi \sim N a (t/w)^2$ where $a$ is a lattice constant.

For the bare topological coupling constant one finds 
$\tilde\chi = i \langle G_{-+}^R(q) v_+(q)\rangle_q$ (with $G^R=(i\Sigma_0 - \hat H_0)^{-1}$ 
being the retarded SCBA Green's function), which can be also written in the symmetrized form 
\begin{equation}
\label{eq:bare_chi_Z}
\tilde \chi = \frac{i}{2}\langle{\rm tr}(G^R(q) P \partial_q H)\rangle_q.
\end{equation}
It shows the relation of the bare topological  topological number to the fictitious 
'chiral' persistent current, defined by the operator $\hat j_c(q) = P \hat v(q)$. 

With the use of the $q$-representation the SCBA, Eq.~(\ref{eq:SCBA_AIII}), takes the form
\begin{equation}
1 = 2w^2 \int_{-\pi}^\pi\frac{dq}{2\pi} \,\frac{1}{\Sigma_0^2 + t^2 + \mu^2 + 2\mu t \cos q}.
\end{equation}
Performing the integration with the help of the residue theorem in a complex plane of a variable $z=e^{iq}$, 
the SCBA is reduced to $1=2 w^2/{\cal D}$, where
\begin{equation}
\label{eq:Denominator}
{\cal D} = \left(\Sigma_0^2 +( t-\mu)^2 \right)^{1/2} \left(\Sigma_0^2 +( t+\mu)^2 \right)^{1/2}.
\end{equation}
From here $\Sigma_0$ can be explicitly found by resolving the quadratic equation. In the limit
of small staggering and relatively weak disorder, i.e. at $|t-\mu| \ll w \ll t$,  it reads 
$\Sigma_0 = 1/{2\tau} \simeq w^2/t$. 
Introducing the velocity $v=a t$ one finds for the bare localization length 
$\tilde\xi \sim N v \tau \sim N a (t/w)^2$ as stated in the main text.

Finally, substituting the SCBA Green's function into Eq.~(\ref{eq:bare_chi_Z}), one obtains the bare
topological number 
\begin{equation}
\tilde\chi = \frac{N}{2}\left( 1+ (t/w)^2 - \sqrt{1+t^2 \mu^2/w^4} \right).
\end{equation}
In the limit of small staggering it is simplified to
\begin{equation}
\tilde\chi = \frac{N}{2}\left( 1 +  \frac{(t-\mu)t}{w^2}\right), \quad \quad |t-\mu| \ll w \ll t.
\end{equation}
This bare topological number $\tilde\chi$ can be used to find the critical lines/surfaces of phase transition from the equation 
$\tilde\chi(\mu,t,w)=n+1/2$ with $n \in \mathds{Z}$. The corresponding lines are plotted in Fig.~\ref{fig:Diagram_AIII_N3} for $N=3$.

\subsection{Class BDI}
\label{app:BDI}

Derivation of the sigma-model action in the class $\mathrm{BDI}$ system proceeds along the same lines as for the class $\mathrm{AIII}$. We start from the Hamiltonian~(\ref{eq:BDI_model_H})
and transform it to the 'chiral' basis where it has the block off-diagonal form, $\hat H= \hat H_0 +  \hat V \sigma_1$. 
Here $\hat H_0 = \hat h_+ \sigma_+ + \hat h_- \sigma_-$ and the operators $\hat h_\pm $
in the $q$-space read $\hat h_\pm(q) = - (t + \cos q) \pm  i \Delta \sin q$. 

To construct the path-integral representation of $Z$ we consider the Gaussian
action $S[\psi,\bar\psi]=\bar\psi( i0 - \hat H)\psi$ and following the doubling procedure detailed in sec.~\ref{sec:FT_classD}
represent the former in the form, 
$S[\Psi]=\frac{1}{2}\bar\Psi(i0 - \hat H)\Psi$, where in class $\mathrm{BDI}$ the spinor 
$\bar\Psi = (\bar\psi, \sigma^{\rm bf}_3 \psi^T)$, $\Psi = \tau \bar\Psi^T$ and the charge-conjugation matrix 
$\tau= {\cal P}^0 \otimes \tau_1 + {\cal P}^1 \otimes i\tau_2$. Subsequent disorder averaging over the
Gaussian random matrices $\hat V_s$ at each site $s$ produces the quartic term
\begin{equation}
\tilde S_{\rm dis} = \frac{ w^2}{2 N}\sum_s {\rm str} \left( \sigma_1 g_s \right)^2, 
\quad g_s = \sum_k \Psi^k_s \otimes \bar\Psi^k_s.
\end{equation}
This quadratic in $g_s$ form can be now decoupled by four $2\times 2$ super-matrix fields $P_{1,2}$ and $Q_\pm$
acting in the $\mathrm{cc}$-space, which is useful to combine into the single matrix $R$ 
with the structure in the chiral space,
\begin{equation}
R=\left( \begin{array}{cc} Q_+ & P_1 \\ P_2  & Q_-  \end{array}\right).
\end{equation}
By doing so we obtain the following representation for the disorder averaged partition sum, 
$ Z  = \int {\cal D}(\Psi; R) \exp\left( - {\cal S}[\Psi,R] \right)$ with the action
\begin{equation}
{\cal S}[\Psi,R] = \frac{N}{16 w^2} {\rm str} (\sigma_1 R)^2  - \frac{i}{2}\bar\Psi ( i0 -\hat H_0 - R )\Psi.
\end{equation}
This action is gauge invariant under the $T$-rotations from the super-group $G=\mathrm{GL}(2|2)$
operating in the $\mathrm{cc}$ space.  Namely, the transformation of the spinors 
\begin{equation}
\Psi \rightarrow \left(\begin{array}{cc}
\bar T^{-1} & \\
&  T
\end{array} \right) \Psi, \qquad 
\bar\Psi \rightarrow \bar\Psi \left(\begin{array}{cc}
T^{-1} & \\
& \bar T
\end{array} \right),
\end{equation}
with the simultaneous transformation of matrix fields, $Q_+\to T Q_+ \bar T$ and $Q_-\to \bar T^{-1} Q_- T^{-1}$ 
(we remind that the involution is defined as $\bar T = \tau T^T \tau^T$) leaves the action ${\cal S}$ invariant. 

Next we perform the Gaussian integral over the $\Psi$-fields and reduce the partition function to 
the form $Z=\int {\cal D} R  \,\exp(- S[R])$, where
\begin{equation}
\label{eq:SR_BDI}
S[R] = 
\frac{N}{16 w^2} {\rm str} (\sigma_1 R)^2  + \frac{1}{2}{\rm str}\ln( i0 -\hat H_0 - R ).
\end{equation}
Extremizing this action, one obtains the saddle point $R=\hat \Sigma \equiv i\Sigma_0 + \Sigma_1 \sigma_1$,
where two components of the self-energy $\Sigma_{0,1} \in \mathds{R}$ are to be found from the SCBA equation,  
\begin{eqnarray}
i \Sigma_0 &=& w ^2 {\rm tr} \langle s| ( i 0 -  \hat H_0 - \hat \Sigma)^{-1}|s\rangle, \\
\Sigma_1 &=& w^2 {\rm tr} \langle s|  (  i 0 - \hat H_0 - \hat \Sigma)^{-1} \sigma_1| s\rangle.
\end{eqnarray}
We give its solution in the following subsection. The self-energy $\hat \Sigma$ is just one particular
saddle of the action $S[R]$. Other possible extrema follow from the gauge invariance of the action
under $T$-rotations. They generate the manifold of saddle points parametrized as
\begin{equation}
R=\left(\begin{array}{cc}
i \Sigma_0\, T \bar T & \Sigma_1\, \mathds{1}^{cc}\otimes\mathds{1}^{\rm bf} \\
\Sigma_1\, \mathds{1}^{cc}\otimes\mathds{1}^{\rm bf} & i \Sigma_0 \,  (T \bar T)^{-1}
\end{array}\right),
\end{equation}
The super-matrix $T$ is an element of the linear supergroup 
$G={\rm GL}(2|2)$. We see, however, that the subgroup of matrices $T'$ satisfying the constraint
$\bar T' T' = 1$
does not affect the trivial saddle point. All matrices $T'$ form the complex supergroup $H={\rm OSp}(2|2)$. 
Factoring out this subgroup we conclude that essential rotations are $T \in G/H = {\rm GL}(2|2)/{\rm OSp}(2|2)$, 
which is the manifold of the soft (Goldstone) modes of the supersymmetric  $\sigma$-model in the $\mathrm{BDI}$
symmetry class.

Similar to class $\mathrm{AIII}$ the final form of the $\sigma$-model~(\ref{eq:DBIAction}) 
follows from the gradient expansion of the action ${\cal S}[T]$ 
with a smoothly varying matrix field $T(x)$ where
\begin{eqnarray}
&&{\cal S}[T]=\\
&&\frac{1}{2} {\rm str} \ln\left( 
\begin{array}{cc}
-i \Sigma_0\quad  & \!\!\! \hat h_+ + \Sigma_1 + \Delta h_+\\
 \hat h_- + \Sigma_1 + \Delta h_- & \!\!\! -i \Sigma_0 
\end{array} \!\!\right). \nonumber
\end{eqnarray}
Here we have defined $\Delta h_+ = T^{-1} [\hat h_+, T]$ and $\Delta h_- = \bar T [\hat h_-, \bar T^{-1}] $,
cf.  Eq.~(\ref{eq:ST_AIII}). 
As a result of the gradient expansion one obtains 
the action~\eqref{eq:DBIAction} which is the functional of the field $Q=T\bar T$ 
with the same estimate for $\tilde \xi$ and 
the same formal result~(\ref{eq:bare_chi_Z}) for the bare topological number as in the case of class $\mathrm{AIII}$
system.

We now discuss the SCBA and derive the bare topological coupling constant $\tilde\chi$ 
for the model of disordered multi-channel p-wave superconducting wire.
Transforming the self-consistent Green's function to the momentum representation, using the explicit
model form of $\hat H_0$ and introducing the notation $\tilde \mu = \mu - \Sigma_1$
one obtains two coupled SCBA equations (we limit ourselves to the special point $t= \Delta$ only)
\begin{eqnarray}
\Sigma_0 &=& 2 w^2 \int_{-\pi}^\pi \frac{d q}{2\pi} \frac{\Sigma_0}{\Sigma_0^2 + \tilde \mu^2 + t^2 + 2\tilde \mu t \cos q}, \nonumber \\
\Sigma_1 &=& 2  w^2 \int_{-\pi}^\pi \frac{d q}{2\pi} \frac{\tilde \mu + t\cos q}{\Sigma_0^2 + \tilde \mu^2 + t^2 + 2\tilde \mu t \cos q}.
\label{eq:SCBA_k_BDI}
\end{eqnarray}
After $q$-integration, with the use of the function ${\cal D}$ defined above in Eq.~(\ref{eq:Denominator}) 
(where the substitution $\mu \to \tilde\mu$ is assumed),
these equations are reduced to the coupled system of algebraic ones,
\begin{eqnarray}
\Sigma_0 &=& \frac{2 w^2 \Sigma_0}{{\cal D}}, \\
\Sigma_1 &=& \frac{ w^2}{\mu-\Sigma_1}\left(1 +\frac{(\mu-\Sigma_1)^2 - \Sigma_0^2 - t^2}{{\cal D}}\right). \nonumber
\end{eqnarray}
It follows from the 1-st equation that ${\cal D} = 2w^2$, which then can be used to express $\Sigma_0$
via $\Sigma_1$ using the 2-nd equation. After these steps, employing the relation ${\cal D} = 2w^2$ once again, 
one arrives to the quartic polynomial equation for $\Sigma_1$ which can be solved numerically.
The self-energy $\hat \Sigma$ can be further used to find the SCBA topological number $\tilde\chi$. For
the given model of p-wave disordered wire we have found from Eq.~(\ref{eq:bare_chi_Z})
\begin{equation}
\label{eq:chi_BDI_full}
\tilde\chi(w,\mu) = \frac{N}{2} \left( 1 + \frac{t^2 - \Sigma_0^2 - (\mu -\Sigma_1)^2}{2 w^2} \right).
\end{equation}

The above results can be simplified for the chemical potential in a close vicinity of the band edge, $|\mu - t| \ll t$,
and at weak disorder, $w \ll t$. In this limit one can set ${\cal \tilde D}\simeq 1$.
By defining the scattering rate as $1/\tau = N w^2/t$ 
and the detuning from criticality $\bar\mu = \mu -t$, we have found
\begin{eqnarray}
\label{eq:Gapless_S}
\Sigma_0 &=& \frac{\sqrt{3}}{2} \left( \frac{1}{\tau} - \frac{\bar \mu}{3}\right)^{1/2}\left( \frac{1}{\tau} + \bar\mu\right)^{1/2},  \nonumber \\ 
\Sigma_1 &=& -\frac{\bar\mu}{2} - \frac{1}{2\tau}
\end{eqnarray}
in the range $\bar \mu \in (-1/\tau, 3/\tau)$ which is the interval of a non-vanishing
mean DoS on the level of SCBA. Under the same assumptions the result~(\ref{eq:chi_BDI_full}) for
the bare topological number is simplified to its approximate value~(\ref{eq:chi_BDI_1}) stated in the main text.

\section{Field theory of class $\mathrm{D}$}
\label{app:D}

\subsection{The Gaussian representation~\eqref{eq:Zsum_D} }

For the sake of completeness, we here briefly describe how the symmetries characterizing the Gaussian integral representation~\eqref{eq:Zsum_D} are derived. The starting point is a `plain' Gaussian super integral representation
\begin{align}
	\label{eq:Zsum_D-app}
	Z=\left\langle \int d(\bar \psi,\psi)\,e^{i\bar \psi (G^+)^{-1}\psi
	}\right\rangle,
\end{align}
where $\psi=(\psi^\mathrm{b},\psi^\mathrm{f})$  is a vector comprising bosonic and Grassmann variables with components
$\psi^\alpha=\{\psi^\alpha_{s,k}\}$. Notice that the symmetry  $H^T = - H$  implies  $(G^+)^T = - G^-$. To see, how these symmetries entail an effective symmetry of the integration variables, we write 
\begin{align}
	&\bar \psi (G^+)^{-1}\psi=\frac{1}{2}(\bar \psi (G^+)^{-1}\psi+\bar \psi (G^+)^{-1}\psi)=\crcr
	&=\frac{1}{2}\left(\bar \psi (G^+)^{-1}\psi+(\bar \psi (G^+)^{-1}\psi)^T\right)=\crcr
	&=\frac{1}{2}\left(\bar \psi (G^+)^{-1}\psi- \psi^T\sigma_3^{\mathrm{bf}} (G^-)^{-1}\bar\psi^T\right)\equiv\crcr
	&\equiv \Bar \Psi  \left(\begin{matrix}
		(G^{+})^{-1}&\crcr &(G^{-})^{-1}
	\end{matrix}\right) \Psi\equiv \bar\Psi G^{-1}\Psi,
\end{align}
where 
\begin{align}
	\label{eq:DPSIDef}
	\bar\Psi\equiv \frac{1}{\sqrt2}(\bar\psi,-\psi^T\sigma_3^\mathrm{bf}),\qquad \frac{1}{\sqrt2}\Psi=\left(\begin{matrix}
		\psi\crcr
		\bar\psi^T
	\end{matrix}\right).
\end{align}
In the second line, we used that the number $\bar \psi (G^+)^{-1}\psi$ is
equal to its transposed, and in the third that the transposition of a
Grassmann bilinear form $\bar \psi^1
(G^+)^{-1}\psi^1=-\psi^{1T}(G^{+})^{-1T}\bar\psi^{1T}$ introduces an extra
minus sign. The two component structure introduced in Eq.~\eqref{eq:DPSIDef}
defines the space of $\tau$-matrices. From the structure of $\Psi$ and $\bar
\Psi$, one reads out the symmetry~\eqref{eq:DPsiSym}, and that  $G^{-1}=i0
\tau_3-H$.

\subsection{Jacobians} 

\label{app:D_1}
Here we calculate Jacobians on the two parts of the  manifold. To this end we write 
$Q^{(\pm)} = e^{\cal W} {\tilde Q}^{(\pm)} e^{-{\cal W}}$, Eq.~(\ref{eq:Q_par}), and differentiate over parameters. In doing it we choose to stay near zero Grassmanns  for simplicity, since Jacobians are expected to be Grassmann-independent 
$ dQ^{(\pm)} =[d{\cal W}, {\tilde Q}^{(\pm)}] + d{\tilde Q}^{(\pm)} $ and $ dQ^{(\pm)-1} =[d{\cal W}, {\tilde Q}^{(\pm)-1}] + d{\tilde Q}^{(\pm)-1} $.
The metric is given by
\begin{eqnarray}
dg&=&-{1\over 8}{\rm str}\left( dQ dQ^{-1} \right) \\\nonumber &=&- {1\over 8}{\rm str}\left( [d{\cal W}, {\tilde Q}]  [d{\cal W}, {\tilde Q}^{-1}]  \right)-
{1\over 8}{\rm str}\left( d{\tilde Q} d{\tilde Q}^{-1} \right)
\end{eqnarray} 
Substituting the parametrization (\ref{eq:Q_par}-\ref{eq:B_blocks}), one finds for the  ``plus'' manifold 
\begin{eqnarray}
dg^{(+)}&=&dy^2+\sinh ^22y \,\, d\alpha^2   +2\eta \xi   \sinh ^2 y 
\end{eqnarray}  
and therefore 
\begin{equation}
J^{(+)}=\sqrt{\mathrm{Sdet} g^{(+)}}=\, \frac{ \sinh 2y  }{\sinh ^2 y }=2\coth y
\end{equation}
In a similar way, for ``minus'' part of the manifold one finds 
\begin{eqnarray}
dg^{(-)} = dy^2+\sinh ^22y \,\, d\alpha^2   - 2\eta \xi   \cosh ^2 y. 
\end{eqnarray}  
The minus sign in the last term implies the opposite sign of the Grassmann measure.
\begin{equation}
J^{(-)}=\sqrt{\mathrm{Sdet} g^{(-)}}=\, \frac{ \sinh 2y  }{\cosh ^2 y }=2\tanh y
\end{equation}

\subsection{Transfer Matrix}
\label{app:D_2}

Our goal is to calculate transfer matrix element between two neighboring  grains $e^{-S(Q^{(\alpha)},\tilde Q^{(\beta)})}$, where 
$\alpha,\beta=\pm$ denote two parts of the manifold, and
\begin{equation}
                                        \label{eq:actionD}
 S(Q,\tilde Q) = \frac{1}{4}\sum_{k=1}^{2N} {\rm str}\ln\left[1+\frac{ t_k^2}{4}(\{Q,\tilde Q\}_+ - 2)\right].
\end{equation} 
This operator acts on the two-component spinor wave function $\Psi(\tilde Q)=\Psi^{(\beta)}(\tilde y)$, which 
is assumed to depend only on the radial variable $\tilde y$, but not on the bosonic angle $\tilde \alpha$ or Grassmann variables $\tilde \eta, \tilde \xi$. The entire angle/Grassmann dependence is thus restricted to the transfer matrix element and may be integrated out explicitly. We thus define the radial part of the transfer  operator as
\begin{equation}
{\cal R}^{(\alpha\beta)}(y;\tilde y) = \int \frac{d\tilde \alpha}{2\pi}\, d\tilde \eta\, d\tilde\xi\, (\beta)  J^{(\beta)}(\tilde y) \, 
e^{-S(Q^{(\alpha)},\tilde Q^{(\beta)} )},
\end{equation}
where factor $(\beta)$ reflects opposite sign of the Grassmann measure on the $(-)$ manifold.
As a result of the angular-rotational invariance 
the element $Q^{(\alpha)}=Q^{(\alpha)}(y)$ may be assumed to be  
pure radial as well (i.e. taken e.g. at zero angles).
We now substitute such $Q^{(\alpha)}(y)$ and $\tilde Q^{(\beta)}(\tilde y,\tilde \alpha, \tilde \eta,\tilde\xi)$ into the action (\ref{eq:actionD}) and find:
\begin{equation}
S=g\big[S_0^{(\alpha\beta)}(y;\tilde y,\tilde\alpha) + \tilde \eta\, \tilde \xi\, F^{(\alpha\beta)}(y;\tilde y,\tilde \alpha)\big],
\end{equation} 
where $S_0^{(\alpha\beta)}$ is the  part of the action which does not contain Grassmanians and $g=\sum_k  t_k^2\gg 1$.
We then expand  $e^{g\tilde \eta\,\tilde \xi F^{(\alpha\beta)}}$ to the first order, integrate over Grassmann variables and obtain 
\begin{equation}
{\cal R}^{(\alpha\beta)}(y;\tilde y) = g\int \frac{d\tilde \alpha}{2\pi}\, (\beta) J^{(\beta)}(\tilde y) \, F^{(\alpha\beta)} \, e^{-gS_0^{(\alpha\beta)}  }.
\end{equation}
 The straightforward, though lengthy, calculation yields:
\begin{eqnarray}
S_0^{(\pm\pm)}&=& \frac{1}{4}\cosh 2y  \cosh 2\tilde y  \\
&-& \frac{1}{4} \cos 2 \tilde \alpha  \sinh 2y  \sinh 2\tilde y -\frac{1}{4} \nonumber \\
&\simeq&  \frac{1}{2}(\tilde y-y)^2 +\frac{\tilde \alpha^2}{2} \sinh 2y  \sinh 2\tilde y ,  \nonumber 
\end{eqnarray}
and
\begin{equation}
S_0^{(\pm\mp)}= -\ln \tilde\chi + S_0^{(\pm\pm)} ,
\end{equation}
where we have defined the kink fugacity,
\begin{equation}
\tilde\chi=\pm \prod_{k=1}^{2N}(1- t_k^2)^{1/2}
\end{equation}
In the limit $g\gg 1$ the angular integration is dominated by the saddle point at $\tilde \alpha=0$, while the radial one by $\tilde y = y$. 
We thus restrict ourselves to the vicinity of these saddle points, where 
\begin{eqnarray*}
F^{(++)}(y;\tilde y,0) &=& \sinh y\sinh \tilde y\cosh(\tilde y-y),\\
F^{(--)}(y;\tilde y,0) &=& -\cosh y\cosh \tilde y\cosh(\tilde y-y),\\
F^{(+-)}(y;\tilde y,\tilde \alpha) &=&   \cosh \tilde y \sinh y \sinh(\tilde y-y)  \\
&-&  {\tilde \alpha^2\over 2} \sinh 2y\sinh 2\tilde y + {\cal O}(\tilde \alpha^3)  ,  \\
F^{(-+)}(y;\tilde y,\tilde \alpha) &=&  - \cosh y \sinh \tilde y \sinh(\tilde y-y) \\
& - & {\tilde \alpha^2\over 2} \sinh 2y\sinh 2\tilde y  + {\cal O}(\tilde \alpha^3), 
\end{eqnarray*}
and in the off-diagonal terms we kept a term $\propto\tilde \alpha^2$. 

We first evaluate $\tilde \alpha$ integral in the Gaussian approximation near $\tilde \alpha=0$:
\begin{eqnarray}
{\cal R}^{(++)}  =\sqrt{\frac{g}{2\pi}}\,\,\sqrt{\frac{\coth \tilde y}{\coth y}} \cosh(\tilde y-y) e^{-g(\tilde y-y)^2/2}.\nonumber 
\end{eqnarray}
Integrating over $\tilde y$ in the saddle point approximation yields unity in agreement with SUSY normalization. 
Similar calculation works for ${\cal R}^{(--)}$.  Going beyond this approximation,
requires expanding pre-exponential factors, including wave factions, to second order in $\delta y=\tilde y-y$. This leads to standard 
Laplace-Beltrami operators $\frac{1}{2}(J^{(\pm)})^{-1}\partial_y(J^{(\pm)} \partial_y)$. 

We turn now to the off-diagonal parts.  The off-diagonal fermionic $F$-factors, being calculated at the saddle point $\tilde \alpha=0$ and $\tilde y=y$ 
yield zero $ F^{(\pm\mp)}(y;y,0)=0$. This is again a manifestation of the SUSY normalization. One has to go thus beyond the saddle point approximation, keeping  the   deviations from the saddle point. 
This way one finds for the off-diagonal components of the transfer operator 
\begin{eqnarray}
\label{eq:+-}
{\cal R}^{(+-)}  (y;\tilde y) 
 &\simeq&
-\tilde\chi \sqrt{\frac{g}{2\pi}}\, e^{-g(\delta y)^2/2}  \Bigl[-{1\over 2g}\frac{1}{\cosh^2y}  \nonumber \\
&+&  \tanh y\,
 \delta y + \frac{(\delta y)^2}{2\cosh^2y} \Bigr] ,\\
 {\cal R}^{(-+)}(y;\tilde y) 
 &\simeq&
-\tilde\chi{\sqrt\frac{g}{2\pi}}\, e^{-g(\delta y)^2/2}  \Bigl[{1\over 2g} \frac{1}{\sinh^2y} \nonumber \\
&+&  \coth y\,
 \delta y - \frac{(\delta y)^2}{2\sinh^2y} \Bigr] ,
 \label{eq:-+}
\end{eqnarray}
where we kept the expansion up to the second order in $\delta y=\tilde y-y$.    
We now consider how these operators act on the radial wave function 
\begin{equation}
\label{eq:Psi_expansion}
\Psi^{(\pm)}(\tilde y) = \Psi^{(\pm)}(y)+\partial_{y} \Psi^{(\pm)}(y)\,\delta y + {\cal O}(\delta y^2).
\end{equation}
In view of the relation 
\begin{equation*}
\int d\tilde y e^{-g(\delta y)^2/2}(\delta y)^2/\int d\tilde y e^{-g(\delta y)^2/2}= g^{-1}, 
\end{equation*}
the zeroth order term $\Psi^{(\pm)}(y)$
does not contribute. This is again a manifestation of SUSY normalization, which manifests itself as a cancellation of subleading terms originating from $\tilde \alpha^2$ and $(\delta y)^2$. The remaining terms come solely from $\partial_{y} \Psi^{(\pm)}(y)\,\delta y$ and term linear in $\delta y$ in Eqs.~(\ref{eq:+-}) and (\ref{eq:-+}). Upon Gaussian integration over $\delta y$, this yields
\begin{eqnarray*}
{\cal R}^{(+-)}\Psi^{(-)} = - \frac{\tilde \chi}{g}\, \tanh y\, \,\partial_y  \Psi^{(-)}= 
- \frac{\tilde\chi}{g}\, \sqrt{\frac{J^{(-)}}{J^{(+)}}}\,\partial_y  \Psi^{(-)} ,\\
 {\cal R}^{(-+)}\Psi^{(+)} = - \frac{\tilde \chi}{g}\, \coth y\, \, \partial_y  \Psi^{(+)}= 
- \frac{\tilde\chi}{g}\, \sqrt{\frac{J^{(+)}}{J^{(-)}}}\,\partial_y  \Psi^{(+)}.
\end{eqnarray*}
 Introducing continious space derivative as $\partial_x \Psi =
\Psi_{s+1}-\Psi_s=[\hat {\cal R}-1] \Psi$, one may write in the matrix form:
\begin{equation}
-\tilde\xi \partial_x \hat\Psi  = \left(\begin{array}{cc}
 {1\over 2J^{(+)}} \partial_y J^{(+)} \partial_y  &    -\tilde \chi \sqrt{\frac{J^{(-)}}{J^{(+)}}} \, \partial_y  \\
 -\tilde \chi \sqrt{\frac{J^{(+)}}{J^{(-)}}}\, \partial_y   & {1\over 2 J^{(-)}} \partial_y J^{(-)} \partial_y
 \end{array}\right)\!
 \left(\begin{array}{c}
 \Psi^{(+)}\\  \Psi^{(-)}
 \end{array}\right).
 \end{equation}
 Notice that $\hat\Psi = {\rm const}$ manifestly nullifies the right hand side in accord with the SUSY normalization.

After the Sutherland substitution: $\Phi^{(\pm)}=\sqrt{J^{(\pm)}}\Psi^{(\pm)}$ one finds: 
\begin{equation}
                                      \label{eq:sutherland-D}
\tilde\xi\partial_x \hat\Phi = \left(\begin{array}{cc}
-\frac12 \partial_y^2  + V^{(+)}(y)  &    \tilde \chi\left[\partial_y - A^{(-)}(y)\right] \\
\tilde \chi \left[\partial_y - A^{(+)}(y) \right]        &  -  \frac12\partial_y^2 + V^{(-)}(y)
 \end{array}\right)
\hat\Phi,
 \end{equation}
where $A^{(\pm)}(y)=\partial_y \sqrt{J^{(\pm)}} /\sqrt{J^{(\pm)}} = \mp 2\lambda /\sinh 2y $ and 
\begin{equation}
\label{eq:PT}
V^{(\pm)}(y)=\frac{\partial^2_y \sqrt{ J^{(\pm)} } }{2\sqrt{ J^{(\pm)}} }=  - \frac{\lambda(\lambda\mp 1)}{2\cosh^2y}+  \frac{\lambda(\lambda\pm 1)}{2\sinh^2y}
\end{equation}
are modified P\"oschel-Teller potentials\cite{Poschl-Teller} with $\lambda=1/2$. Equation (\ref{eq:sutherland-D}) may be rewritten in the manifestly supersymmetric form (\ref{eq:transfer-matrix-D}), (\ref{eq:Dirac-D}). 

Looking for the ``stationary'' solutions in the form $\hat\Phi(y,x)=\hat\Phi_l(y) e^{-\epsilon(l) x/\tilde\xi}$, one finds for the spectrum of the transfer matrix operator $ \epsilon(l)=l^2/2+i\, \tilde\chi\, l$,                                
where $l$ labels the eigenfunctions according  to their asymptotic behavior  $ \hat\Phi_l  \sim e^{ily}$. 
The supersymmetric form (\ref{eq:transfer-matrix-D}), (\ref{eq:Dirac-D}) of the transfer matrix operator insures that the eigenfunctions do not depend on the fugacity $\tilde\chi$ and are those of the modified P\"oschel-Teller potentials (\ref{eq:PT}). Their explicit form is given by\cite{Poschl-Teller}
\begin{eqnarray*}
\Phi^{(+)}_l&=& \frac{1}{\sqrt{N^{+}_l}}\, _2F_{1}\Bigl(1\!-\!\frac{il}{2},1\!-\!\frac{il}{2},2; 1\!-\!z \Bigr) 
 (1-z)^{3/4}z^{-il/2} \label{eq:Phi_plus} \\
\Phi^{(-)}_l&=& \frac{-i}{\sqrt{N^{-}_l}}\, _2F_{1}\Bigl(1-\frac{il}{2},-\frac{il}{2},1; 1\!-\!z \Bigr) 
(1-z)^{1/4}z^{-il/2},
\end{eqnarray*}
where $z=1/\cosh^2y$ and $N^{\pm}_l$ are $l$-dependent normalization constants. 
To find $N^{+}_l$ one should explore the asymptotic of $\Phi^{(+)}$ at $y \to +\infty$,
which corresponds to $z \to +0$, where 
\begin{equation*}
_2F_{1}\Bigl(1-\frac{il}{2},1-\frac{il}{2},2; 1-z\Bigr) \sim \frac{\Gamma(il)}{\Gamma^2(1+il/2)} +  
\frac{\Gamma(-il)z^{il}}{\Gamma^2(1-il/2)},
\end{equation*}
and since $z \sim 4e^{-2y}$ at $y \to +\infty$ one finds
\begin{equation}
\Phi^{(+)}_l(y) \sim \frac{1}{\sqrt{N^{+}_l}}\left\{
\frac{\Gamma(il)e^{il(y-\ln2)}}{\Gamma^2(1+il/2)} + {\rm c.c.}
\right\}.
\end{equation}
Requiring the asymptotic condition  
\begin{equation}
\Phi^{(\pm)}_l(y) \sim \cos (ly + \delta^{\pm}_l),
\label{eq:Phi_asym_p}
\end{equation}
at infinity with $\delta^{\pm}_l$ being some phase shifts, one finds 
\begin{eqnarray*}
N^{+}_l &=& \left| \frac{2\,\Gamma(il)} {\Gamma^2(1+il/2)} \right|^2 = \frac{8 \tanh \left(\frac{\pi  l}{2}\right)}{\pi  l^3}, \\ 
N^{-}_l &=& \left| \frac{2\,\Gamma (i l)}{\Gamma \left(\frac{i l}{2}+1\right) \Gamma \left(\frac{i l}{2}\right)} \right|^2
= \frac{2\tanh \left(\frac{\pi  l}{2}\right)}{ \pi  l}.
\end{eqnarray*} 
where $N^{-}_l$ is found in the similar way. We note that the phase shifts satisfy  
$\delta^-_l = \delta^+_l + \pi/2$ and hence
\begin{equation*}
\Phi^{(-)}_l(y) \sim -i\cos (ly + \delta^-_l) \sim i\sin (ly + \delta^+_l), \quad y \to +\infty. 
\label{eq:Phi_asym_m}
\end{equation*}
The latter asymptote is consistent with the mutual relation between two solutions $\Phi_l^{+}$ and $\Phi_l^{-}$
of P\"oschel-Teller potential, namely: 
\begin{equation*}
(B\Phi^{(+)}_l) /(-il) \sim \partial_y \Phi^{(+)}_l/(il) \sim i \sin(ly + \delta_l^+) \sim \Phi^{(-)}_l, 
\end{equation*}
as it should be.

\subsection{Observables}
\label{app:D_3}
We start by evaluating the  measure,
\begin{equation}
\mu(l) = -\langle (1,0)|\Psi_l\rangle = 
\int_0^{+\infty}\!\!dy\,\, \sqrt{J^{(+)}(y)}\, \Phi_l^{(+)}(y)
\end{equation} 
Using the explicit form of $\Phi^{(+)}(y)$ and changing the variable of integration to $u=\tanh^2y$, one obtains
\begin{eqnarray*}
\mu(l)\! &=&\! \frac{-1}{\sqrt{2 N^+_l}} \int_0^1\!\! du\,  (1-u)^{-1-il/2} {_2F_1}\!\Bigl(\!1\!-\!\frac{il}{2},1\!-\!\frac{il}{2},2 ; u\!\Bigr)\\ 
&=&
 \frac{-1}{\sqrt{2 N^+_l}}\times\frac{4}{l^2} = -\sqrt{\frac{\pi}{l\tanh\frac{\pi l}{2}}}.
\end{eqnarray*}
The wave function with the proper initial, i.e. $x=0$, conditions is  given by
\begin{eqnarray*}
\Psi(y,x)=   \left(\!
\begin{array}{c}
1\\ 0 \end{array}\!
\right) +
\int\!\! \frac{dl}{2\pi} \mu(l) \left(\!
\begin{array}{c}
\Phi^{(+)}_l(y)/\sqrt{J^{(+)}} \\ \Phi^{(-)}_l(y)/\sqrt{J^{(-)}}\end{array}\!
\right)\, e^{-\epsilon(l) x/\tilde\xi}
\label{eq:ZD_phi_x}
\end{eqnarray*} 
Let us first focus on the $(-)$ component and evaluate the renormalized topological number $\chi(L) = \Psi^{(-)}(0,L)$.
Evaluating the limiting value
\begin{equation*}
\lim_{y\to 0}\Phi_l^{(-)}(y)/\sqrt{J^{(-)}(y)} = -{i}/{\sqrt{2N_l^-}},
\end{equation*}
we obtain
\begin{equation}
\chi(L) = \int\limits_0^{+\infty}\!\! dl\,\coth\left(\frac{\pi l}{2}\right) \sin\left(\frac{\tilde\chi lL}{\tilde\xi}\right)e^{- l^2L/(2\tilde\xi)} .
\end{equation}
In the limit of long system, $L\gg \xi = 2\tilde\xi\, \tilde\chi^{-2}$, one finds 
\begin{equation}
\chi(L) \simeq \mathrm{sign} \,\tilde\chi\left(1- \sqrt\frac{\xi}{\pi L}\, e^{-{L/\xi}}\right). 
\end{equation}
Therefore 
the topological number approaches exponentially the quantized limiting value $\pm 1$, which indicates topologically trivial/non-trivial  phase.

To evaluate the renormalized conductance we use  
$g(L) = \partial_y^2\Psi^{(+)}(y,L)\bigl|_{y=0}$ to obtain 
\begin{eqnarray}
g(L)  &=& \frac{1}{4}
\int\limits_0^{+\infty}  dl\,l\coth\left(\frac{\pi l}{2}\right) \cos\left(\frac{\tilde\chi lL}{\tilde\xi}\right) 
e^{- l^2L/ (2\tilde\xi) }\nonumber \\
 &\propto& \frac{1}{\sqrt{L}}\, e^{-L/\xi},
\end{eqnarray}
in the limit $L\gg \xi$. Therefore, unless $\tilde \chi=0$, the conductance is exponentially small, indicating the Anderson insulator nature of the class $\mathrm{D}$ quasi-1d system.
 
\vspace{1cm}

\section{Field theory of class $\mathrm{DIII}$}

\subsection{Jacobians}

We start by evaluating Jacobians on two disconnected parts of the group manifold. 
Following Eq.~(\ref{eq:Q_par_DIII}) we parametrize the group element as 
$T^{(\pm)} =e^{{\cal W}}{\tilde T}^{(\pm)}e^{-{\cal W}}$ and further find the metric on the group 
in the chosen coordinates. 
As in class $\mathrm{D}$ we choose to stay near zero Grassmans, 
since in what follows we will need only the Grassmann independent part of Jacobians. Using
$dT^{(\pm)} =[d{\cal W}, {\tilde T}^{(\pm)}] + d{\tilde T}^{(\pm)} $ and $ dT^{(\pm)-1} =[d{\cal W}, {\tilde T}^{(\pm)-1}] + d{\tilde T}^{(\pm)-1} $, the metric is given by
\begin{eqnarray}
dg&=&-{1\over 2}{\rm str}\left( dT dT^{-1} \right) \\
&=& - {1\over 2}{\rm str}\left( [d{\cal W}, {\tilde T}]  [d{\cal W}, {\tilde T}^{-1}]  \right)-
{1\over 2}{\rm str}\left( d{\tilde T} d{\tilde T}^{-1} \right) \nonumber
\end{eqnarray} 
For ``plus'' manifold one finds
\begin{eqnarray}
dg^{(+)}&=&dy_0^2+dy_1^2+\sinh ^2y_0  \left(d\alpha^2 \sin ^22 \rho +4 d\rho^2\right)\nonumber \\ 
&-&4 (1-\cos y_1  \cosh y_0 ) (\eta \xi +\mu \nu ) \nonumber\\
&-&4 \sin y_1  \sinh y_0  \Bigl(\sin 2 \rho  
(e^{i\alpha}\eta \mu + e^{-i\alpha}\nu \xi ) \nonumber\\
&+&i\cos 2 \rho (\eta  \xi -\mu  \nu )\Bigr)
\end{eqnarray}  
which results in 
\begin{equation}
J^{(+)}=\sqrt{\mathrm{Sdet} g^{(+)}}={1\over 2}\, \frac{\sin 2 \rho \sinh ^2y_0  }{(\cosh y_0-\cos y_1)^2}
\end{equation}
Here one should regard $\rho$ as an angular variable, while $y_0$ and $y_1$ are radii. The reason the Jacobian depends on the angle $\rho$ is the chosen parameterization, where the commuting angles $b,b^*$ 
are treated separately from the  Grassmann ones. Here $\sin 2\rho d\rho\approx 2\rho d\rho\sim d b^* db$ 
can be understood as a part of the plane angular measure.

In a similar way in the case of ``minus'' part of the manifold one finds 
\begin{eqnarray}
dg^{(-)}&=&dy_0^2+dy_1^2+\sinh ^2y_0  \left(d\alpha^2 \sin ^22 \rho +4 d\rho^2\right)\nonumber\\ &-&  
4 \left(\eta\xi +\mu \nu -  \cosh y_0  \left(e^{-i y_1 }\mu \xi +e^{ i y_1 } \eta  \nu \right) \right)
\end{eqnarray}  
and the Jacobian reads
\begin{equation}
J^{(-)}=\sqrt{\mathrm{Sdet} g^{(-)}}={1\over 2}\, \sin 2 \rho 
\end{equation}

\subsection{Transfer matrix}

Our goal here is to evaluate a transfer matrix element between two neighboring dots $e^{-S(T^{(\alpha)}, \tilde T^{(\beta)})}$, 
where  $\alpha,\beta=\pm$ refer to the two disconnected parts of the manifold, and
\begin{equation}
                                        \label{eq:actionST_DIII}
 S(T,\tilde T) = \frac{1}{4}\sum_{k=1}^{2N} {\rm str}\ln\Bigl[1+\frac{t_k^2}{4}(T^{-1}\tilde T + \tilde T^{-1}T-2)\Bigr],
\end{equation}
which is a discrete version of the continuous action ${\rm str}\left(\partial_x T^{-1}\partial_x T\right)$. 
Similar to class $\mathrm{D}$, this operator acts on the wave function $\Psi^{(\beta)}(\tilde T)$, which in turn
is assumed to be angle-independent. We thus define the radial transfer operator
\begin{equation}
{\cal R}^{(\alpha\beta)}(y;\tilde y) = \int d \tilde \Omega^{(\beta)} J^{(\beta)}(\tilde y,\tilde \rho) \, 
e^{-S(T^{(\alpha)},\tilde T^{(\beta)} )},
\end{equation}
where $d\tilde \Omega^{(\beta)} = {\beta}(2\pi)^{-2} d\tilde \alpha d\tilde \rho d\tilde \nu d\tilde \xi d\tilde \mu d\tilde \nu$ is the angular measure.
Because of the invariance of the action under $T$-rotations
the group element $T^{(\alpha)}=T^{(\alpha)}(y)$ may be chosen to be   
pure radial and independent on angles $\alpha, \rho$ and Grassmanns. 
We now evaluate the action~(\ref{eq:actionST_DIII}) on such $T^{(\alpha)}$ and $\tilde T^{(\beta)}$. 
The result has the structure 
\begin{eqnarray}
S=g\Bigl[S_0^{(\alpha\beta)}(y; \tilde y,\tilde \Omega) 
+ F_2^{(\alpha\beta)}(y; \tilde y,\tilde \Omega) +
F_4^{(\alpha\beta)}(y; \tilde y,\tilde \Omega)  \Bigr] \nonumber
\end{eqnarray}
where $S_0$ contains only commuting variables, while $F_2$ is bilinear and $F_4$ is quartic in Grassmanns
and $g = \sum_k t_k^2$ is the dot-to-dot conductance.
We then expand  
\begin{equation}
e^{-g(F_2 + F_4)} = 1  - g (F_2 + F_4)+ \frac{g^2}{2} F_2^2
\end{equation}
up to the non-vanishing second order.
Because of our assumption $g\gg 1$, only the last term of this expansion should be retained. We denote $\frac{1}{2}F_2^2 = F \times \tilde\nu \tilde\xi \tilde\mu \tilde\nu$,  
which happens to be independent of angle $\tilde \alpha$.
Integrating further over Grassmanns one arrives at
\begin{align}
{\cal R}^{(\alpha\beta)}(y;\tilde y) = \frac{g^2}{2\pi}
\int d\tilde \rho  (\beta)  J^{(\beta)}(\tilde y,\tilde\rho)\\
\times\, F^{(\alpha\beta)}(y;\tilde y,\tilde\rho)\, e^{- g S_0^{(\alpha\beta)} (y;\tilde y,\tilde \rho) }. \nonumber
\end{align}
Direct evaluation leads to the following result for the diagonal actions,
\begin{eqnarray}
S_0^{(\pm,\pm)}&=& \frac{1}{4}\cosh y_0  \cosh \tilde y_0  \\
&-& \frac{1}{4}\cos 2 \tilde\rho  \sinh y_0  \sinh \tilde y_0  -  \frac{1}{4} \cos (y_1 -\tilde y_1) \nonumber \\
&\simeq&  \frac{(y_0-\tilde y_0)^2 + (y_1 -\tilde y_1)^2}{8} +  \frac{\tilde \rho^2}{2}\,\sinh y_0 \sinh \tilde y_0,   \nonumber
\end{eqnarray}
and the kink's actions
\begin{eqnarray}
S_0^{(\pm, \mp)}&=& -\ln \tilde\chi \\
&+ &\frac{1}{4}\cosh y_0  \cosh \tilde y_0-\frac{1}{4}\cos 2 \tilde\rho  \sinh y_0  \sinh \tilde y_0 \nonumber \\
&\simeq& 
-\ln \tilde\chi + \frac{(y_0-\tilde y_0)^2}{8} +  \frac{\tilde \rho^2}{2}\,\sinh y_0 \sinh \tilde y_0. \nonumber
\end{eqnarray}
Notice that, there is always a strong Gaussian confinement for angle $\tilde\rho\approx 0$ and non-compact radius $\tilde y_0\approx y_0$. On the other hand the compact radius is confined for diagonal elements $\tilde y_1\approx y_1$, but is {\em not confined at all} for the off-diagonals. In fact the action in this case
is altogether independent of compact radii. Due to such confinement one may evaluate the diagonal fermionic parts  $F^{(\pm\pm)}(y;\tilde y,\tilde \rho)$ at 
$\tilde y_0=y_0$, $\tilde y_1=y_1$ and $\tilde \rho=0$. This yields:
\begin{eqnarray}
F^{(++)}(y; y,0) &=& \frac{1}{4}(\cosh y_0-\cos y_1)^2 \nonumber\\ 
F^{(--)}(y; y,0)&=& - \frac{1}{4}\sinh^2y_0.
\end{eqnarray}
Notice that $J^{(+)}F^{(++)} = -J^{(-)}F^{(--)}$.  The relative minus sign suggests that 
the Grassmann measure on the 'minus' manifold comes with the relative minus sign with respect to 'plus' part.  
Calculating then $\tilde \rho$, and $\tilde y$  integrals in the saddle point approximation, one finds that 
both diagonal operators are identities, 
${\cal R}^{(++)}(y;\tilde y)={\cal R}^{(--)}(y;\tilde y)=1$. 
This is a manifestation of SUSY normalization. 
Going beyond this approximation, requires expanding pre-exponential factors, including wave functions, 
to second order in deviations. This leads to standard Laplace-Beltrami operators 
$(J^{(\pm)})^{-1}\partial_\nu (J^{(\pm)} \partial_\nu)$, where $\nu =(y_0,y_1)$, on the two sub-manifolds.  

We turn now to the off-diagonal parts. 
 The off-diagonal fermionic factors, being calculated at the saddle point $\tilde y_0=y_0$ and $\tilde \rho=0$ (and arbitrary $y_1$, $\tilde y_1$) yield zero $ F^{(\pm\mp)}(y;\tilde y,0)=0$. This is again a manifestation of the SUSY normalization. One has to go thus beyond the saddle point approximation, expanding both 
 $F^{(\pm\mp)}$-factor and the wave function to the first order in deviations from the saddle point (expanding $F^{(\pm\mp)}$-factor to the second order does not help, in view of SUSY normalization, while expanding the wave-function  to the second order does not help in view of $F^{(\pm,\mp)}(y;\tilde y,0)=0$). 
The only non-zero first order deviation in $ F^{(\pm,\mp)}$ is in $\tilde y_0-y_0$ direction, which is found from 
\begin{eqnarray}
F^{(-+)}(y;\tilde y,0) &=& -\frac{1}{4}(\cosh \tilde y_0-\cos \tilde y_1)\nonumber\\
&\times &\sinh y_0 \sinh(\tilde y_0-y_0), \nonumber \\ 
F^{(+-)}(y;\tilde y,0) &=& \frac{1}{4}(\cosh y_0 -\cos y_1) \nonumber \\
&\times&\sinh \tilde y_0 \sinh(\tilde y_0-y_0). \qquad\,
\end{eqnarray}
Keeping only the linear variation $\delta y_0= \tilde y_0-y_0$ 
one finds from here the off-diagonal components of the transfer operator,
\begin{eqnarray}
{\cal R}^{(+-)}(y;\tilde y) &=&  -\frac{\tilde\chi g}{4\sqrt{
{\cal J}(y)}}\, 
 e^{-g(\delta y_0)^2/8} \, \delta y_0, \\
 {\cal R}^{(-+)}(y;\tilde y) &=&  - \frac{\tilde\chi g}{4}
 \sqrt{
{\cal J}(\tilde y)}\,  e^{-g(\delta y_0)^2/8}\,  \delta y_0, \nonumber
\end{eqnarray}
where we have introduced
\begin{equation}
{\cal J}(y)=\frac{\sinh^2y_0}{(\cosh y_0-\cos y_1)^2}.
\end{equation}

We now consider how these operators act on the radial wave function $\Psi^{(\beta)}(\tilde y)$. Expanding  
to the first order in $\delta y_0$ and integrating over $\tilde y_0$ (Gaussian integration) and over $\tilde y_1$,  
one finds
\begin{eqnarray}
{\cal R}^{(+-)}\Psi^{(-)} &=& -\frac{\tilde\chi (2/\pi g)^{1/2}}{\sqrt{
{\cal J}(y)}}\, \partial_{y_0} \!\!\int\!\! d\tilde y_1   \Psi^{(-)}(\tilde y), \\
 {\cal R}^{(-+)}\Psi^{(+)} &=& - \tilde\chi (2/\pi g)^{1/2} \, \partial_{y_0} \!\!\int\!\! d\tilde y_1   \sqrt{
{\cal J}(\tilde y)}  \Psi^{(+)}(\tilde y) . \nonumber
\end{eqnarray}
Notice that if $ \Psi^{(+)}={\rm const}$ then $\int d\tilde y_1  \sqrt{
{\cal J}(\tilde y)} =\pi$ 
is $\tilde y_0$-independent and thus ${\cal R}^{(-+)}{\rm const} ={\cal R}^{(+-)}{\rm const} =0$, 
in agreement with SUSY normalization. Introducing finally the space derivative, 
$\partial_x \Psi = \Psi_{s+1}- \Psi_s = [{\cal R}-1]\hat\Psi$,  one deduces the transfer equation in the
matrix form,
\begin{eqnarray}
\tilde \xi \partial_x \hat\Psi\! =\! 
\left(\!\begin{array}{cc} 
-{\cal J}^{-1}\partial_\nu ({\cal J}\partial_\nu)  &  \frac{v }{\sqrt{\cal J}(y)}\, \partial_{y_0} \int \frac{d\tilde y_1}{2\pi} \\
  v \partial_{y_0} \int \frac{d\tilde y_1}{2\pi} \sqrt{{\cal J}(\tilde y)} &  -\partial_\nu\partial_\nu  
 \end{array}\!\right)\!
 \left(\!\begin{array}{c} \Psi^{(+)}\\ \Psi^{(-)}
 \end{array}\!\right), \nonumber
\end{eqnarray}
where $\tilde\xi = g/2$, $\nu =y_0,y_1$, and $v=\tilde\chi\sqrt{2\pi g}$. The off-diagonal operator is differential in non-compact radius $y_0$ and integral (minus first derivative) in compact radius $y_1$. After performing the Sutherland transformation,
$\Phi^{(+)}=\sqrt{\cal J}\Psi^{(+)}$ and $ \Phi^{(-)}=\Psi^{(-)}$, this transfer equation is reduced to Eq.~(\ref{eq:TM_DIII})
discussed in Sec.~\ref{sec:DIII}.

\subsection{Observables}
\label{app:DIII_3}

The measure is given by 
as 
\begin{equation}
\mu(l)=-\langle \Psi_{l}|(1,0)\rangle.
\end{equation}
Here $\Psi_{l}$ is the eigenfunction of the transposed transfer matrix operator,
the former playing a role of the 'bra' vector orthogonal to the 'ket' state $\hat\Psi_{l}$, as it is not
difficult to verify. One finds
\begin{equation}
\mu(l)=\frac{2il_0}{l_0^2+l_1^2}; \quad \mu(l_0, 0)=-{\sqrt{2}i}/{l_0}.
\end{equation}
The $l_1\neq 0$ components are not affected by kinks and leads to exponentially decaying with $x$ terms. We thus focus exclusively on $l_1=0$ term. With the help of spectral decomposition~(\ref{eq:ZL_DIII}) one finds
\begin{eqnarray}
&&\Psi_0(y,L) 
=(1,0)^T  \\
&& +\, i \int \frac{dl_0}{\pi l_0} \,\,  e^{il_0y_0}
\left[   \left(\begin{array}{c} {\cal J}^{-1/2}(y) \cos(  v l_0 L/\tilde\xi) \\-i\sin(v l _0L/\tilde\xi) \end{array}\right)  \right]\, e^{- l_0^2 L/\tilde\xi}.\nonumber
\end{eqnarray}
The  fugacity at scale $L$ is given by $\chi(L)=\Psi^{(-)}(0,L)$,
\begin{equation} 
\chi(L)= \int \frac{dl_0}{\pi l_0} 
\sin\left(\frac{\displaystyle\sqrt{2\pi g}\,\tilde\chi l_0 L}{\tilde\xi} \right) \, e^{-l_0^2 L/\tilde\xi} = \mathrm{erf} \left(\tilde\chi\sqrt{\pi L}\right).
\end{equation}
We keep now the (+) part of the wave function and proceed with evaluation of the generating function,
\begin{eqnarray}
{\cal F}(\phi_0,L)&=&\left.\partial_{\phi_1} \Psi^{(+)}(2\phi,L)\right|_{\phi_1=-i\phi_0} \\
&=&  2 \int \frac{dl_0}{\pi}  \frac{1}{l_0} \,\,  e^{2 il_0 \phi_0}\cos( vl_0L/\tilde\xi)\,e^{-l_0^2 L /\tilde\xi}. \nonumber
\end{eqnarray}
Based on the relation~(\ref{eq:rho_lambda}), it gives the average density of the Lyaponov exponents, 
$\rho(\phi_0,L)$, in accordance with Eq.~(\ref{eq:rho_DIII}). To evaluate the conductance one can
use two complementary relations,
\begin{equation}
 g(L) = \int d\phi_0\, \frac{\rho(\phi_0,L)}{\cosh^2\phi_0} = -i\partial_{\phi_0} {\cal F}(\phi_0,L)\Bigl|_{\phi_0=0}.
\end{equation}  
Both representations lead to the result~(\ref{eq:g_DIII}),
\begin{equation}
g(L) = 4 \int \frac{dl_0}{\pi}  \cos( vl_0L/\tilde\xi)\,e^{-l_0^2L/\tilde\xi}=4\,\sqrt{\frac{g}{\pi L}}\, e^{-\pi\, \tilde\chi^2 L }.
\end{equation}

\section{Scattering theory}

\label{app:Scattering}

In this appendix, using the analog of quasiclassical Eilenberger method of superconductivity,
we show how the partition function $Z(\phi)$ is related to the set $\{\lambda_n\}$ of Lyapunov exponents.

\subsection{Chiral classes AIII and BDI}

To define the scattering matrix of the chain we choose to connect it to two leads, which are described 
by the gapless Hamiltonian of the same symmetry class as the random Hamiltonian of the disordered chain.
For the class $\mathrm{BDI}$ 
chain described by the random Kitaev's model it can be achieved by setting $\Delta=0$ in the leads.
Essentially the same model of the leads can be also used in case of class $\mathrm{AIII}$ if one 
identifies the p/h grading of the spinors with the $+/-$ grading due to the bipartite unit cell of the 
$\mathrm{AIII}$ chain. 
In the lattice representation such model reads
\begin{eqnarray}
{\cal H}_{\rm lead} &=& {\mu}\sum_l( \bar\psi_{+,l} \psi_{-,l} + \bar\psi_{-,l} \psi_{l,+} ) \\
&-&\frac{t}{2}\sum_l( \bar\psi_{+,l+1} \psi_{-,l} + \bar\psi_{+,l-1} \psi_{-,l}  + {\rm h.c.}). \nonumber
\end{eqnarray}
Assuming $\mu<t$ 
we accept the long wave approximations in the leads and approximate
\begin{equation}
{\cal H}_{\rm lead}^{L,R} \simeq \int_{\mp \infty}^{0,L}\, dx\, \bar\psi_+ \left( \mu-t - \partial_x^2/{2m} \right)\psi_- + {\rm h.c.}, 
\end{equation} 
here $b$ is a lattice constant and a mass $m = {1}/{(t b^2)}$. 
Introducing the spinor structure $\psi=(\psi_+, \psi_-)^T$ in the sublattice space ($+/-$), the parity operator becomes $\hat P=\sigma_3^{\pm}$.
The leads' Hamiltonian is exactly the same as in class BDI with the only difference that Pauli matrices operate
in Majorana basis. The subsequent discussion will be more transparent if we rotate the basis in AB subspace so that
Pauli matrices are permuted according to cyclic rule, $\sigma_1^{\pm} \to \sigma_3^{\pm}$ and etc. 
In this new basis $\hat P = \sigma_1^{\pm}$. 
One further linearizes the leads' Hamiltonian around Fermi-momentum 
$k_F =  \bigl( 2m(\mu-t)\bigr)^{1/2}$ by representing 
$\psi_\pm(x) ~ \sim \psi_{\pm}^R e^{\pm ik_F x} + \psi^L_{\pm} e^{\mp ik_F x}$
and doubles the number of spinor's components, 
$\psi=(\psi_+^R, \psi_+^L, \psi_-^R, \psi_-^L)^T$, in order to accommodate right and left modes. 
In this way we obtain
\begin{equation}
{\cal H}_{\rm lead}^{L,R} \simeq -i v \int_{\mp \infty}^{0,L}\, dx\, \bar\psi\,  \sigma_3^{RL} 
\otimes \mathds{1}^{\pm}\,\partial_x \psi,
\end{equation}
(with velocity $v=k_F/m$) and at the same time the parity operator transforms into ${\cal  P} = \hat P\otimes\sigma_1^{RL}$.
Let us now subject ${\cal H}$ to the gauge 
transform, ${\cal H} \to e^{i\xi {\cal P}} {\cal H} e^{ i\xi {\cal P}}$.
Introducing the gauge field $A_x = \partial_x \xi$ associated with the
the parity current, the transformed Hamiltonian of the leads takes the form
\begin{equation}
{\cal H}_{\rm lead}^{L,R}[\xi] = v \int_{\mp \infty}^{0,L}\, dx\, \bar\psi\,  \sigma_3^{RL} 
\,\bigl( -i \partial_x \otimes \mathds{1}^{\pm} + {\cal P} A_x \bigr)\,\psi.
\end{equation}
From here the (second quantized) parity current operator $\hat{\cal I}_P$ can be found in accordance with the standard definition, 
$\hat{\cal I}_P=\delta {\cal H}/\delta A_x$, which yields to
\begin{equation}
\hat{\cal I}_P = v \bar \psi \sigma_3^{RL} {\cal P} \psi.
\end{equation}

We now aim to find the general form of the expectation value of the parity current $\hat{\cal I}_P$ 
in terms of the transfer matrix $M$. 
Our method is adapted from the Nazarov's 'circuit theory'~\cite{Nazarov-book}  (see also Ref.~[\onlinecite{Neven13}]). 
Introducing  the Green's function in the  leads, 
$G_{nm}(x,x') = -i \langle \psi_n(x) \bar\psi_m(x') \rangle$, where $m,n$ are channel indices, 
the current becomes 
\begin{equation}
\label{eq:jP_Def_G}
{\cal I}_P(\theta) = - iv \lim_{x\to x'}  {\rm tr} \left( \sigma_3^{RL} {\cal P} G(x,x') \right)
\end{equation}
In what follows we use the quasiclassical approach and introduce the Green's function $Q(x)$ at the coinciding spatial points,
\begin{eqnarray}
\label{eq:Qx_def}
Q_{nm}(x) &=& \lim_{x\to x'} \Bigl\{ 2i (v_n v_m)^{1/2} G_{nm}(x,x') \sigma_3^{RL} \nonumber \\
&-&  {\rm  sign} (x-x') \delta_{nm}\Bigr\},
\end{eqnarray}
where we have taken into account that each  channel can be characterized by its own velocity.  The $Q$-function is normalized, $Q(x)^2=\mathds{1}$,
and satisfies the Eilenberger equation,
\begin{equation}
\label{eq:Eil_eq}
 i \partial_x Q (x) + \left[ i0^+\,\sigma_3^{RL}+ {\cal P} A_x, Q(x) \right] = 0, 
\end{equation}
with the boundary conditions $Q(x)\Bigl|_{x\to \pm \infty} = \sigma_3^{RL} \otimes \mathds{1}^{\pm}$.

\begin{figure}[t]
\includegraphics[width=7.5cm]{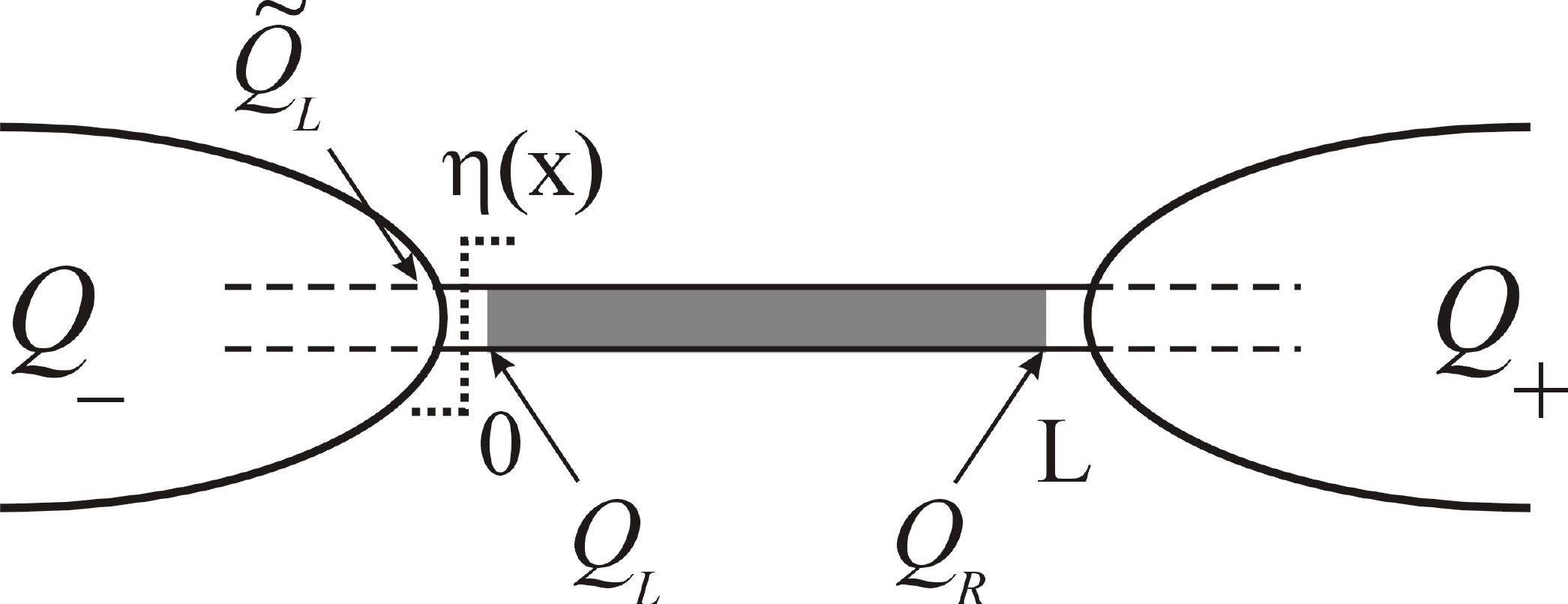}
\caption{Disordered wire of symmetry class AIII connected to two normal terminals 
which are described by the Eilenberger functions $Q_\pm$. 
The $Q$-matrix at the boundaries between the wire and terminals is denoted by $Q_L$ and $Q_R$.
The counter $\eta(x)$ which 'measures' the parity current $j^P$ is located close to the left end
of the wire.}
\label{fig:Wire_AIII}
\end{figure}

The role of infinitesimal term which breaks the chiral symmetry is to provide the boundary conditions at infinities. 
The parity gauge field $A_x$ can be now eliminated at the expense of twisted boundary conditions in the left lead. Using the freedom in the choice of $A_x$ we set $A_x = \theta\times\eta'(x)$, where
$\eta(x)$ denotes any smooth step function on atomic scale at the crossection where the parity  current is 'measured'
as shown in Fig.~\ref{fig:Wire_AIII}. Abbreviating by $\tilde Q_L = Q(-0)$ the Green's function before the 'counter' 
and by $Q_L = Q(+0)$ the one at the left end of the wire after the 'counter', we see that
the Eilenberger equation~(\ref{eq:Eil_eq}) yields
\begin{equation}
\label{eq:QL}
Q_L = e^{  i \theta{\cal P}}\, \tilde Q_L \,e^{ -i\theta {\cal P}}.
\end{equation}
Let us also denote by $Q_R=Q(L)$ the Green's function at the right end of the wire. It is related with the 
left configuration $Q_L$ by the transfer matrix,
\begin{equation}
\label{eq:MQM}
Q_R = M Q_L M^{-1}.
\end{equation}
This relation is central to the whole discussion. It obviates the need of solving the complicated
Shr\"odinger equation with a disorder potential in the wire substituting the latter by the `black box' 
characterized by the transfer matrix. 
As shown in Refs.~[\onlinecite{Nazarov-book,Neven13}], the configurations $Q_{R,L}$ satisfy the following
boundary conditions
\begin{eqnarray}
(\mathds{1} + Q_-)(\mathds{1} - \tilde Q_L) &=& 0,  \\
(\mathds{1} - Q_R)(\mathds{1} + Q_+) &=& 0, \qquad Q_\pm = \sigma_3^{RL} \otimes \mathds{1}^{\pm},\nonumber
\end{eqnarray}
where $Q_\pm$ are asymptotic configurations.  If one further expresses $\tilde Q_L$ in terms of $Q_L$, see Eq.~(\ref{eq:QL}),
one arrives at the 'twisted' boundary conditions,
\begin{eqnarray}
\label{eq:BC_tw}
\left(\mathds{1} + Q_-(\theta)\right)(\mathds{1} - Q_L) &=& 0,  \\
(\mathds{1} - Q_R)\left(\mathds{1} + Q_+\right) &=& 0. \nonumber
\end{eqnarray}
We have introduced here the rotated asymptotic configuration at the left
\begin{eqnarray}
\label{eq:Q_minus_Z}
\qquad Q_-(\theta) &=& e^{  i \theta {\cal P}}\,\left (\sigma_3^{RL} \otimes \mathds{1}^{\pm}\right) \,
e^{ - i \theta {\cal P}} \nonumber \\
 &=& 
\left(\begin{array}{cc}
\cos 2\theta &  -i\hat P\sin 2\theta \\
 i\hat P\sin 2\theta  & 
 -\cos 2\theta 
\end{array}\right)_{RL}
\end{eqnarray} 
and used that ${\cal P} = \hat P\otimes\sigma_1^{RL}$. 
To proceed one can use the relation~(\ref{eq:MQM}) and the new boundary conditions~(\ref{eq:BC_tw}) in order
to find the general expression for $Q_{R,L}$ in terms of the transfer matrix $M$ and asymptotic configurations $Q_\pm(\theta)$, 
\begin{eqnarray}
Q_R &=& \mathds{1} + \frac{2}{Q_+ + M Q_-(\theta) M^{-1}}(\mathds{1}-Q_+), \nonumber \\
\label{eq:QL_M}
Q_L &=& \mathds{1} + \left(\mathds{1} - Q_-(\theta)\right)\frac{2}{Q_-(\theta) + M^{-1} Q_+ M}.\quad
\end{eqnarray}
The derivation of this result can be found in Ref.~[\onlinecite{Neven13}] and we do not repeat it here.
The parity current when expressed in terms of $Q$-matrix becomes
\begin{equation}
\label{eq:jP_Q}
{\cal I}_P(\theta) =  
-\frac{1}{2}{\rm tr} \bigl( \hat P \sigma_1^{RL} Q_L \bigr)=
-\frac{1}{2}{\rm tr} \bigl( \hat P \sigma_1^{RL} Q_R \bigr)
\end{equation}
Since this current is conserved the result should not depend in which of the two terminals it is evaluated.

Our next goal is to relate ${\cal I}_P(\theta)$ with the Lyapunov exponents $\lambda_n$. We observe that $Q_-(\theta)$
can be factorized as $Q_-(\theta) = {\cal R} \tilde Q_-(\theta) {\cal R}$ where
$\tilde Q_-(\theta) = e^{2 i\theta \sigma_1^{\rm RL}}\,\sigma_3^{\rm RL} $ and 
${\cal R} = {\rm diag}(1, \hat P)_{\rm RL}$.
A similar decomposition holds for the transfer matrix if one takes into account Eq.~(\ref{eq:M_DMPK}),
$M={\cal R} \tilde M {\cal R}$ with $\tilde M = V e^{\lambda \sigma_1^{RL}} U'$. 
We finally note that the parity operator in Eq.~(\ref{eq:jP_Q}) can be represented as  
${\cal P} = P \otimes \sigma_1^{\rm RL} = {\cal R}  \sigma_1^{\rm RL} {\cal R}$. 
Hence such factorization enables us to write
${\cal I}_P(\theta) =  
-\frac{1}{2}{\rm tr} \bigl( \sigma_1^{RL} \tilde Q_L \bigr),
$ 
and the parity operator $\hat P$ acting in the sublattice space ($\pm$) can be dropped out from the subsequent manipulations. 
Noting that $\tilde Q_-(\theta)$ can be found via $\tilde M$ by means of the relation~(\ref{eq:QL_M}),
with $M$ being substituted for $\tilde M$, and using the unitarity of matrices $V, U'$, we find
\begin{equation}
\label{eq:jP_theta}
{\cal I}_P(\theta) =  \sum_{k=1}^{N'} \tanh\left(\lambda_k+i\theta\right).
\end{equation}
Comparing now two definitions~(\ref{eq:F_phi_1}) and (\ref{eq:jP_Def_G}), we can read off from here 
the generating function ${\cal F}(\phi_0) = i {\cal I}_P(-i\phi_0)$ and arrive to the 
result~(\ref{eq:F_phi_Z}). 

\subsection{Class D}

\label{app:Scattering_D}

We outline here the proof of relations~\eqref{eq:Part_sum_Z2} and \eqref{eq:chi_L_Z2} for the class $\mathrm{D}$ 
disordered wire. 

The way to derive the 'plus' partition sum $Z^{(+)}(\phi)$, Eq.~\eqref{eq:Part_sum_Z2},
is completely analogous to the one for chiral classes $\mathrm{AIII}$ and $\mathrm{BDI}$ considered above 
and we only sketch the main steps.  The generator of the conserved symmetry current is given by $\tau_1^{\rm cc}$.
Hence the Eilenberger equation corresponding to the gauge transformed Hamiltonian reads
\begin{equation}
\label{eq:Eil_eq_classD}
 i \partial_x Q (x) +\left[ i0^+\,\sigma_3^{RL} \otimes \tau_3^{cc} +  \tau_1^{\rm cc} A_x, Q(x) \right] =0, 
\end{equation}
where we set $A_x = \theta \times \delta(x)$.
At variance with $\mathds{Z}$-classes
the $Q$-matrix here is defined in the direct product of the chiral- and the cc-spaces. The
twisted asymptotic left configuration here assumes the form 
\begin{equation}
\label{eq:Qm_theta_D}
Q_-(\theta) = \sigma_3^{RL} \otimes \left( e^{i\tau_1^{\rm cc} \theta} \tau_3^{cc} e^{-i\tau_1^{\rm cc} \theta}
\right),
\end{equation} 
cf. with Eq.~\eqref{eq:Q_minus_Z}, while the right one reads $Q_+ =\sigma_3^{RL} \otimes \tau_3^{cc}$
(the appearance of the matrix $\tau_3^{cc}$ here is rooted in the construction of the field integral 
for class $\mathrm{D}$ system, see sec.~\ref{sec:FT_classD}). The generating 
function~\eqref{eq:F_phi_1} at the specific disorder realization can be further found from the relation 
${\cal F}(\phi_0) = (i/4){\rm tr} (\tau_1^{cc} Q_L)\bigl|_{\theta= -i\phi_0}$ 
with the matrix $Q_L$ at the left end of the wire 
given by Eq.~(\ref{eq:QL_M}).  When evaluating this expression one can use the fact that configurations
$Q_-(\theta)$ and $Q_+$ have the trivial diagonal structure in the channel space. This means that the 
transfer matrix in Eq.~(\ref{eq:QL_M}) can be taken in the block-diagonal form by omitting $U$ and $V$
orthogonal rotations which mix the eigen-channels. 
In this way one arrives at the final result for the generating function depending only on the set of $\lambda$'s, 
\begin{equation}
\label{eq:F_phi0_D}
{\cal F}(\phi_0) = i\sum_{k=1}^{N'} \frac{\sinh 2\phi_0}{ \cosh 2\lambda_k + \cosh 2\phi_0},
\end{equation}
in agreement with Ref.~[\onlinecite{Lamacraft2004}]. It enables one to construct the bosonic 
partition sum using the relation $i\partial_{\phi_0} \ln Z_B^{(+)}(\phi_0) = {\cal F}(\phi_0)$. 
Choosing the normalization ${\cal Z}_B^{(+)}(0)=1$, it can be cast into the form
\begin{equation}
{\cal Z}^{(+)}_B(\phi_0) = \prod_{k=1}^{N'} \left(1 + t^2_k \sinh^2 \phi_0\right)^{1/2}, \quad t_k = 1/\cosh \lambda_k,
\end{equation}
from where the fermion partition sum follows as ${\cal Z}^{(+)}_F(\phi_1) = {\cal Z}^{(+)}_B(i\phi_1)$. 
With the help of basic trigonometric identities one then finds for their ratio, 
${\cal Z}^{(+)}(\phi) = {\cal Z}^{(+)}_F(\phi_1)/{\cal Z}^{(+)}_B(\phi_0)$, the result~(\ref{eq:Part_sum_Z2}).

Let us now evaluate the `minus' partition function  ${\cal Z}^{(-)}$. It was defined in sec.~\ref{sec:FT_classD}
as a response of the system to the insertion of the $\mathds{Z}_2$ gauge `flux' $\tau_1^{cc}$ on a single bond
$(0,1)$. We have also stressed that the gauge transformation $\psi^1\to \tau_1\psi^1,\bar \psi^1
\to \bar \psi^1 \tau_1$ can be used to shift such source to any other link. In particular, applying it
to all cites $s$ with $s \leq 0$ the $\tau_1$-flux can be shifted to the left infinity. This transformation
also changes the sign of the infinitesimal convergence factor in the fermionic sector of the path integral, 
$i 0\,\tau_1^{cc} \to (-i 0)\tau_1^{cc}$, which is important to keep in mind as long as one considers
an open system (a wire connected to the left/right lead). On the level of Eilenberger equation this 
swaps the left asymptotic $Q$-matrix in the ff-sector, i.e. $Q_- = - \, Q_+ = - \sigma_3^{RL} \otimes \tau_3^{cc}$.
As one can now see from Eq.~(\ref{eq:Qm_theta_D}) such boundary conditions are equivalent to setting 
$\phi_1=\pi/2$ and thereby the identity~(\ref{eq:Z_kink_D}) is proved. 

\subsection{Class DIII}

\label{app:Scattering_DIII}

Let us now extend the quasiclassical treatment introduced in the previous two sections  to the
class $\mathrm{DIII}$ system. In doing so we consider a more general model where impurity scattering
is present not only in the wire but also in the leads. It will serve us twofold purpose. First, we
will justify the general form of the grained action $S[T]$, see Eq.~\eqref{eq:action-DIII} or
\eqref{eq:actionST_DIII}. Second, we will obtain the partition functions $Z^{(\pm)}(\phi)$ in terms of Lyapunov exponents 
using the action $S$ calculated for a specific choice of matrix $T$.

To this end we consider two dots or two leads (we refer them later as `terminals'), connected by a scattering
region (`junction') which is assumed to be completely defined by its transfer matrix $\hat M$ obeying all required
symmetries specific for the class $\mathrm{DIII}$, see Sec.~\ref{sec:Scattering} for details. 
Depending on the situation, by the 'junction' we understand the wire of a length $L$ itself or just a 
contact between two dots. For a given $\hat M$ we then perform a disorder averaging 
in the terminals in the framework of SCBA. Introducing further two matrices $T_{1,2} \in \mathrm{SpO}(2|2)$,
which parametrize the Goldstone fluctuations in the terminals, and following the standard root
outlined in Appendix~\ref{app:BDI},  one arrives at the following action
\begin{equation}
\label{eq:S_T12_DIII}
S[T_1,T_2] = \frac{1}{2}{\rm str} \ln \left( 
\begin{array}{cc}
i\Sigma_0\, \tau_3^{cc} \, T(x) & - D \\
- {D}^\dagger & i\Sigma_0 T^{-1}(x)\,\tau_3^{cc}
\end{array}
\right),
\end{equation}
Here $T(x)=T_{1,2}$ depending on whether $x$ lies in the L/R terminal,
$\Sigma_0$ is the imaginary part of a SCBA self-energy ($\Sigma_0 = 0$ inside the junction), 
and the operator $D$ defined in Eq.~\eqref{DIIIgeneric}
should be renormalized by the real part of a self-energy. The precise form of SCBA equations will not be important
for the subsequent discussion. We only comment here that we imagine the SCBA scheme being performed separately for each
terminal along the route of random matrix theory (RMT) approach~\cite{Ivanov:2002}, and adding afterwards an inter-grain non-random hopping
matrix $W$ to the operator ${\cal D}$, see Eq.~\eqref{DDotHamiltonian}. By construction, the matrix $T$ enters
the low-energy action as the element of the coset space $T=T^{-1}_R T_L$, see Eq.~\eqref{eq:T_LR_DIII}. 
In other words $T \in \mathrm{SpO}(2|2)\otimes \mathrm{SpO}(2|2)/ \mathrm{SpO}(2|2) \simeq \mathrm{SpO}(2|2)$,
thus the Goldstone manifold becomes isomorphic to the single copy of the group $\mathrm{SpO}(2|2)$.

Subjecting the Hamitonian $H$ 
to the gauge transform one obtains the phase dependent action $S_\phi[T_1, T_2]$.
It is given by Eq.~(\ref{eq:S_T12_DIII}) with the phase dependent Hamiltonian $H_\phi= H' + W_\phi'$
(we refer the reader to our previous discussion in Sec.~\ref{sec:DIII}).
By virtue of the gauge invariance the flux dependence can be removed from the junction to the left terminal, 
which yields 
\begin{equation}
\label{eq:S_phi_T_DIII}
S_\phi[T_1,T_2] = \frac{1}{2}{\rm str} \ln \left( 
\begin{array}{cc}
i\Sigma_0 \tau_3^{cc} \, T_\phi(x) & - D' \\
- {D'}^\dagger & i\Sigma_0 T^{-1}_\phi(x)\, \tau_3^{cc}
\end{array}
\right),
\end{equation}
where the rotated field $T_\phi(x)$ is defined as
\begin{equation}
T_\phi(x) = e^{-i\tau_3^{cc}\phi(x)}T(x)e^{-i\tau_3^{cc}\phi(x)}, 
\end{equation}
with $\phi(x)={\rm diag}(-i\phi_0,\phi_1)_{\rm bf}$ if $x$ belongs to the left terminal and $\phi(x)=0$ otherwise.
Let us also denote by
\begin{equation}
T_1(\phi) = e^{-i\tau_3^{cc}\phi}\,T_1 e^{-i\tau_3^{cc}\phi} = T_\phi(x)\bigl|_{x \in L}
\end{equation}
the rotated configuration in the left terminal.

In fact, the action $S_\phi[T_1,T_2]$ depends only on the group element $T_{12}=T_1(\phi) T_2^{-1}$ 
but not on each of the two fields separately.  To see that we introduce the Green's function 
${\cal G}_\phi[T_1, T_2]$ such that $S_\phi = \frac{1}{2}{\rm str} \ln {\cal G}^{-1}_\phi$ 
and subject the former to a global similarity transformation,
\begin{equation}
G^{-1}_\phi = \left( \begin{array}{cc} \tau_3^{cc} & \\ & T_2  \end{array}\right)
{\cal G}^{-1}_\phi \left( \begin{array}{cc}  T^{-1}_2 & \\ & \tau_3^{cc}  \end{array}\right),
\end{equation}
which preserves the structure of the action since the rotations involved have unit super-determinant.
Explicitly, the transformed Green's function takes the form
\begin{equation}
\label{eq:G_phi_T_DIII}
G_\phi[T_{12}] = \left( 
\begin{array}{cc}
i\Sigma_0\, T_{12}(x) & - D'  \\
- D'^\dagger  & i\Sigma_0\, T_{12}^{-1}(x)
\end{array}
\right)^{-1},
\end{equation} 
with $T_{12}(x)$ being the step-like in space field defined by relations
$T_{12}(x)|_{x \in R}=\mathds{1}$ and $T_{12}(x)|_{x \in L}=T_{12}$.
The above form~\eqref{eq:G_phi_T_DIII} of the Green's function 
is valid in the chiral basis where
the parity operator $P=\sigma_3$. It is advantageous to rewrite $G_\phi$ in the basis independent form. Introducing
the (step-like) element $\hat\Omega(x) = -i \ln T_{12}(x)$ from the Lie superalgebra $\mathfrak{g}=\mathfrak{gl}(2|2)$
one obtains
\begin{align}
\label{eq:T_12_Sigma_DIII}
&G_\phi[T_{12}] =\left( \hat \Sigma(x) - H\right)^{-1},\nonumber \\
&\hat\Sigma(x) = i\Sigma_0\exp\{ 2 i P  \otimes \hat\Omega(x)\}.
\end{align}
We have used here that $P^2 = \mathds{1}$ and matrices $P$ and $\Omega$ act in different subspaces. 

To evaluate the action $S_\phi[T_{12}] = \frac{1}{2} \ln {\rm Sdet}\, G_\phi[T_{12}]$ 
we introduce the auxiliary parameter $t \in [0,1]$ and define the field 
$T_{12}^t(x) = e^{i t \hat\Omega(x)}$. 
By rescaling the algebra element $\Omega$ in Eq.~(\ref{eq:T_12_Sigma_DIII})
in the same way one obtains the $t$-dependent Green's function $G_\phi[T_{12}^t]$ and the
action $S_\phi[T_{12}^t]$, where $T_{12}^t= T_{12}^t(x)|_{x\in L}$ is the $T$-field in the 
left terminal. Our subsequent strategy to find this new action will be the same as for other symmetry classes. 
First of all, using the quasiclassical approach, we find
the average $t$-dependent symmetry current ${\cal I}_\Omega(t)$, and 
later on with the help of this current reconstruct the action. 
To this end we put the gauge source on the 
link $0\leftrightarrow 1$ (it can be thought of as the boundary between the left terminal and the junction)
by changing the corresponding hopping matrix,
\begin{equation} 
W' \to W'_t =\left( \begin{array}{cc} &    e^{i t\hat\Omega} w'  \\ 
  w'^\dagger e^{i t\hat\Omega} & \end{array} \right),
\end{equation}
where $\hat \Omega = \hat \Omega(x)|_{x\in L} = -i\ln T_{21}$ is the `angle' in the left terminal. We use it to define the average symmetry current according to the relation
\begin{equation}
\label{eq:IP_DIII_lattice}
{\cal I}_\Omega(t): =  i\partial_t S_\phi[T_{12}^t] = 
-\frac{i}{2}{\rm str}\Bigl( (\partial_t H_t)\bigl|_{t=0}\,G_\phi[T_{12}^t]\Bigr),
\end{equation}
with $H_t = H' + W'_t$. The first equality here is a definition and the second one 
follows from the gauge invariance.

Following the logic of the quasiclassical approach let us now linearize the Hamitonian
in the terminals around zero energy, $H \to {\cal H}_{L,R}$, 
and reduce the equation of motion for the Green's function $G_\phi[T_{12}^t]$ to the Eilenberger equation. 
In the Majorana basis one has ${\cal H}_{L,R} = - i v \sigma_3^{RL} \otimes \mathds{1}^{cc}\, \partial_x$, 
and the definition of Eilenberger $Q$-function is given by Eq.~\eqref{eq:Qx_def}.
The Eilenberger equation~\eqref{eq:Eil_eq_classD} itself is get modified because of the presence of 
the self-energy in the terminals and takes the form
\begin{equation}
\label{eq:Eil_eq_classDIII}
 i \partial_x Q (x) + i (\Sigma_0/v)\, \left[  \hat\Xi(x), Q(x) \right] = 0, 
\end{equation}
where ${\cal P} =  P \otimes \sigma_1^{RL}$ is the parity operator and  
\begin{equation}
\hat \Xi(x) = \sigma_3^{RL}\, e^{2 i t {\cal P} \otimes \hat\Omega(x)} = 
e^{- i t {\cal P} \otimes \hat\Omega(x) } \sigma_3^{RL} e^{ i t  {\cal P} \otimes \hat\Omega(x)}.
\end{equation}
The last representation here is valid since $\{{\cal P},\sigma_3^{RL} \}_+ =0$.
The presence of $\sigma_3^{RL}$ Pauli matrix in the self-energy $\hat\Xi$ stems from the original definition of 
the $Q$-matrix~\eqref{eq:Qx_def} and leads to the normalization $\hat \Xi^2(x) = \mathds{1}$. 
It is clear that the matrix $\hat\Xi$ will fix the $t$-dependent boundary conditions 
in the terminals for such Eilenberger equation, 
\begin{equation}
\label{eq:Q_Omega}
Q_-(t)  = e^{-i t {\cal P} \otimes \hat\Omega } \sigma_3^{RL} e^{ i t {\cal P} \otimes \hat\Omega},
\qquad Q_+ = \sigma_3^{RL}.
\end{equation}
With this understanding we proceed further with the evaluation of a symmetry current ${\cal I}_\Omega(t)$
in terms of the field $T_{12}^t$ and the Lyapunov exponents $\lambda_k$.
The quasiclassical approximation of the lattice representation~(\ref{eq:IP_DIII_lattice}) for this  current has the form
\begin{equation}
\label{eq:jP_Omega}
{\cal I}_\Omega(t) = -\frac{1}{2}{\rm str}\bigl( ({\cal P}\otimes\hat\Omega)\, Q_L \bigr),
\end{equation} 
with $Q_L$ being the Green's function~\eqref{eq:QL_M} right at the boundary of the left terminal and the junction. 
We note that the SCBA self-energy $\Sigma_0$ does not enter into $Q_L$ and $Q_R$. 
In analogy with $\mathds{Z}$-class calculations we observe that the configuration $Q_-(t)$
admits the factorization $Q_-(t) = {\cal R}\, \tilde Q_-(t) {\cal R}$ where
$\tilde Q_-(t) = \sigma_3^{\rm RL} e^{2 i (\sigma_1^{RL}\otimes \hat\Omega) t }$ and 
${\cal R} = {\rm diag}(1, \hat P)_{\rm RL}$.
The transfer matrix $M$ can be put in the same form if one takes into account its  
DMPK decomposition~(\ref{eq:M_DMPK}) and the class $\mathrm{DIII}$ symmetries of the rotation matrices,
cf. Sec.~\ref{sec:Scattering} for these details. We thus write
$M={\cal R} \tilde M {\cal R}$ where $\tilde M = V e^{ \lambda P \sigma_1^{RL}} U'$. 
We finally note that the symmetry current generator in Eq.~(\ref{eq:jP_Omega}) can be represented as  
${\cal P}\otimes\hat\Omega = P \otimes \sigma_1^{\rm RL} \otimes\hat\Omega 
= {\cal R}  (\sigma_1^{\rm RL} \otimes\hat\Omega ) {\cal R}$. Thereby we are able to write
${\cal I}_\Omega(t) =  \frac{1}{2}{\rm str}\bigl( (\sigma_1^{RL}\otimes\hat\Omega)\, \tilde Q_L \bigr)$,
where $\tilde Q_L$ has to be found using Eq.~(\ref{eq:QL_M}),
with $M$ being changed to $\tilde M$ and $Q_-(\theta)$ being changed to $\tilde Q_-(t)$. 
Obviously, the orthogonal rotations $V$ and $U'$ do not enter the final result for the current. 
Proceeding in basis where $P$ is diagonal we obtain
\begin{equation}
\label{eq:jOmega_t_DIII}
{\cal I}_\Omega(t) =   \frac{1}{2}\sum_{\sigma=\pm}
\sum_{k=1}^{N'/2}{\rm str} \Bigl( \hat\Omega \tanh\left(\sigma\lambda_k-i\hat\Omega t\right)\Bigr).
\end{equation}
The sum over index $\sigma$ is due to $\pm 1$ eigenvalues of the parity operator $P$ and the sum over $k$
extends over the set of Lyapunov exponents without taking into account their Kramers degeneracy. 
Using trigonometry this result can be cast into the form
\begin{equation}
{\cal I}_\Omega(t) =   -\frac{1}{2}\sum_{k=1}^{N'} {\rm str} \left( 
\frac{i\hat\Omega \sin(2\hat\Omega t)}{\cosh 2\lambda_k + \cos(2\hat\Omega t)}
\right),
\end{equation}
which is of the same type as our preceding class $\mathrm{D}$ result~(\ref{eq:F_phi0_D}).
From here the action is found via relation $S_\phi[T_{12}] = -i\int_0^1 dt\, {\cal I}_\Omega(t)$ that yields
\begin{equation}
S_\phi[T_{12}] = \frac{1}{4}\sum_{k=1}^{N'}{\rm str}\ln \left( \openone - t_k^2 \sin^2 \hat\Omega  \right).
\end{equation}
With the help of ansatz 
\begin{equation*}
4 \sin^2 \hat\Omega = 2( 1 - \cos 2\hat \Omega)= 2 - T_1(\phi)T_2^{-1} - T_2 T_1^{-1}(\phi)
\end{equation*}
the final form of the action reads
\begin{align}
&S_\phi[T_{12}]= \\
&\quad{1\over 4} \sum\limits_{k=1}^{N'}  \mathrm{str}\ln \left[\openone +
\frac{t_k^2}{4} \left(T_1(\phi)T_2^{-1} + T_2 T_1^{-1}(\phi)-2\right)\right], \nonumber
\end{align} 
cf. Eq.~(\ref{eq:action-DIII}).

With this result the evaluation of the partition functions $Z^\pm = e^{-S^\pm}$ becomes particularly simple.
Setting $T_1 = T_2 = \openone$ and keeping a non-vanishing angle $\phi$ 
one obtains the relation~\eqref{eq:Part_sum_Z2}. On other hand,
the choice $T_1={\cal P}^\mathrm{b}  + {\cal P}^\mathrm{f} \tau_1^{cc}$ and $T_2 = \openone$ gives the desired result~\eqref{eq:chi_L_Z2}
for the kink's action.

\bibliography{library,Remote}
\end{document}